\documentclass[12pt]{article}
\pdfoutput=1
\usepackage{amssymb}
\usepackage{amsmath}
\usepackage{graphicx}

\newcommand{\be}{\begin{equation}}
\newcommand{\ee}{\end{equation}}
\newcommand{\bea}{\begin{eqnarray}}
\newcommand{\eea}{\end{eqnarray}}

\begin{document}

\title{\textbf{FRW Cosmology with \\Non-positively Defined Higgs Potentials}}

\author{I.~Ya.~Aref'eva$^1$\footnote{arefeva@mi.ras.ru}, N.~V.~Bulatov$^2$\footnote{nick\_bulatov@mail.ru},
R.~V.~Gorbachev$^1$\footnote{rgorbachev@mi.ras.ru}\\[2.7mm]
\small{${}^1$Steklov Mathematical Institute, Russian Academy of Sciences,}\\
\small{Gubkina st. 8, 119991, Moscow, Russia}\\
\small{${}^2$Department of Quantum Statistic and Field Theory, Faculty of Physics}\\
\small{Moscow State University, Leninskie Gory 1, 119991, Moscow, Russia}
}

\maketitle

\begin{abstract}

We discuss the classical aspects of dynamics of scalar models with non-positive Higgs potentials
in the FRW cosmology. These models appear as effective local models in non-local models related with string field theories. After a suitable field
redefinition these models have the form of local Higgs models with a negative extra
cosmological term and  the  total Higgs potential is non-positively defined and has rather small coupling constant. The non-positivity of the potential leads to the fact that on some stage of evolution the expansion mode gives place to the mode of contraction, due to that the stage of reheating is absent. In these models the hard regime of inflation  gives place to inflation near the hill top and  the area of the slow roll inflation is very
small. Meanwhile one can obtain enough e-foldings before the contraction to make the
model under consideration admissible to describe  inflation.
\end{abstract}
\newpage

%%%%%%%%%%%%%%%%%%%%%%%%%%%%%%%%%%%%%%%
\section{Introductory remarks}
%%%%%%%%%%%%%%%%%%%%%%%%%%%%%%%%%%%%%%%

Scalar fields play a central role in current models of the early
Universe -- the majority of present models of inflation requires an
introduction of an additional scalar -- the "inflaton"
\cite{Linde,Mbook,Weinberg,Rubakov,inflation}. Though the
mass of the inflaton  is not fixed, typical considerations deal with
a heavy scalar field with mass $~ 10^{13}$
 GeV \cite{Linde-1}. The interaction of the inflaton (self-interacting and interacting with matter fields)
  is not fixed too, moreover,  in typical examples the self-interaction potential
energy density of the inflaton  is undiluted by the expansion
of the universe and it acts as an effective cosmological constant.
However the detailed evolution depends on the specific form of the potential
$V$. In the last years, in particular,  Higgs inflation has attracted a lot of attention
\cite{HI}-\cite{HI5}. The kinetic term usually has the standard form, however models with nonstandard
kinetic terms also have attracted attention in the literature
\cite{k-essence}-\cite{IA05}.
The present paper  deals with models mainly related with a nonlocal kinetic term
of a special form \cite{IA05}-\cite{GK:2010}.

The distinguish property of these models is their string field (SFT)
origin \cite{ABGKM}. The original motivation to deal with this type of
nonlocal cosmological models was related with the dark energy problems \cite{IA05}.
The relevance of this type of models to the early Universe problems has been pointed out
in \cite{Lidsey,BarCline}.  In this case the scalar field is the tachyon of the NSR
fermion string and the model has the form of a nonlocal Higgs model.
These nonlocal factors  produce essential changes in cosmology. Due to the
effect of stretching the potential with the kinetic energy advocated by
 Lidsey\cite {Lidsey}, Barnaby and Cline\cite{BarCline} nonlocality destroys
the relation  between the coupling constant in the  potential, the mass term and the value of the vacuum energy
(cosmological constant) and produces the total Higgs potential with the negative extra
cosmological term. This total Higgs potential is non-positively defined.

We discuss the classical aspect of dynamics of scalar models with non-positive Higgs
potentials in the FRW cosmology. As it has been mentioned above these models appear as effective local models in nonlocal models
of a special form related with string field theory.

Since nonlocality can produce an effective local theory with rather small coupling constant
some stages of evolution can be described by the free tachyon approximation.
Due to this reason  we start from  consideration of dynamics of the free tachyon in
the FRW  metric. We show that generally speaking there are 4 stages of evolution of the tachyon
in the FRW metric: reaching the perturbative vacuum from the past cosmological singularity;
rolling from the perturbative vacuum; transition from inflation to contraction and in the ending
contraction to the  future cosmological singularity.
We present approximations suitable to study evolution  near the past cosmological
singularity,  evolution near the top of the hill and
evolution near the transition from expansion to contraction.

As to the case of the quartic
potential the situation depends on the constant relations.
For very small coupling constant  the hard regime of inflaton gives place to inflation
near the hill top  making the area of the slow-roll inflation very small.
 Meanwhile the model can perform  enough e-foldings to make the model under consideration
 admissible do describe  inflation.

The plan of the paper is the following. We start, Sect. 2.1, from the short exposition why the nonlocality of the SFT type produces
non-positively defined potentials.  In Sect. 2.2  we  remind the known
facts about massive scalar field and the positively defined Higgs potential.

In Sect.3 we consider dynamics of the free tachyon in
the FRW  metric. First, we show the presence of the cosmological singularity \cite{BGXZ}.
 Then we show that starting from the top of the hill the
tachyon  first evolves   as an inflaton, i.e. it keeps the inflation regime.
 After performing a finite number of e-foldings $N_{max}$ the tachyon reaches
  the boundary of the forbidden domain at some point $(\phi_{max},\dot\phi_{max})$
  of the phase diagram.   At this point
  the Hubble $H$ becomes equal to zero and the tachyon
  continues to evolve  with negative $H$, i.e. the tachyon becomes in fact  a ``contracton".
  We study the phase portrait of the system and conclude that near the top
  we can use the de Sitter approximation and, moreover, we can restrict ourself to
  consideration of  one mode of the tachyon evolution, so called $C_1$-mode, and find
  corrections to the dS approximation. This regime has been previously studied in
  \cite{Hill-top,top-hill}. The $C_1$-mode is a more viable mode and in
  the case of the zero cosmological constant describes exponentially increasing
  solution at $t\to \infty$.
To study  dynamics of the system near the boundary of the accessible
area in more details we use a special parametrization, that is in
fact a generalization of a parameterizations that one has used
previously to study dynamics of the inflaton with  a positively
defined potential in the FRW metric \cite{BGXZ} and refs.
therein. Using this parametrization we find asymptotic behavior near the boundary of the accessible region.

In Sect.4 we consider dynamics of the  Higgs field in
the FRW  metric in the presence of an extra  negative cosmological
constant. This constant makes the total potential non-positively
defined and produces the forbidden area in the phase space of the
Higgs field.

In conclusion we briefly summarize   new cosmological features for
Higgs potentials with extra negative cosmological constant. The
perturbation analysis will be the subject of forgoing paper.

\setcounter{equation}{0}
%%%%%%%%%%%%%%%%%%%%%%%%%%%%%%%%%
 \section{Setup}
%%%%%%%%%%%%%%%%%%%%%%%%%%%%%%%%
\subsection{Non-positively defined potentials as a consequence  of nonlocality}
\label{Higgs-nonlocal-model}
%%%%%%%%%%%%%%%%%%%%%%%%%%%%%%%%
 \setcounter{equation}{0}

 In this section we make few remarks why scalar matter nonlocality leads to
 essential changes of properties of  corresponding cosmological models in comparison with those
 of pure local cosmological models (we mean early cosmological models).
 These changes appear due to an effective stretch  of the kinetic part of the matter Lagrangian.

We deal with the nonlocal action originated from SFT~\cite{IA05}
\begin{equation}
\label{SFT-action} S =\int d^4x \sqrt{-g} \left[ \frac{m_{p}^2}{2} R +
\left( \frac{1}{2} \phi (\xi ^2\Box+\mu^2) e^{-\lambda\Box }\phi
-V(\phi)-\Lambda \right) \right]   ,
\end{equation} where $\Box $ is  the
D'Alembertian operator, $\Box \equiv \frac{1}{\sqrt{-g}}
\partial^{\mu} (\sqrt{-g} \partial_{\mu} )$, $\phi$ is a dimensionless
scalar field and  all constants are dimensionless.

The spatially isotropic dynamics of this field in the FRW metric, with the
interval:
\begin{equation*}
 ds^2={}-dt^2+a^2(t)\left(dx_1^2+dx_2^2+dx_3^2\right),
\end{equation*}
is given by nonlocal equation of motion \bea \label{nl-nl-H} e^{\lambda
(\partial^2+3H\partial)}\left(\xi ^2(\partial^2+3H\partial)
-\mu^2\right)\phi(t)&=&-V'_{\phi},\\
\label{cal-E} 3H^2&=&8\pi G\,{\cal E},\eea
where the Hubble parameter
$H=\dot a/a$, ${\cal E}$ is a sum of nonlocal modified kinetic and
potential energies, that are
 explicit functionals of $\phi$ and $\dot\phi$.
${\cal E}$  is given by \cite{IA05}-\cite{LJ:2008}
(see also \cite{MZ}-\cite{AJK})
\be \label{TEV}
{\cal E}=2E_{00}-g_{00}\left(g^{\rho\sigma}E_{\rho\sigma}+W\right),\nonumber
\ee
where
\bea \label{E-munu} E_{00}&\equiv&\frac{1}{2}\sum_{n=1}^\infty
\frac{\lambda^n}{n!}\sum_{l=0}^{n-1}\partial_0\Box_g^l\phi\partial_0\Box_g^{n-1-l}\phi,\nonumber\\
\label{W} W&\equiv&\frac{1}{2}\sum_{n=2}^\infty
\frac{\lambda^n}{n!}\sum_{l=1}^{n-1}\Box_g^l\phi\Box_g^{n-l}\phi+\frac{f_0}{2}\phi^2+V(\phi).\nonumber
\eea
In the main consideration we deal with
$V(\phi)=\epsilon \phi^4$ and $f_0=-\mu^2$ (the tachyon case).

In comparison with the local scalar field dynamics in the FRW metric
\bea
\label{e.o.m.} \ddot{\phi}+3H\dot{\phi}&=&-V'_{\phi},\\
\label{H2} 3H^2&=&8\pi G (\,\frac12 \dot{\phi}^2+V(\phi)),\eea
 there
are two essential differences: there is a nonlocal kinetic term in the
equation for scalar field (\ref{nl-nl-H}) as well as the form of the
energy is different. Generally
speaking the system (\ref{nl-nl-H}), (\ref{cal-E}) is a complicated
system of nonlocal equations and its study is a rather nontrivial
mathematical problem. There are several approaches to study this
system. Let us first  mention  the results of study of equation
(\ref{nl-nl-H}) in the flat metric, i.e. equation (\ref{nl-nl-H}) with
$H=0$. In the flat metric for the quartic potential it is known that for $\xi<\xi_c$ there exists an
interpolating solution of (\ref{nl-nl-H}) between two vacua\footnote{There is no
interpolating solution for the case of cubic potential \cite{MZ}.}.
There is a numerical evidence for existence of oscillator solutions for
$\xi>\xi_c$. In the LHS of  (\ref{nl-nl-H}) there is a nonlocal
term that is related with the diffusion equation. This relation
provides rather interesting physical consequences \cite{AV11}.

The case of the FRW metric in the context of the DE problem has been
studied in \cite{IA05,AI-KAS:2006,Joukovskaya:2007nq,AJV:2007b}. In the context of the inflation problem this
system (especially in the case $\xi=0$) has been studied in
\cite{Lidsey,BarCline}. In \cite{Lidsey} it has been pointed out that
nonlocal operators lead to the effects that can be quantified in terms of a local
field theory with a potential whose curvature around the turning point
is strongly suppressed.

In the case of a quadratic potential a cosmological model with one nonlocal
scalar field can be presented as a model with local scalar fields and
quadratic potentials~\cite{AI-IV-NEC,AK:2007,AJV,AJV:2007b,MN:2008_AIP,Vernov,KoshVe:2010}.
For this model  it has been observed \cite{Lidsey,BarCline} that in the regime suitable for the
Early Universe the effect of the effective stretch of the kinetic terms
takes place (see also \cite{AJV,AJV:2007b}).
 One can generalize  this result to a  nonlinear case and assume that
a nonlocal nonlinear model  can be described by  an effective  local nonlinear theory
with the following energy density
\be \label{cal-E-approx} {\cal E}\approx
e^{\lambda \Omega^2}\left(\frac12 \dot \phi^2-\frac 12
\mu^{2}\phi^2\right) +\frac 14 \epsilon\phi^4 +\Lambda,\ee and the
tachyon dynamics equation becomes \be \label{nor-nl-H}
\partial ^2 \phi+
3H\,\partial  \phi -\mu^{ 2}\phi
=-\epsilon e^{-\lambda \Omega^2}\phi^3,\ee
where
\be
H=\frac{1}{\sqrt{3}\,m_p}\sqrt{\,e^{\lambda \Omega^2}
\left( \frac12 \dot \phi^2-\frac 12 \mu^{2} \phi^2\right)
+\frac 14 \epsilon\phi^4+\Lambda\,}\ee
and we take $\xi =1$ for simplicity. $\Omega$ is a  modified ``frequency" of the asymptotic
expansion  for equation (\ref{nor-nl-H}) in the flat case.
In the first approximation $\Omega=\mu$.

Or in other words, one can say that the higher powers in $\Box$ cause
the factor $e^{\lambda \Omega^2}$,
where $\Omega^2$  in the first approximation is just equal to $\mu^2$.
Therefore, the corresponding effective action becomes
\be
\label{SFT-action-eff}
S =\int d^4x \sqrt{-g} \left[ \frac{m_{p}^2}{2} R +
\left( \frac{1}{2}
\phi (\Box+\mu^2) e^{\lambda\Omega^2 }\phi -\frac{\epsilon \phi^4}{4} -\Lambda\right) \right].
\ee
In term of the canonical field $\varphi\equiv e^{\lambda\Omega^2/2 }\phi$
the action (\ref{SFT-action-eff}) becomes
\be
\label{SFT-action-eff-var}
S =\int d^4x \sqrt{-g} \left[ \frac{m_{p}^2}{2} R +
\left( \frac{1}{2}
\varphi (\Box+\mu^2) \varphi -\frac{\epsilon}{4} e^{-2\lambda\Omega^2 } \varphi^4 -\Lambda\right) \right]   .
\ee
Just this form is responsible for equation (\ref{nor-nl-H}) for
$V(\phi)=-\frac{\epsilon \phi^4}{4}$.
This equation is nothing more than equation for the Higgs
scalar field considered as an inflaton. There is only one difference, that in fact,
 as we will see,  is very important.
The potential for the Higgs model being considered usually in the context of the
cosmological applications is
\be
\label{Higgs}
V(\phi)=\frac{\epsilon}{4} (\phi^2-a^2)^2,\ee
i.e. the cosmological constant $\Lambda=\frac{\epsilon}{4} a^4$
is such that the potential is the positively defined potential. Let us remind that
in the SFT inspired nonlocal model the value of the  cosmological constant
is also fixed by the Sen conjecture \cite{ABGKM,ABKM,Sen} that exactly puts the potential energy
of the system in the nontrivial vacua equal to zero
\be
V(\phi)=-\frac 12 \mu^{2}\phi^2
+\frac 14 \epsilon\phi^4 +\Lambda=\frac 14 \epsilon\left(\phi^2-\frac{\mu^2}{\epsilon}\right)^2,
\ee
that gives us he value of $\Lambda=\frac{\mu^4}{4\epsilon}$.
The appearance of the extra factor
$e^{\lambda \Omega^2}$ destroys this positivity
\be
\label{Omega-shift}
V_{EL}(\phi)=-\frac 12 \mu^{2}e^{\lambda \Omega^2} \phi^2
+\frac 14 \epsilon\phi^4+\frac{\mu^4}{4\epsilon}=
\frac 14 \epsilon\left(\phi^2-\frac{\mu^{2}}{\epsilon}e^{\lambda \Omega^2}\right)^2+
\frac{\mu^4}{4\epsilon}\left(1-e^{2\lambda \Omega^2}\right),
\ee
(here "$EL$" means "effective local model"). Since $\lambda >0$, we have to deal with non-positively defined potential.

%%%%%%%%%%%%%%%%%%%%%%%%%%%%%%%%%%%%%%%%
\subsection{Phase portraits for positively defined potentials}
\setcounter{equation}{0}
%%%%%%%%%%%%%%%%%%%%%%%%%%%%%%%%%%%%%%%%

\begin{figure}[!h]
\centering
\includegraphics[width=5cm]{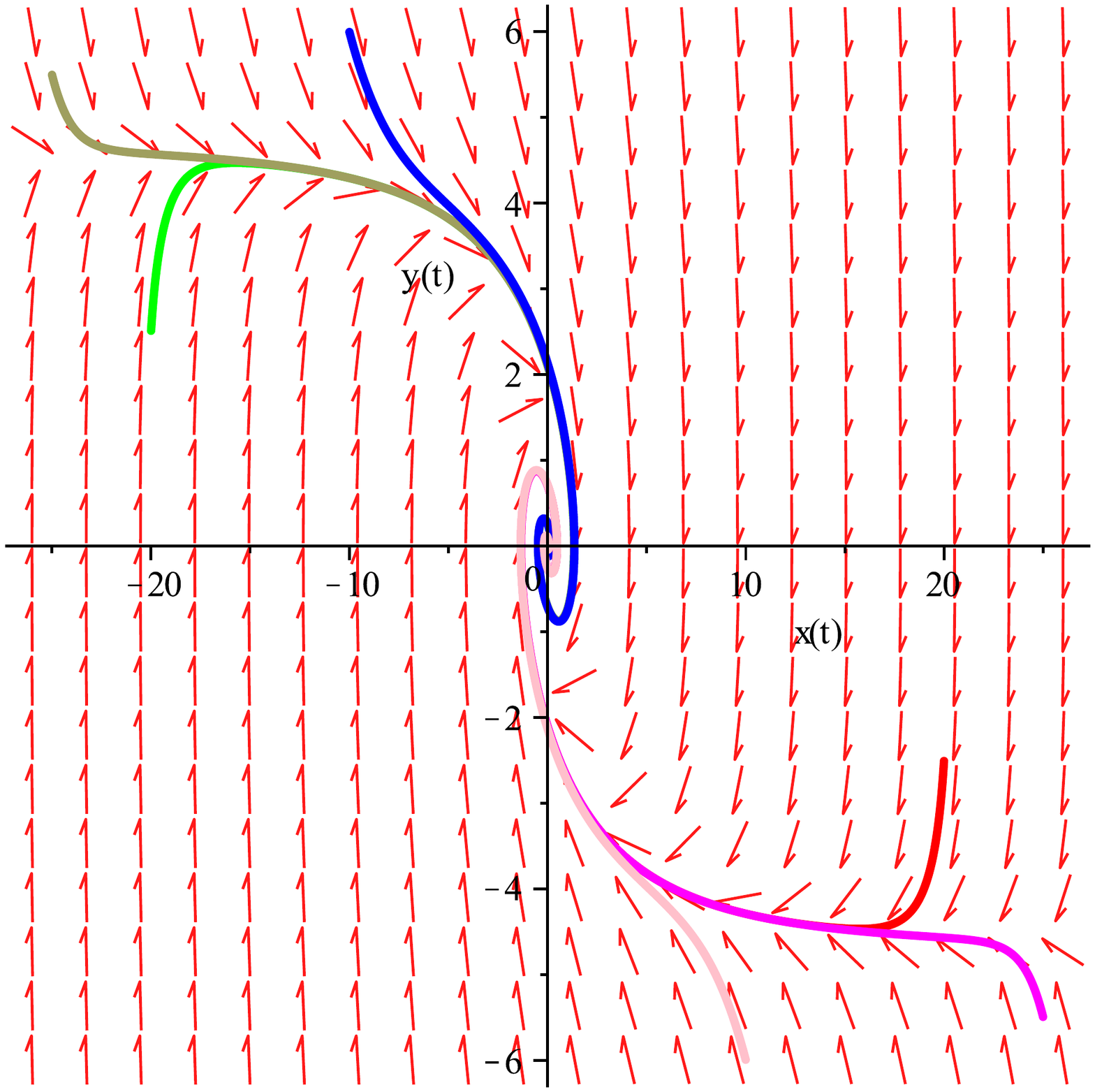}$A.~~~~~$
\includegraphics[width=5cm]{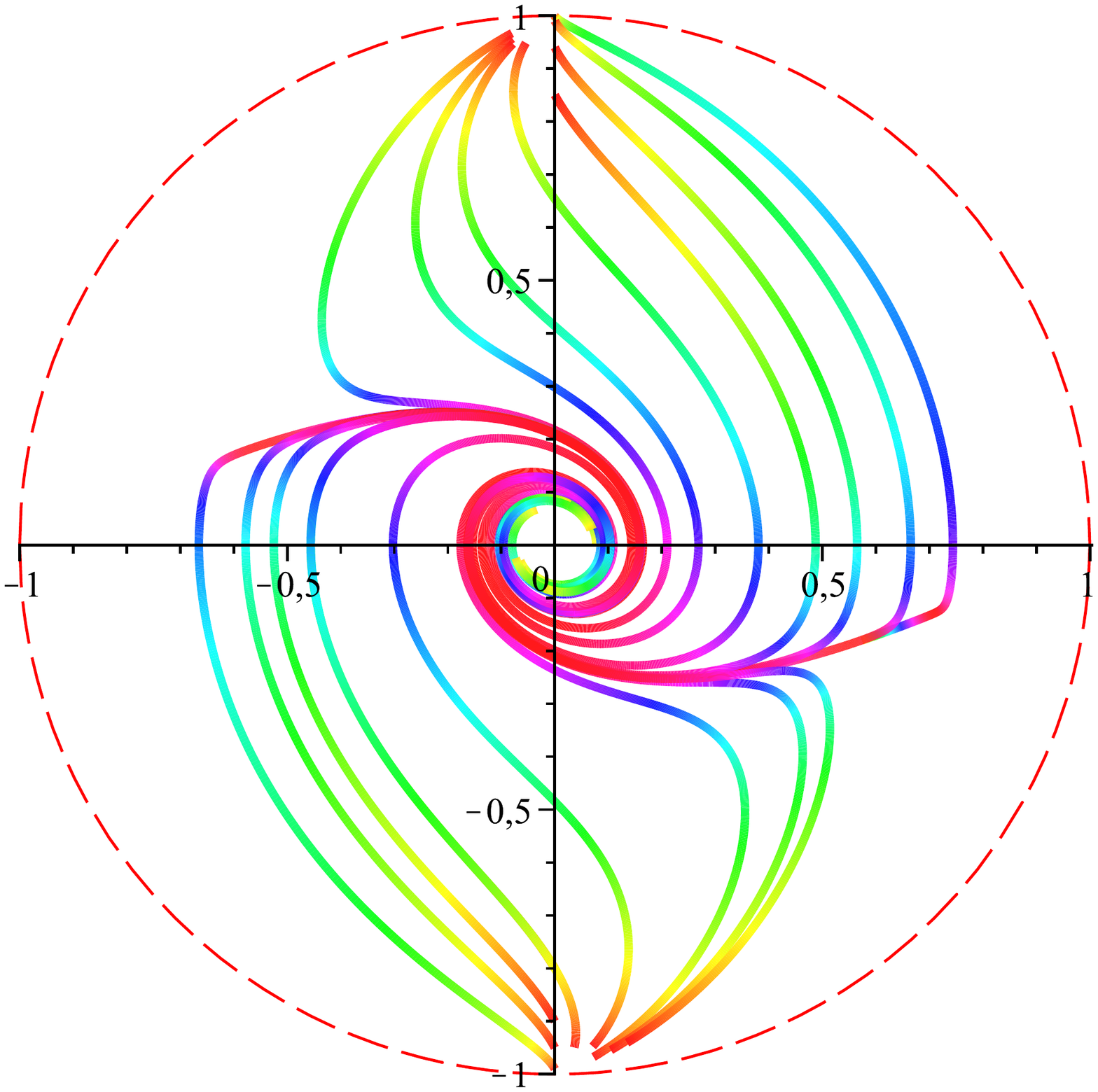}$B.~~~~~~~~~$
\caption{ (color online) A. The phase portrait of  equation
(\ref{EOM-21}). B. Inflaton flows in compactified variables $\psi$ vs $\rho$
are given by the phase portrait of  equations (\ref{rho-s-2}) and (\ref{psi-s-2}).
}
\label{pp-positive}
\end{figure}

Cosmological models with one scalar field with
positively defined potential
are well studied models, see for example \cite{Mbook,BGXZ,Copellan,Macorra,Boyanovsky}.
Let us remind the main features of the phase portraits of these models.
The phase portrait for a free massive field that is a solution of \be
\label{EOM-21}
 \ddot{\phi}+3\sqrt{\frac{8\pi G}{3} \left(\,\frac12\, \dot{\phi}^2+\frac{m^2}{2}\phi^2\right)}\,\dot{\phi}
 =-m^2\phi,\ee
  is presented in Fig.\ref{pp-positive}.A. The same phase portrait is presented in Fig.\ref{pp-positive}.B
in  compactified variables. The advantage of this picture is that we see the existence
of the cosmological singularity.
The phase portrait for the  Higgs field with positively defined potential
  \be
\label{EOM-H}
 \ddot{\phi}+3\sqrt{\frac{8\pi G}{3}\,\left(\frac12\, \dot{\phi}^2+\frac{\epsilon}{4}(\phi^2-a^2)^2\right)}\,\dot{\phi}
 =-\epsilon\phi^3+\epsilon a^2\phi,\ee
 is presented in Fig.2.A.
\begin{figure}[!h]
\centering
\includegraphics[width=5cm]{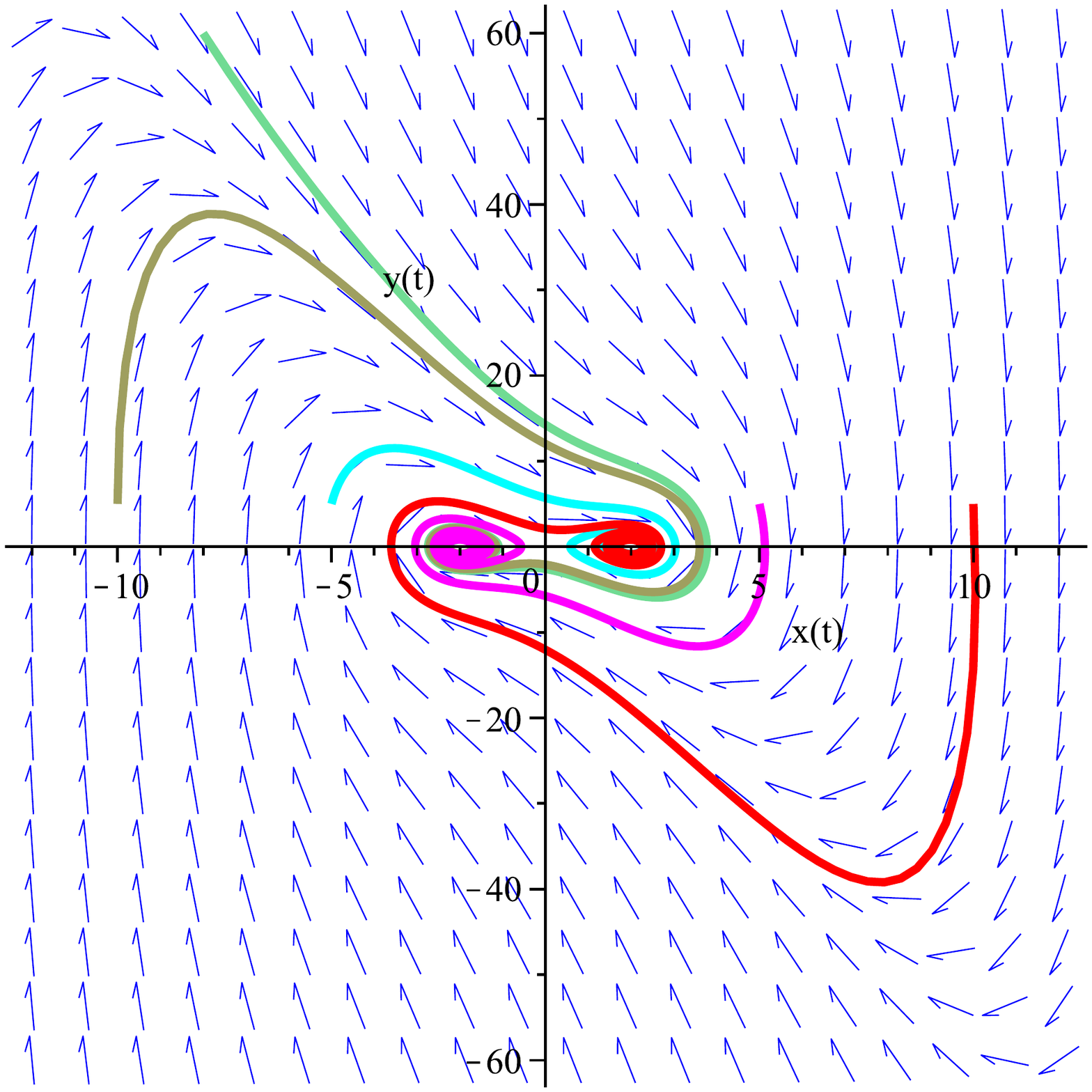}$A.~~~~~$
\begin{picture}(100,140)
\put(-5,-1){
\includegraphics[width=6cm]{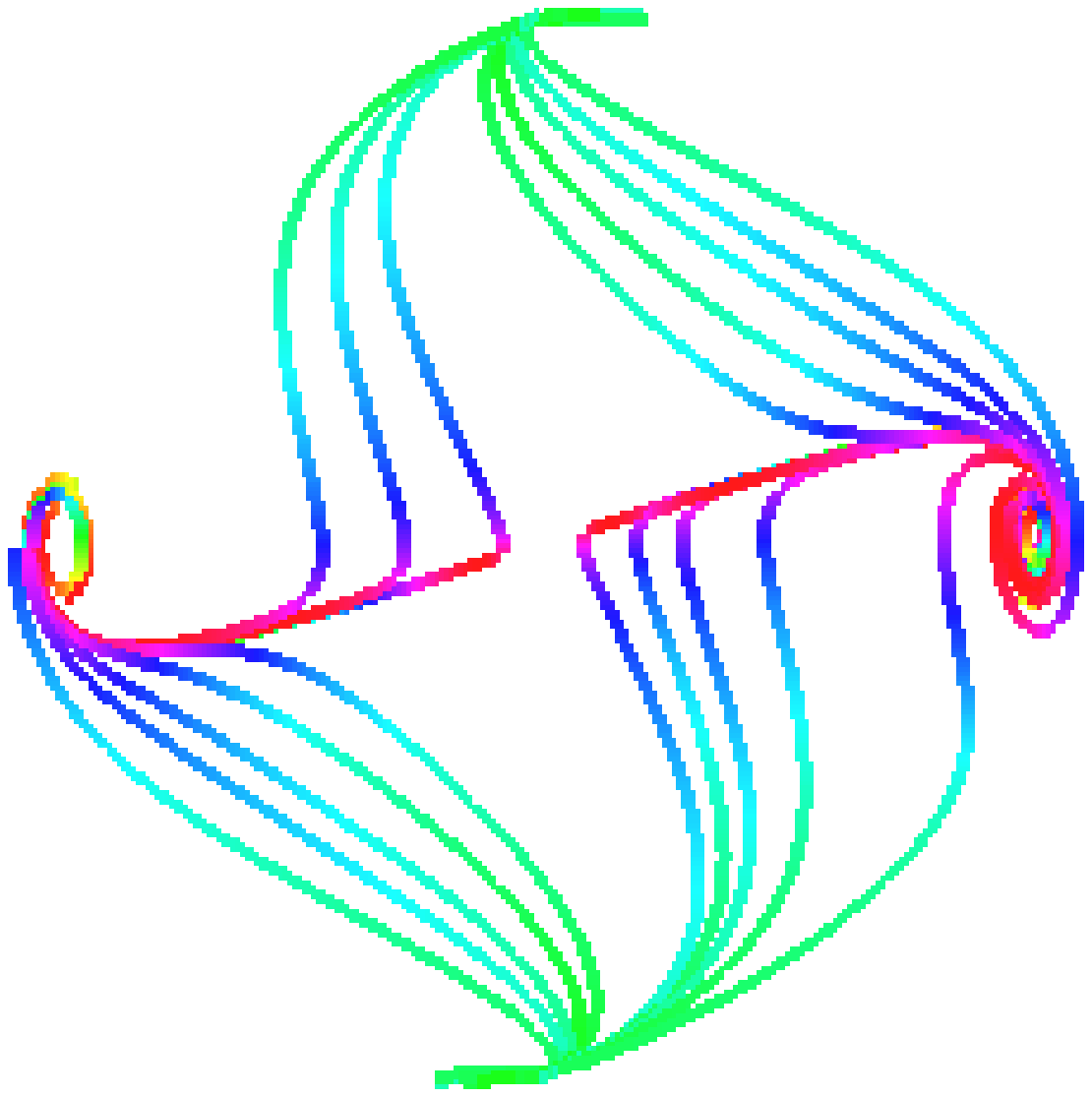}
}
\end{picture}$B.$
\caption{ (color online)
A. The phase portrait of equation (\ref{EOM-H}). B. The Phase portrait of the system of equations (\ref{rho-s-H}), (\ref{rho-s-p}).}
\label{pp-positive-m}
\end{figure}

Comparing the inflaton flows in Fig.\ref{pp-positive}.A, Fig.2.A, we see that there is  no big difference between the phase
 portrait presenting the chaotic inflation (\ref{EOM-21}) and the phase portrait of eq. (\ref{EOM-H}).
This is in accordance with the general statement about a weak dependence of the inflaton flow
 on  the potentials. Only explicit formulae describing the slow-roll regime as well as
 the oscillation regime  are  different.
 In all these cases there are 3 stages (eras) of evolution. Just after cosmological singularity the ultra-hard regime starts. Then in the case of
  any positive/negative  initial position $\phi(0)>0$/$\phi(0)<0$  the inflaton
reaches very fast the ``right''/``left''  attractor. Then the inflaton moves along
the ``right''/``left'' slow-roll line. In particular, in Fig.\ref{pp-positive}.A there is almost horizontal line that corresponds to the ``right'' slow-roll line and it  is seen in the plot as a common part
of the red, pink and magenta lines. Then oscillations near the point
$\phi=0$, $\dot \phi=0$ Fig.\ref{pp-positive}.A, or near one of two nonperturbative vacua Fig.2.A start.

It is well known that it is convenient to present the phase portraits in the compactified variables
\cite{BGXZ} ($G=\frac{1}{m^2_p}$)
\bea
\label{phi-BGZX}
\phi&=&\frac{3m_{p}}{\sqrt{12\pi}}\frac{\rho}{1-\rho}\sin \theta \cos\psi,\\
\label{dot-phi-BGZX}\dot\phi&=&\frac{3m m_{p}}{\sqrt{12\pi}}\frac{\rho}{1-\rho}\sin \theta \sin\psi,\\
\label{H-BGZX}H&=&m\frac{\rho}{1-\rho}\cos \theta.
\eea
With the help of these variables we can rewrite equations (\ref{EOM-21}) and $\dot{H}=-\frac{4\pi\dot{\phi}^2}{m_p^2}$ in the following form (here we use the time variable  $\sigma$ such that $\frac{d\sigma}{d\tau}=\frac{1}{1-\rho}$ where $\tau=m t$)
\bea
\label{rho-s-2}
\rho_\sigma&=&-\frac{3\rho^2(1-\rho)}{\sqrt{2}} \sin ^2\psi,\\
\label{psi-s-2}\psi_\sigma&=&-(1-\rho)-\frac{3\rho}{2\sqrt{2}} \sin 2\psi,\eea
and
equation for the Higgs inflaton (\ref{EOM-H})  has the form
($g=\frac{3m^2_p}{4\pi a^2}$)
\bea \label{rho-s-H}
\rho_{\sigma}&=&(1-\rho)^2\rho\sin^2\theta\sin2\psi-\rho^2(1-\rho)\cos\theta\left(1+4\sin^2\theta\sin^2\psi\right)\\
&+&\frac{1}{2g}(1-\rho)^3\cos\theta+g\rho^3\sin^4\theta\cos^3\psi\left(\frac{\rho}{1-\rho}\cos\psi\cos\theta-\sin\psi\right),\nonumber\\
\psi_{\sigma}&=&-\frac32\rho\cos\theta\sin2\psi+(1-\rho)\cos2\psi-g\frac{\rho^2}{1-\rho}\sin^2\theta\cos^4\psi. \label{rho-s-p}\\
\theta_{\sigma}&=&(1-\rho)\sin2\psi\sin2\theta-\frac{1}{2g}\frac{(1-\rho)^2}{\rho}\sin\theta
+\rho\sin\theta\left[1-3\sin^2\psi\right.\nonumber\\
&+&\left.2\sin^2\psi\sin^2\theta\right]
-g\frac{\rho^2}{1-\rho}\sin^3\theta\cos^3\psi\left[\sin\psi\cos\theta+\frac{\rho}{1-\rho}\sin^2\theta\cos\psi\right]
\eea

Phase portraits for systems of equations (\ref{rho-s-2}), (\ref{psi-s-2}) and (\ref{rho-s-H}), (\ref{rho-s-p}) are presented in Fig.1.B. and Fig.2.B.

%%%%%%%%%%%%%%%%%%%%%%%%%%%%%%%%%%
\section{Free Tachyon in the FRW space}\label{FT}
\setcounter{equation}{0}
%%%%%%%%%%%%%%%%%%%%%%%%%%%%%%%%%
\subsection{The exact phase portrait of the free tachyon}\label{ds-appr}
%%%%%%%%%%%%%%%%%%%%%%%%%%%%%%%%

\begin{figure}[!h]
 \centering
 $\,\,\,\,\,\,\,\,\,\,\,\,\,$
 $\,\,\,\,\,\,\,\,\,\,\,\,\,$
 \setlength{\unitlength}{1mm}
\begin{picture}(50,70)
\put(-25,-14){\includegraphics[width=7cm]{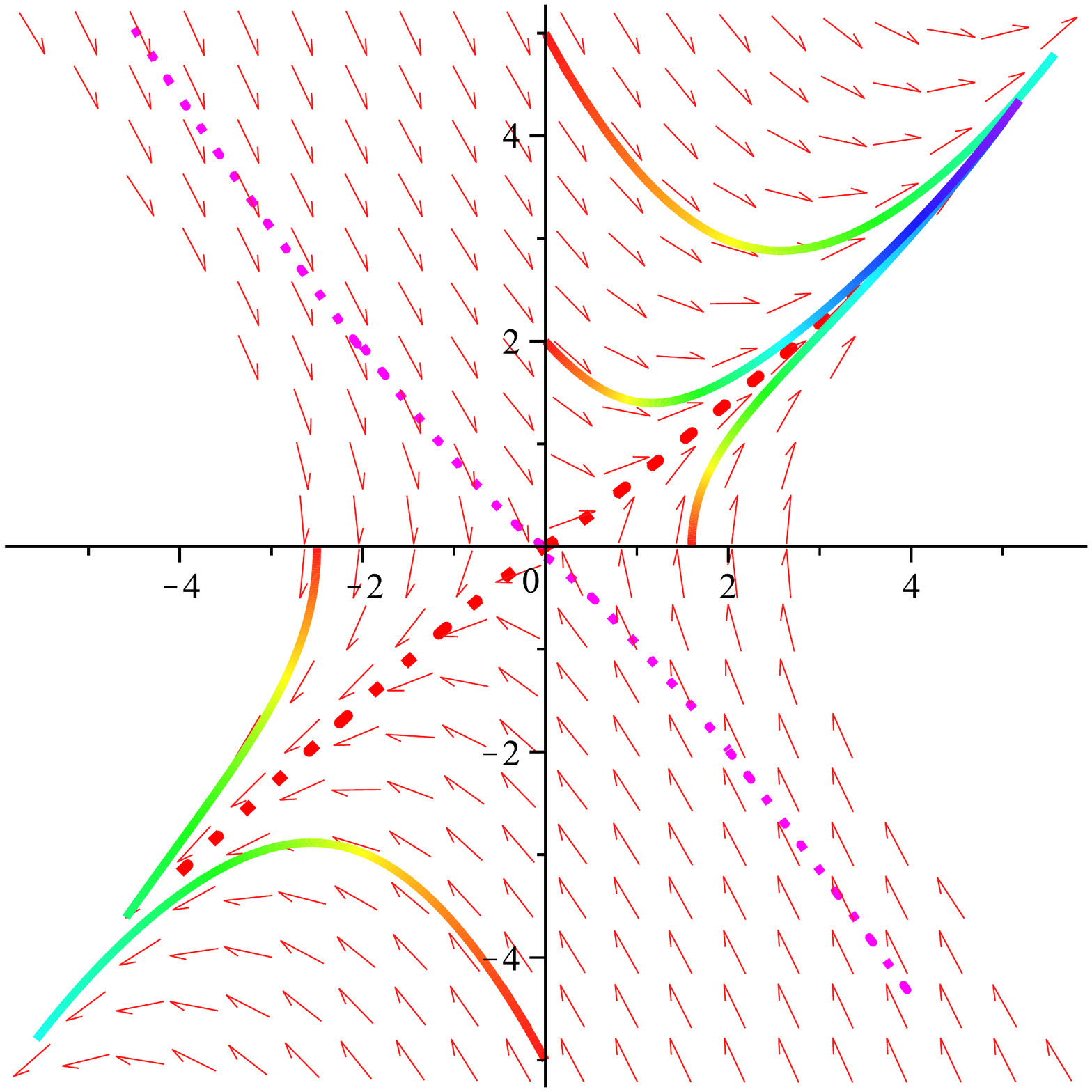}
}
\put(12,38){${\bf S_1}$} \put(-6,55){${\bf S_2}$} \put(-6,5){{\bf
II}} \put(-6,35){{\bf III}} \put(15,45){{\bf I}} \put(20,35){{\bf
IV}} \put(25,55){{\bf 1}} \put(25,47){{\bf 2}} \put(25,40){{\bf 3}}
\put(-12,20){{\bf 4}} \put(-12,8){{\bf 5}} \put(40,35){$\phi$}
\put(5,68){$\dot\phi$}
\end{picture}
 $A.\,\,\,\,\,\,\,\,\,\,\,\,$
 \begin{picture}(50,70)
\put(-10,-10){\includegraphics[width=7cm]{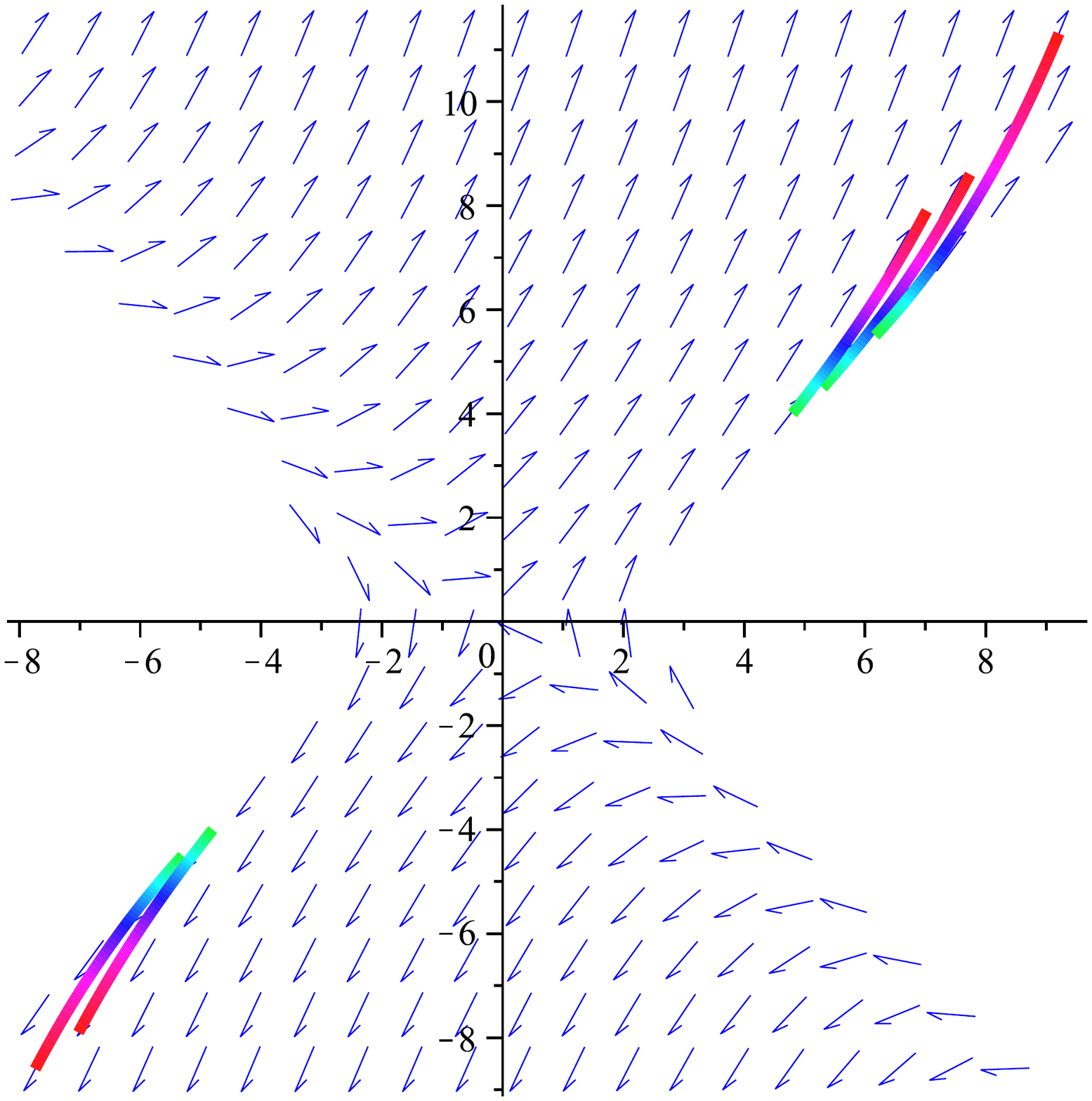}}
\put(5,5){{\bf VI}} \put(10,55){{\bf V}} \put(52,65){$1^\prime$}
\put(47,62){$2^\prime$} \put(42,51){$3^\prime$}
\put(2,20){$4^\prime$} \put(-10,8){$5^\prime$} \put(55,35){$\phi$}
\put(25,71){$\dot\phi$}
\end{picture}$B.$
\caption{ (color online)
 A. The phase portrait of  equation (\ref{EOM-2}).
 The accessible region is divided by 4 domains, that are separated by two
 separatrices:
 $S_1$ (the red dotted line) and $S_2$ (the magenta dotted line).
 Colored lines present trajectories starting from different points of the phase diagram. B. The phase portrait of equation (\ref{EOM-neg-H}). Continuations of the  trajectories presented in the left panel to solutions of the Friedman
equation with negative values of $H$ are indicated by the same numbers with primes added.}
\label{pp-tach-2}
 \end{figure}

The dynamics of the free tachyon in the FRW metric corresponding to the expansion
($H>0$) is described by the following equation
\be
\label{EOM-2}
 \ddot{\phi}+3\sqrt{\frac{8\pi G}{3} \left(\,\frac12\, \dot{\phi}^2-\frac{\mu^2}{2}\phi^2+\Lambda\right)}
 \,\dot{\phi}=\mu^2\phi.
\ee
 The phase portraits of  eq. (\ref{EOM-2}) is presented in Fig.\ref{pp-tach-2}.A. We see that there are two forbidden regions
 in the phase space.
 The distance between these forbidden regions depends on the value of the
 cosmological constant.
 The boundary of the accessible domain is given by the relation $\dot\phi =\pm \sqrt{\mu^2\phi^2-2\Lambda }$.
 The accessible region is divided by 4 domains, I, II, III and IV, that are separated by two separatrices:
 $S_1$ and $S_2$, see  Fig.\ref{pp-tach-2}.A.
  The upper half of the  separatrix $S_1$
 is an  attractor (the ``right'' attractor),
   and the low part of the  separatrix $S_1$ is the other attractor (the ``left''  attractor). The separatrix
 $S_2$ consists of two repulsers.

\begin{figure}[!h]
 \centering
 \includegraphics[width=7cm]{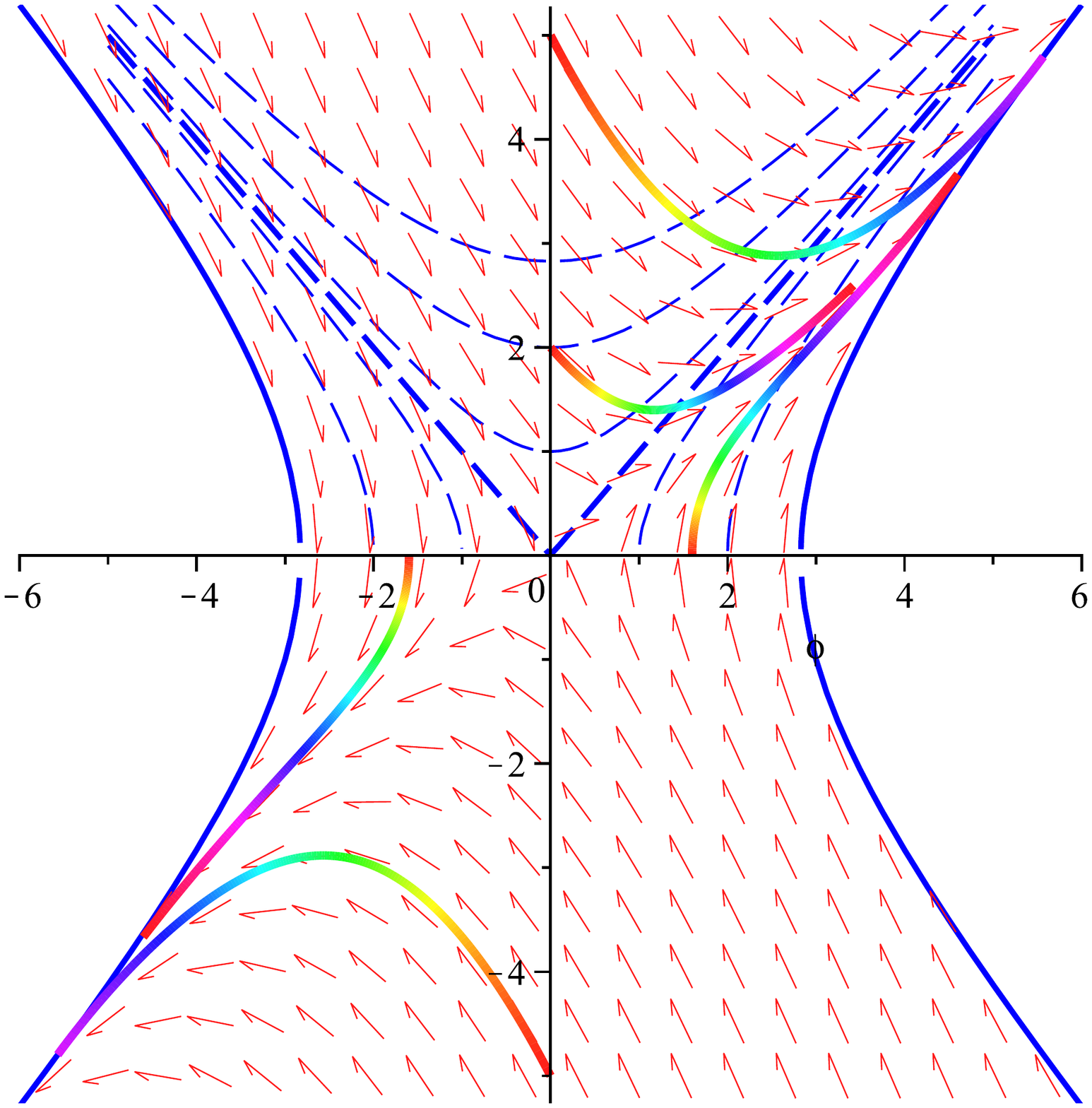}$A.~~~~~$
 \includegraphics[width=7cm]{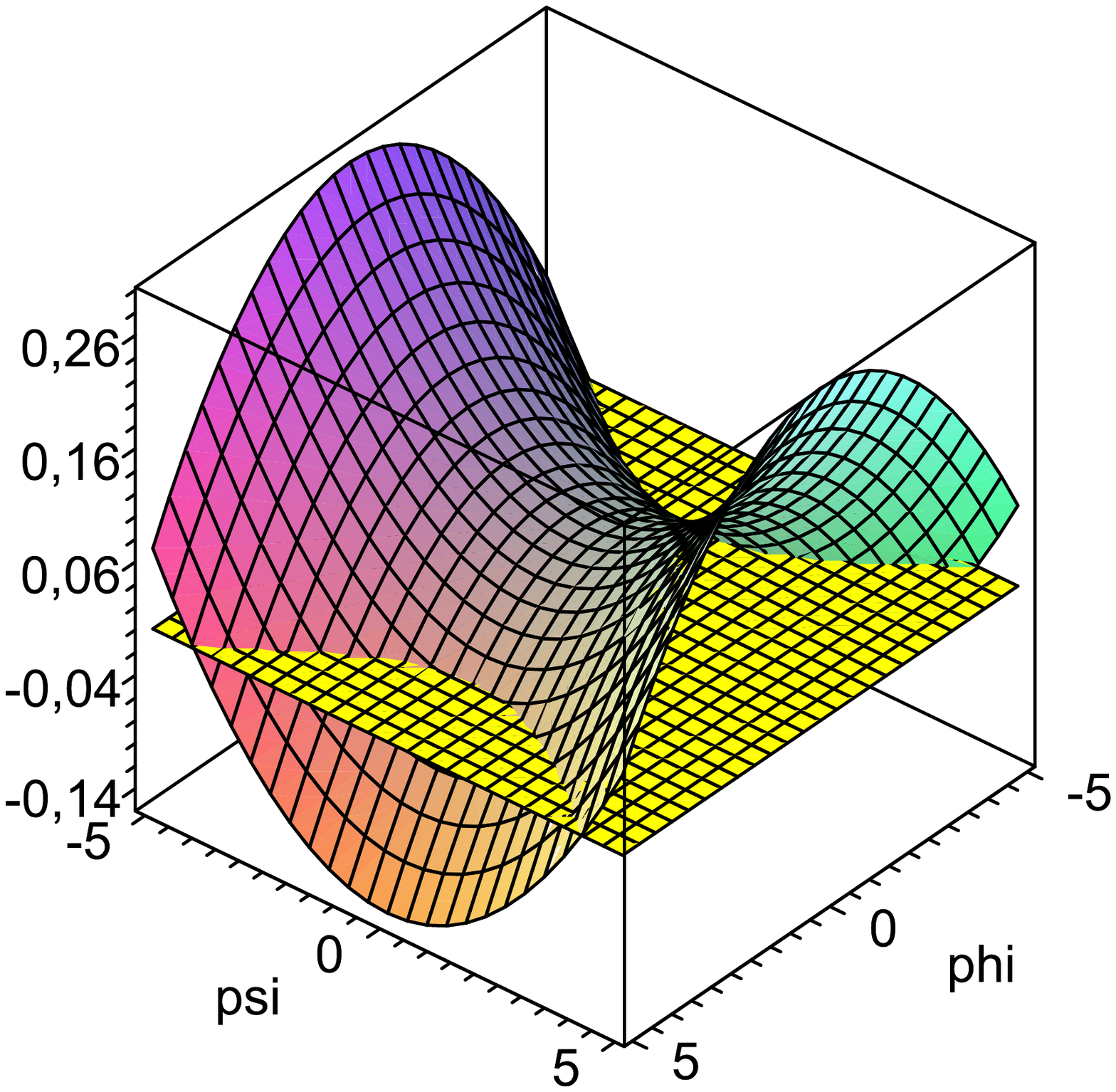}$B.$
 \caption{ (color online)
 A. The same phase
portrait as in Fig.\ref{pp-tach-2}.A. with marked lines (dashed blue lines) of the constant
values of $H$. The thick dashed line corresponds to
$H^2=3\kappa\Lambda$ $(\kappa=\frac{8\pi G}{9})$. The solid blue line corresponds to $H=0$ and
defines the boundary of the accessible domain. B. The
plot of the functional
$\kappa(\frac12\psi^2-\frac{\mu^2}2\phi^2+\Lambda)$ that defines the
value of $H^2$ (here $\dot \phi=\psi$). The intersection with the yellow
plane corresponds to the boundary of the accessible region.}
\label{H-tach-2}
 \end{figure}

 In Fig.\ref{pp-tach-2}.A. we see several eras of the tachyon evolution. Let us consider
 trajectories starting at region I. First,
 the tachyon reaches the area of the right attractor. In particular, starting
from zero initial value $\phi(0)=0$ (the corresponding trajectories are presented in Fig.\ref{pp-tach-2}.A. by colored lines) the tachyon decreasing the value of $H$
reaches  the area near the boundary of the accessible domain,  where $H=0$.
In Fig.\ref{H-tach-2}.A. the lines with fixed positive values of $H$ are denoted by the blue dashed lines. In Fig.\ref{H-tach-2}.B we present the 3d plot of values of H depending on the position in the phase diagram. In the end of this  era of evolution the tachyon reaches the line corresponding to $H=0$. Then
 the tachyon keeps moving in the area of the negative $H$, see Fig.\ref{pp-tach-2}.B, regions V and VI. Therefore, starting at region I the tachyon continues its evolution in region V.
 The tachyon EOM in regions V and VI has the form
 \be
\label{EOM-neg-H}
 \ddot{\phi}-3\sqrt{\frac{8\pi G}{3} \left(\,\frac12\, \dot{\phi}^2-\frac{\mu^2}{2}\phi^2+\Lambda\right)}
 \,\dot{\phi}=\mu^2\phi.
\ee

We see from the phase portrait Fig.\ref{pp-tach-2} that for finite
$\phi$ and $\dot\phi$ there are no critical points. To search for
critical points in the infinite phase space as well as to justify
approximations in different domains of the phase space (in
particular, in neighborhood of the forbidden domains) it is
convenient following to \cite{BGXZ} to rewrite the scalar field
equation in FRW space-time  as a system of equations for
3-dimensional dynamical system. There are two types of slightly
different  recipes how to do this. One is given in  \cite{BGXZ}  and
uses variables (\ref{X})-(\ref{Z}) (see below) and the second one is given in
\cite{Liddle:1994dx}  for  positively defined potentials  and is given by formula
(\ref{x2th1})-(\ref{z2th1}). The first recipe is convenient to study the cosmological
singularity (see sect.\ref{NCS}) and the second one -- to study a
transition from expansion to contraction (see sect.\ref{2-3-dim})

%%%%%%%%%%%%%%%%%%%%%%%%%%%%%%%%%%%%%%%%%%%%%%%%%%%%%%%%%%%
\subsection{Free Tachyon in FRW in the slow-roll regime}
\label{FT-slow-roll}
\setcounter{equation}{0}
%%%%%%%%%%%%%%%%%%%%%%%%%%%%%%%

In the slow-roll regime one usually ignores the second derivative of the field on time in the equation of motion and the kinetic term in the equation for $H$. In this case the tachyon equation becomes
\be
\label{EOM-sr}
 3\sqrt{\frac{8\pi G}{3} \left(\,\Lambda-\frac{\mu^2}{2}\phi^2\right)}
 \,\dot{\phi}=\mu^2\phi.
\ee
This equation obviously has a meaning only for
\be
|\phi|<\frac{\sqrt{2\Lambda}}{\mu}.\nonumber\ee

According to the common approach we can use the slow-roll approximation when slow-roll parameters $\varepsilon$ and $\eta$ are very small
\bea
 \varepsilon&=&\frac{m_{p}^2}{16 \pi}\left(\frac{ V'}{V}\right)^2\ll 1,\nonumber\\
 |\eta|&=&\left|\frac{m_{p}^2}{8 \pi}\left(\frac{ V''}{V}\right)\right|\ll 1.\nonumber
\eea
\begin{figure}
 \centering
\includegraphics[width=4cm]{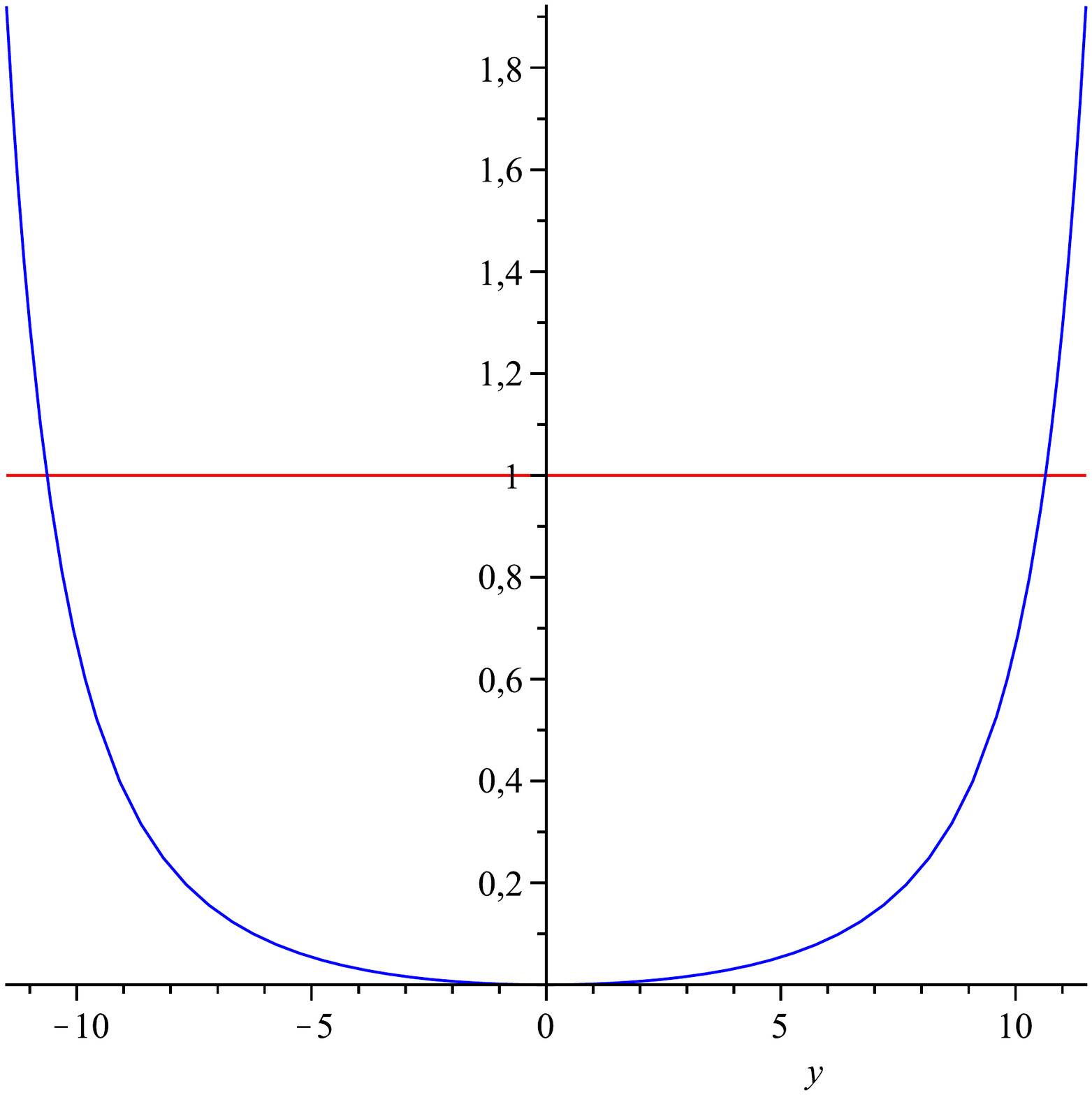}$A.~~~~~~~~~~~~~~~~~~~~~$
\includegraphics[width=4cm]{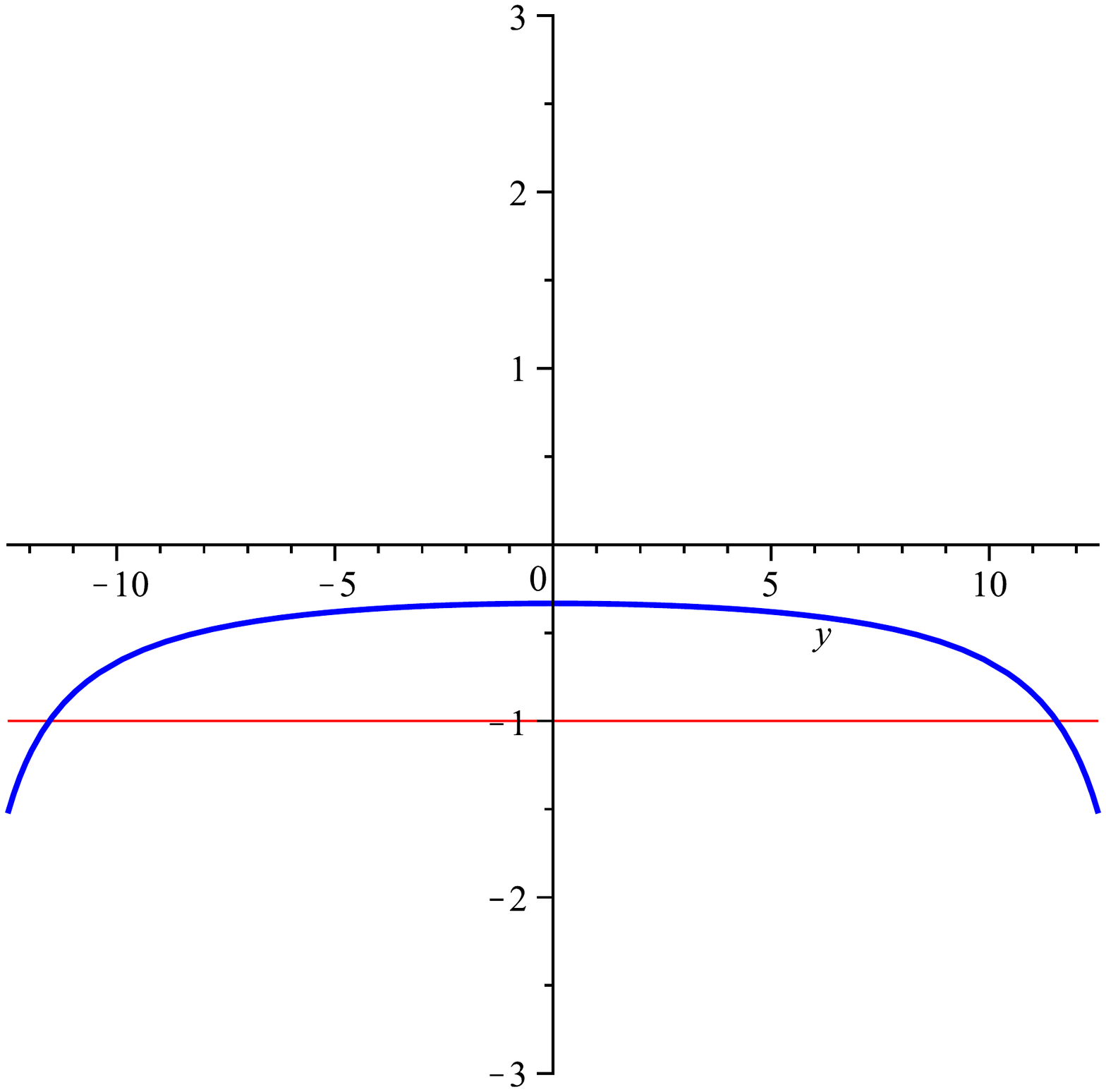}$B.$
\includegraphics[width=7cm]{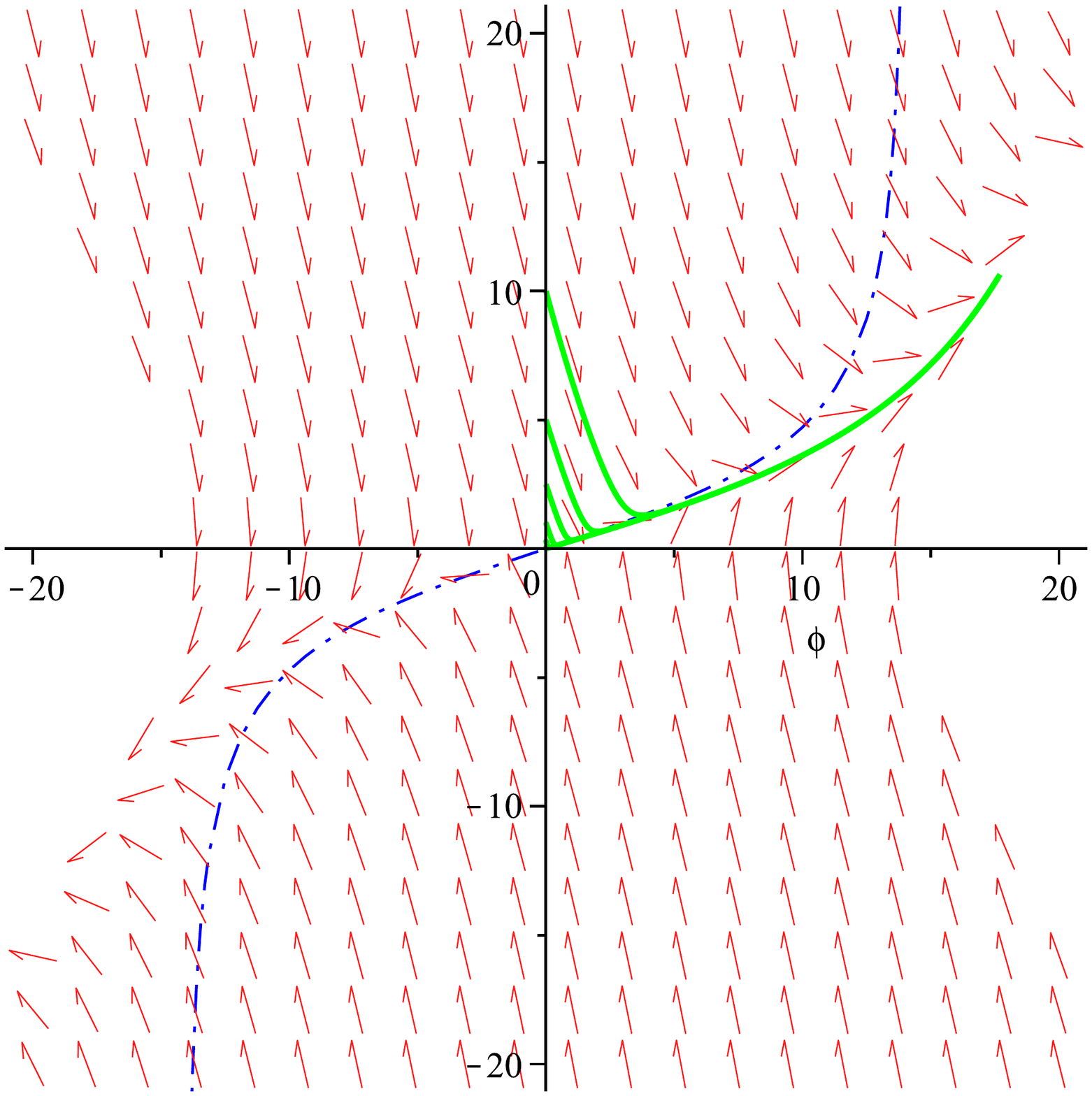}$C.$
\includegraphics[width=7cm]{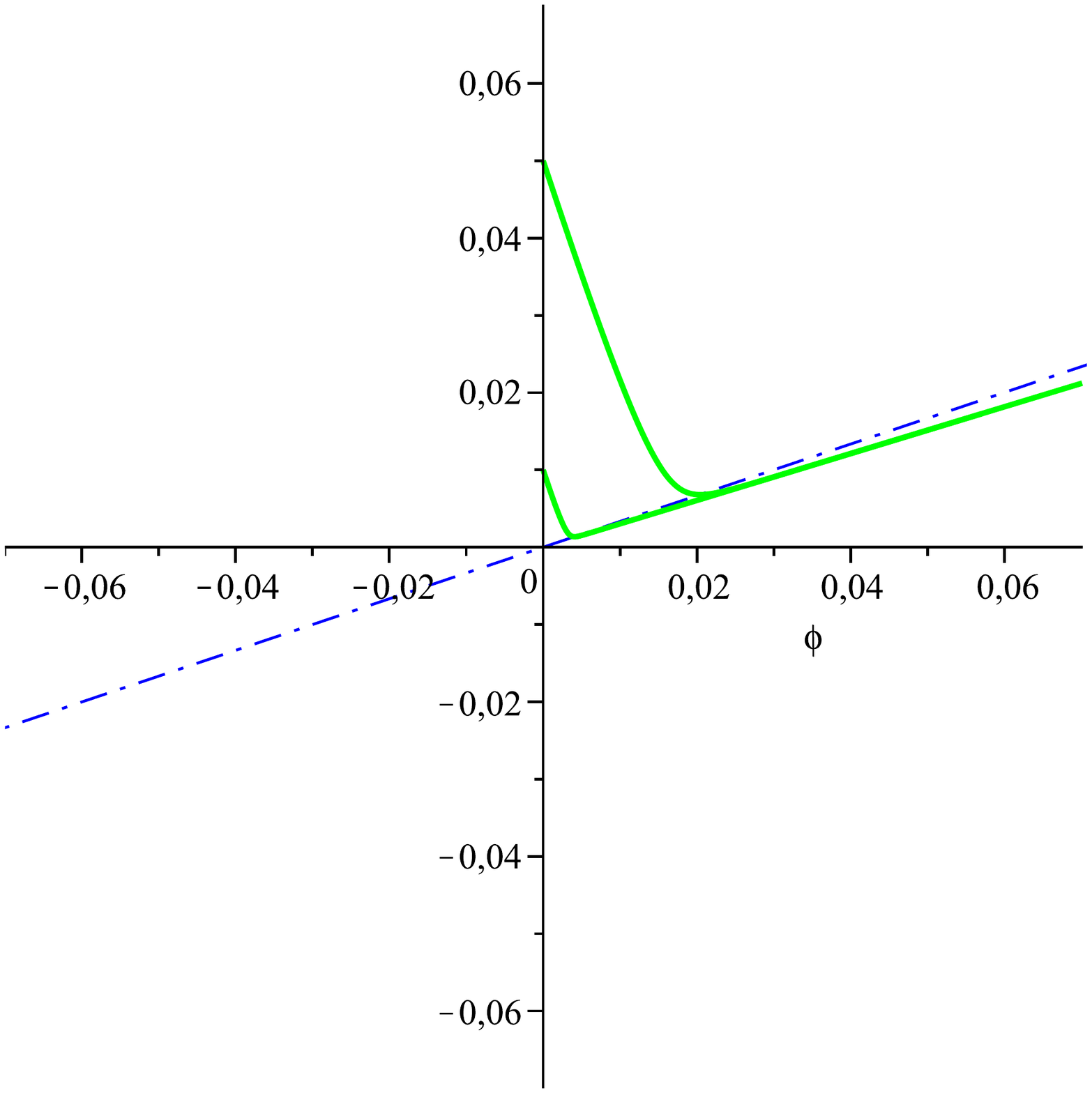}$D.$
 \caption{ (color online)
 A. The dependence of slow-roll parameter $\varepsilon$ on $\phi$ in the case when the slow-roll conditions are satisfied. B. The dependence of slow-roll parameter $\eta$ on $\phi$ in the case when the slow-roll conditions are satisfied. C. The phase portraits for the model when the slow-roll conditions are satisfied on some interval of $\phi$ values near $0$. The green solid lines correspond to different exact trajectories, the blue dashed and dot line corresponds to the slow-roll trajectory. D. The zoomed part of the Fig.C near (0,0) where the slow-roll conditions are highly satisfied.}
\label{eps-et-sl}
 \end{figure}

We can see the typical behaviour of these parameters for the model being considered in Fig.\ref{eps-et-sl} A and B in the case when the slow-roll conditions are satisfied. As we see in some interval of $\phi$ values near $0$ the slow-roll parameters are small enough.  That means that one should expect a good approximation of slow-roll solution for the exact phase trajectories on this interval.

 We can see in the phase portraits shown in Fig.\ref{eps-et-sl} C and D that really in this case on some interval of $\phi$ values near $0$ the exact phase trajectories are close enough to the slow-roll trajectory. When the slow-roll parameters become large enough we see the difference between the exact trajectories and the slow-roll trajectory.
 \begin{figure}
 \centering
 \includegraphics[width=3cm]{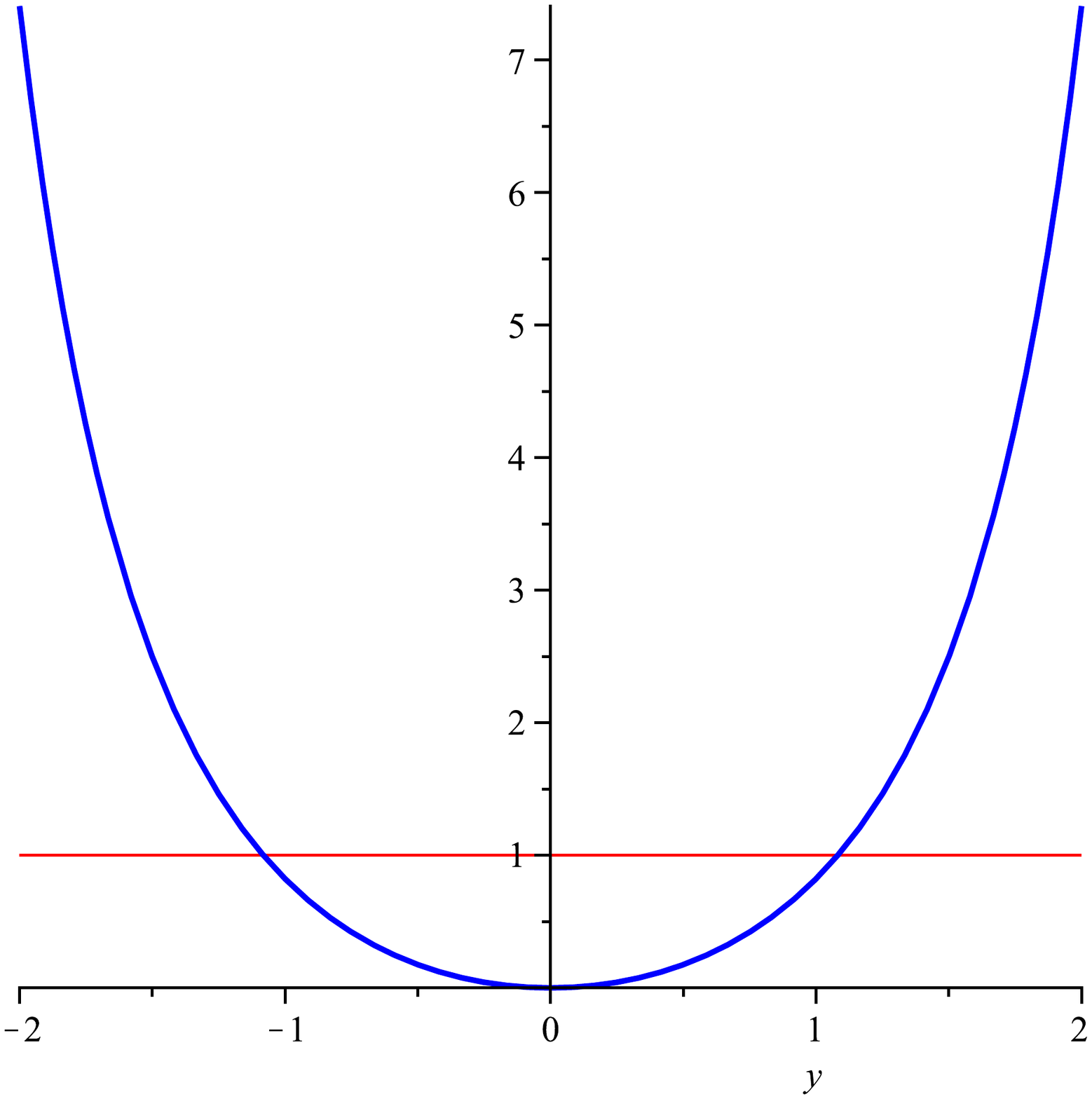}$A.~~~~~~~~~~~~~~~~~~~~~~~~~~~~~$
\includegraphics[width=3cm]{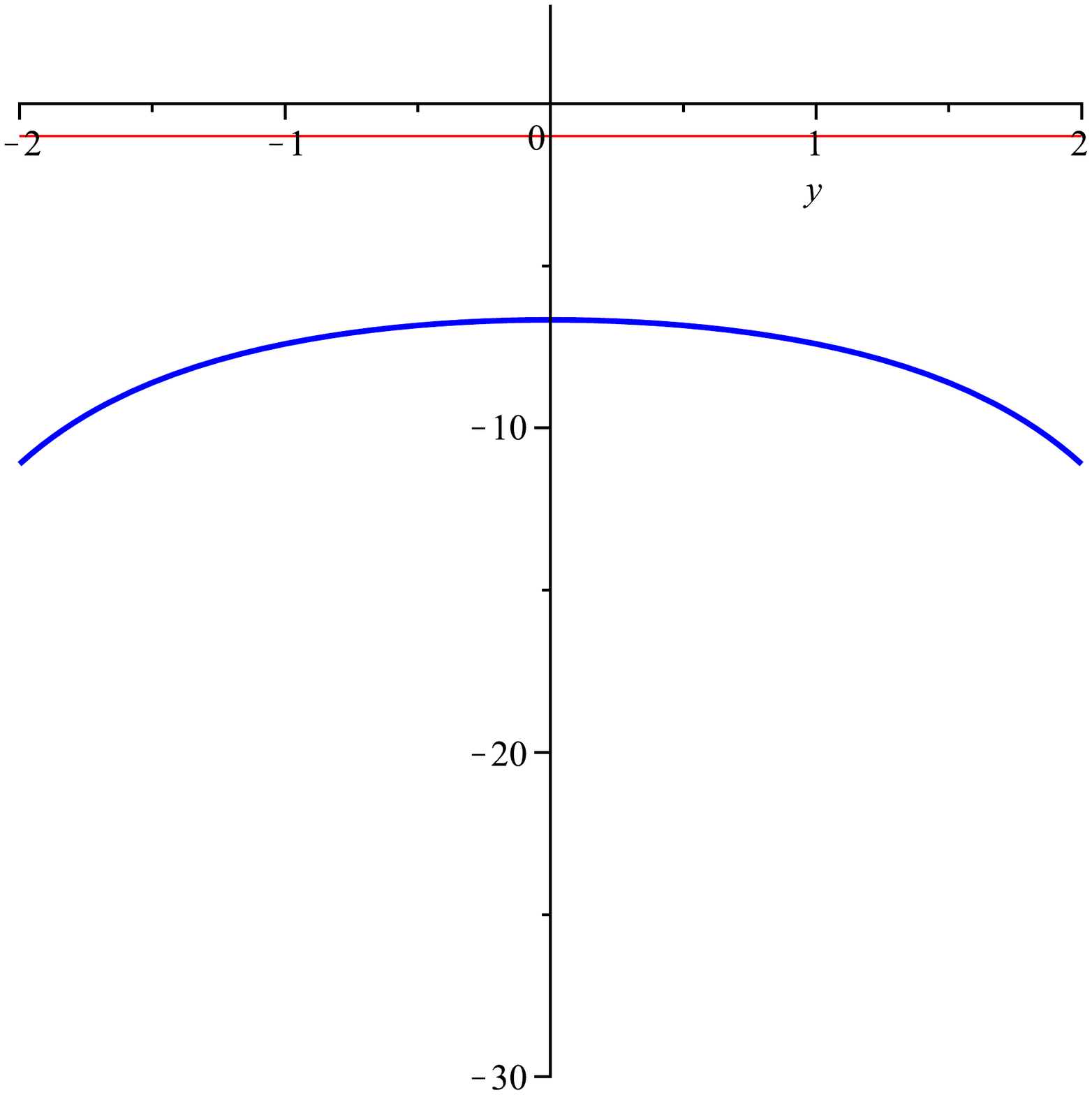}$B.$
\includegraphics[width=7cm]{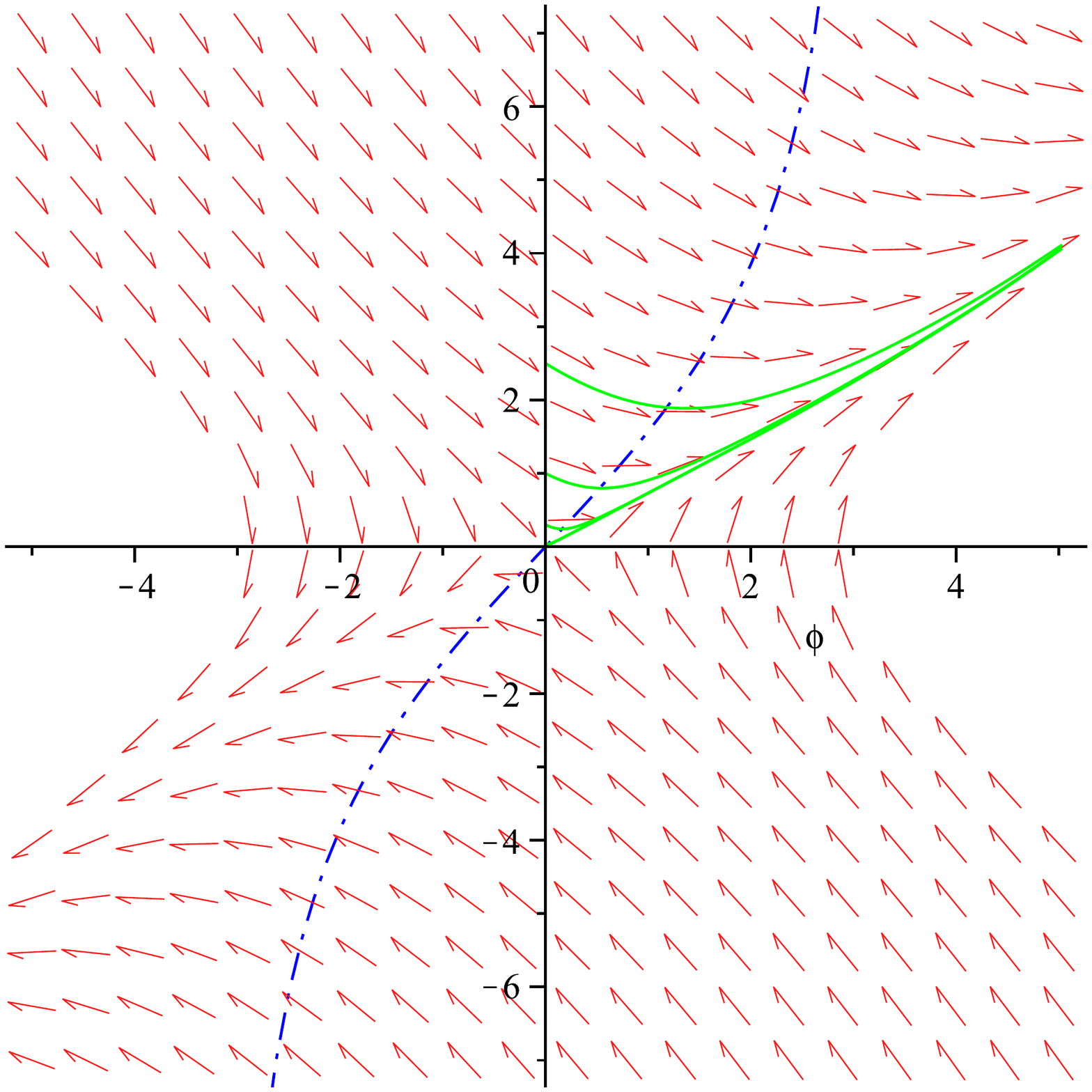}$C.$
\includegraphics[width=7cm]{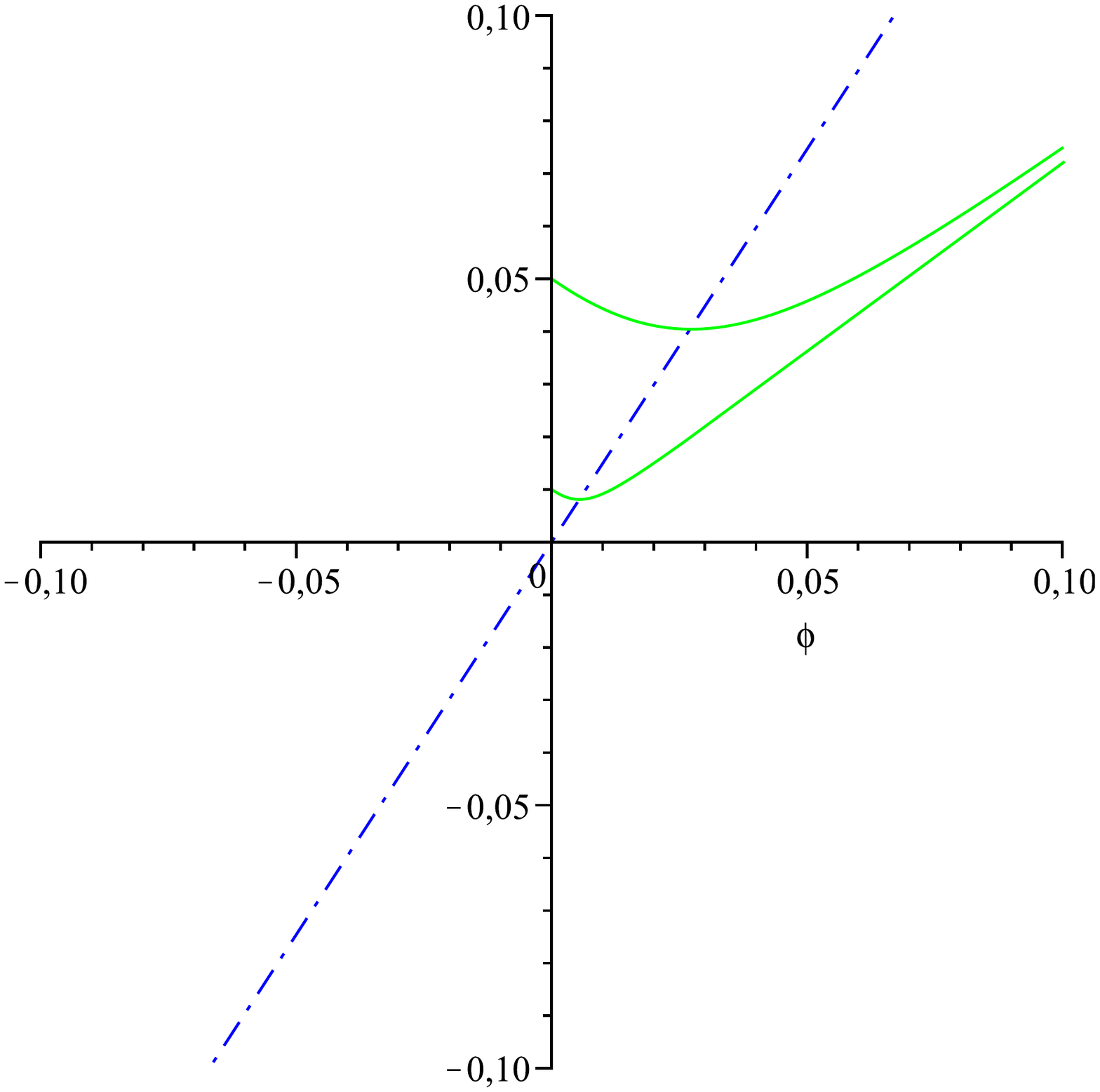}$D.$
 \caption{ (color online)
 A. The dependence of slow-roll parameter $\varepsilon$ on $\phi$ in the case when the slow-roll conditions are not satisfied. B. The dependence of slow-roll parameter $\eta$ on $\phi$ in the case when the slow-roll conditions are not satisfied. C. The phase portraits for the model when the slow-roll conditions are not satisfied. The green solid lines correspond to different exact trajectories, the blue dashed and dot line corresponds to the slow-roll trajectory. D. The zoomed part of the Fig.C near (0,0).}
\label{eps-et-withoutsl}
 \end{figure}

In the case when the slow-roll conditions are not satisfied the slow-roll parameters evolve as it's shown in Fig.\ref{eps-et-withoutsl} A and B. The phase portraits for this case are shown in Fig.\ref{eps-et-withoutsl} C and D. As we see the exact trajectories cross the slow-roll trajectory but don't keep on evolving along it. It means that the slow-roll approximation really cannot be used for the description of the exact trajectories in this case. We can note that in the region where $\dot{\phi}$ is small enough the minima of the exact phase trajectories are situated on the slow-roll trajectory. It is connected with the fact that the minimum of the exact phase trajectory corresponds to the point where $\ddot{\phi}=0$ and if we also can neglect $\dot{\phi}$ in the expression for $H$ then from the exact equation of motion we obtain the point of the slow-roll trajectory.

%%%%%%%%%%%%%%%%%%%%%%%%%%%%%%
\subsection{Tachyon dynamics near cosmological singularity}\label{NCS}
\setcounter{equation}{0}
%%%%%%%%%%%%%%%%%%%%%%%%%%%%%%%

To study tachyon dynamics near the cosmological singularity let us consider for simplicity the case of the zero cosmological constant
\be
\label{EOM-T-zero-Lambda}
 \ddot{\phi}+3\sqrt{\frac{8\pi G}{3} \left(\,\frac12\, \dot{\phi}^2-\frac{\mu^2}{2}\phi^2\right)}\,\dot{\phi}
 =\mu^2\phi.\ee
 This approach is valid when $\phi$ and $\dot{\phi}$ are large enough.

In terms of dimensionless variables
\bea \label{X}
X&=&\frac{\sqrt{12\pi}}{3m_{p}}\phi,\\
\label{Y}Y&=&\frac{\sqrt{12\pi}}{3\mu m_{p}}\dot\phi,\\
\label{Z}Z&=&\frac{H}{\mu},\\
\tau&=&\mu t,\eea
 system (\ref{EOM-T-zero-Lambda}) has the form
\bea
\label{EOM-X-mu}X_\tau&=&Y,\\
\label{EOM-Y-mu}Y_\tau&=&+X-3ZY,\\
\label{EOM-Z-mu}Z_\tau&=&-X^2-2Y^2-Z^2.
\eea
To get compactification of the phase space we  use the coordinates of the unit ball
\bea
\label{X-t}
X&=&\frac{\rho}{1-\rho}\sin \theta \cos\psi,\\
\label{Y-t}Y&=&\frac{\rho}{1-\rho}\sin \theta \sin\psi,\\
\label{Z-t}Z&=&\frac{\rho}{1-\rho}\cos \theta,
\eea
and in these new coordinates the equation of motion takes the form
\bea
\label{rho-s-t}\rho_\sigma&=&
\rho(1-\rho)^2\,\left[\sin2\psi\sin^2 \theta-\frac{\rho}{1-\rho}\cos\theta\left(1
+4\sin^2\psi\sin^2\theta\right)\right],\\
\label{psi-s-t}
\psi_\sigma&=&-\frac32\rho\sin2\psi \cos \theta+(1-\rho)\cos2\psi,\\
\label{theta-s-t}
\theta_\sigma&=&\frac12(1-\rho)\sin2\psi\sin2\theta
+\rho\sin\theta\left(1+\sin^2\psi
-4\, \cos ^{2} \theta
\sin^2 \psi \right),
\label{psi-s-t}\eea
here
\be
\frac{d\sigma}{d\tau}=\frac{1}{1-\rho}.\nonumber
\ee
This system has  4 critical points with $\rho=1$  lying on the cone
\be
Y^2=X^2+Z^2,\nonumber\ee
that  in the spherical coordinates has the form
\bea
\label{theta-psi}
\tan^2 \theta=-\frac{1}{ \cos2\psi}.\eea
These critical points are
\bea
\label{cr-point-cone-1}
 \theta_0  &=&\frac{\pi}{4},\,\,\,\,\, \psi_0 = \frac{\pi}{2},\\
\label{cr-point-cone-2}\theta_0 &=& \frac{\pi}{4},\,\,\,\,\, \psi_0 = -\frac{\pi}{2},\\
\label{cr-point-cone-3}\theta_0 &=& \frac{3\pi}{4},\,\,\,\,\, \psi_0 = \frac{\pi}{2},\\
\label{cr-point-cone-4}\theta_0 &=& \frac{3\pi}{4},\,\,\,\,\, \psi_0 = -\frac{\pi}{2}.\eea
\begin{figure}[!h]
\centering
\includegraphics[width=5cm]{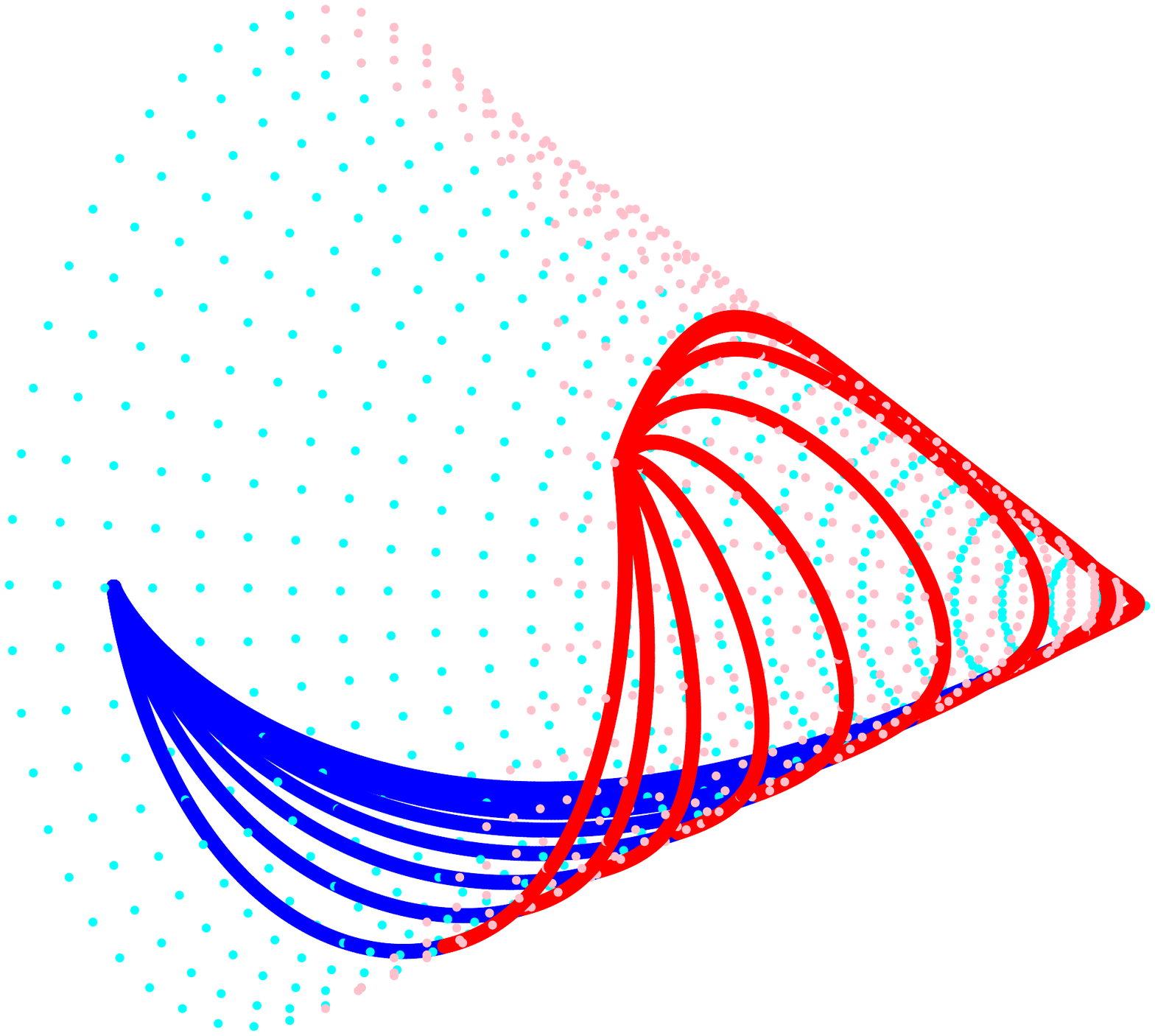}$A.$\,\,\,\,\,\,\,\,\,\,\,\,
\includegraphics[width=5cm]{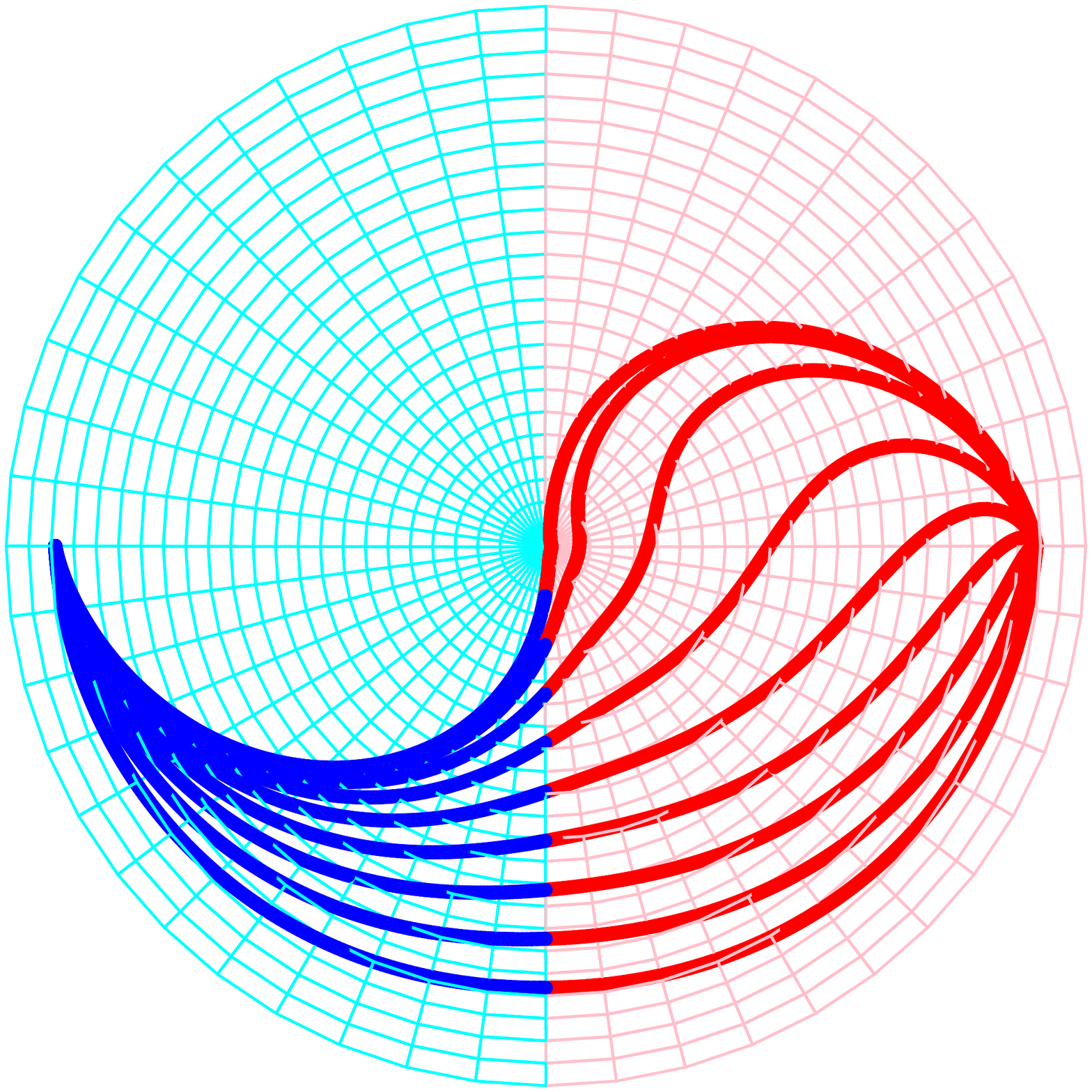}$B.$
\caption{ (color online) A. The 3-dimensional picture of the dynamics of the tachyon (system of equations (\ref{rho-s-t})-(\ref{theta-s-t})). We see that all trajectories starting on the conus keep moving on the conus. The red parts of trajectories correspond to the expansion and the blue part to the contraction. B. The projection of the previous picture on the $XY$-plane. This picture is the tachyon analogue of the picture for the free particle in the FRW presented in the Fig.\ref{pp-positive}}
\label{BGXZ-tach-cone}
\end{figure}

Let us now consider the behavior of the system near the critical point
(\ref{cr-point-cone-1}), i.e. $\rho=1,\,\,
 \theta_0  =\frac{\pi}{4},\,\, \psi_0 = \frac{\pi}{2}$.
 The have near this point in the linear approximation
 \bea
{\frac {d}{d\sigma}}\Delta \rho  \left( \sigma \right) &=&\frac32\,\Delta
\rho  \left( \sigma \right) \sqrt {2},\\
{\frac {d}{d\sigma}}\Delta \psi  \left( \sigma \right) &=&-\Delta \rho
 \left( \sigma \right) +\frac32\,\sqrt {2}\Delta \psi  \left( \sigma
 \right),\\
 {\frac {d}{d\sigma}}\Delta \theta  \left( \sigma \right) &=&2\,\sqrt {2}
\Delta \theta  \left( \sigma \right).
\eea
The eigenvalues are $(\frac32\sqrt{2}, \,\frac32\sqrt{2},\, 2\sqrt{2})$, and
solutions to this system are
\bea
\label{Delta-rho}
\Delta \rho  \left( \sigma \right)& =&C_2\,e^{\frac32\,\sqrt
{2}\sigma},
\\\label{Delta-psi}
\Delta \psi  \left( \sigma \right) & =&(C_1-C_2\sigma)e^{\frac32\,\sqrt {2}\sigma},\\
\label{Delta-theta}
\Delta \theta  \left( \sigma \right) &=&C_3\,e^{2\,\sqrt2\sigma}.
\eea

%%%%%%%%%%%%%%%%%%%%%%%%%%%%%%%
\subsection{Approximation of tachyon dynamics in the FRW by dynamics in the dS space}
\setcounter{equation}{0}
%%%%%%%%%%%%%%%%%%%%%%%%%%%%%%%%

During the evolution of the tachyon between
two dashed lines, see Fig.\ref{H-tach-2}.A, corresponding to $H=H_0\pm \Delta H_0$, we can ignore the dependence of $H$ on time and
we  take an approximate
\be
\label{EOM-H0}
 \ddot{\phi}+3H_0\,\dot{\phi}=\mu^2\phi.\ee
Near the thick blue dashed line we can use
\be
\label{H0-Lambda}H_0=\sqrt{\frac{8\pi G}{3}\Lambda},
\ee
and near the starting point with given $\phi_0$ and $\dot{\phi}_0$ we can use the approximation
\be
H_0(\phi_0^2,\dot\phi_0^2)=\sqrt{\frac{8\pi G}{3} \left(\,\frac12\, \dot{\phi}_0^2-
\,\frac{\mu ^2}2\, \phi_0^2+\Lambda\right)}.\nonumber
\ee

\begin{figure}[h!]
 \centering
\includegraphics[width=5cm]{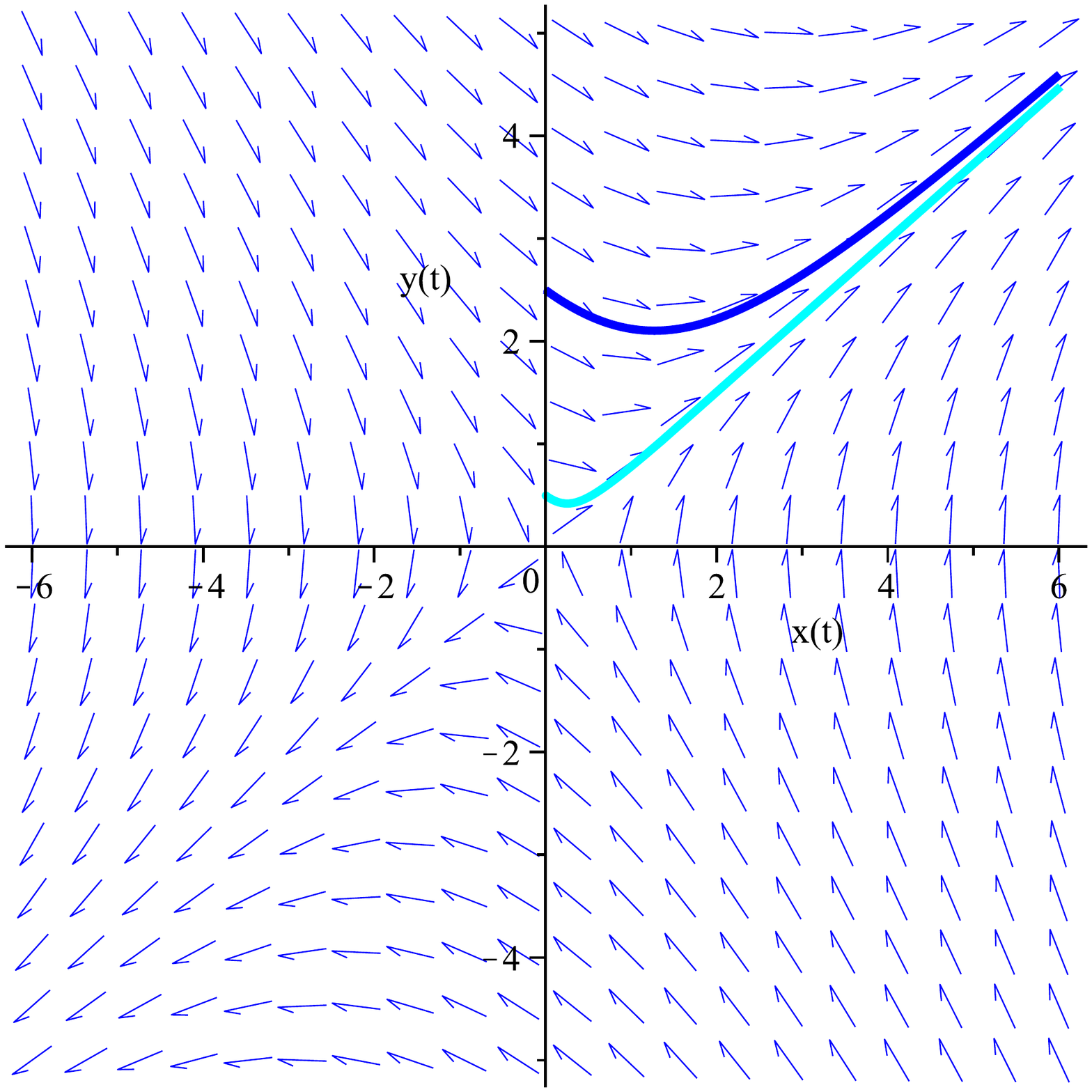}$A.\,\,\,\,\,$
\includegraphics[width=5cm]{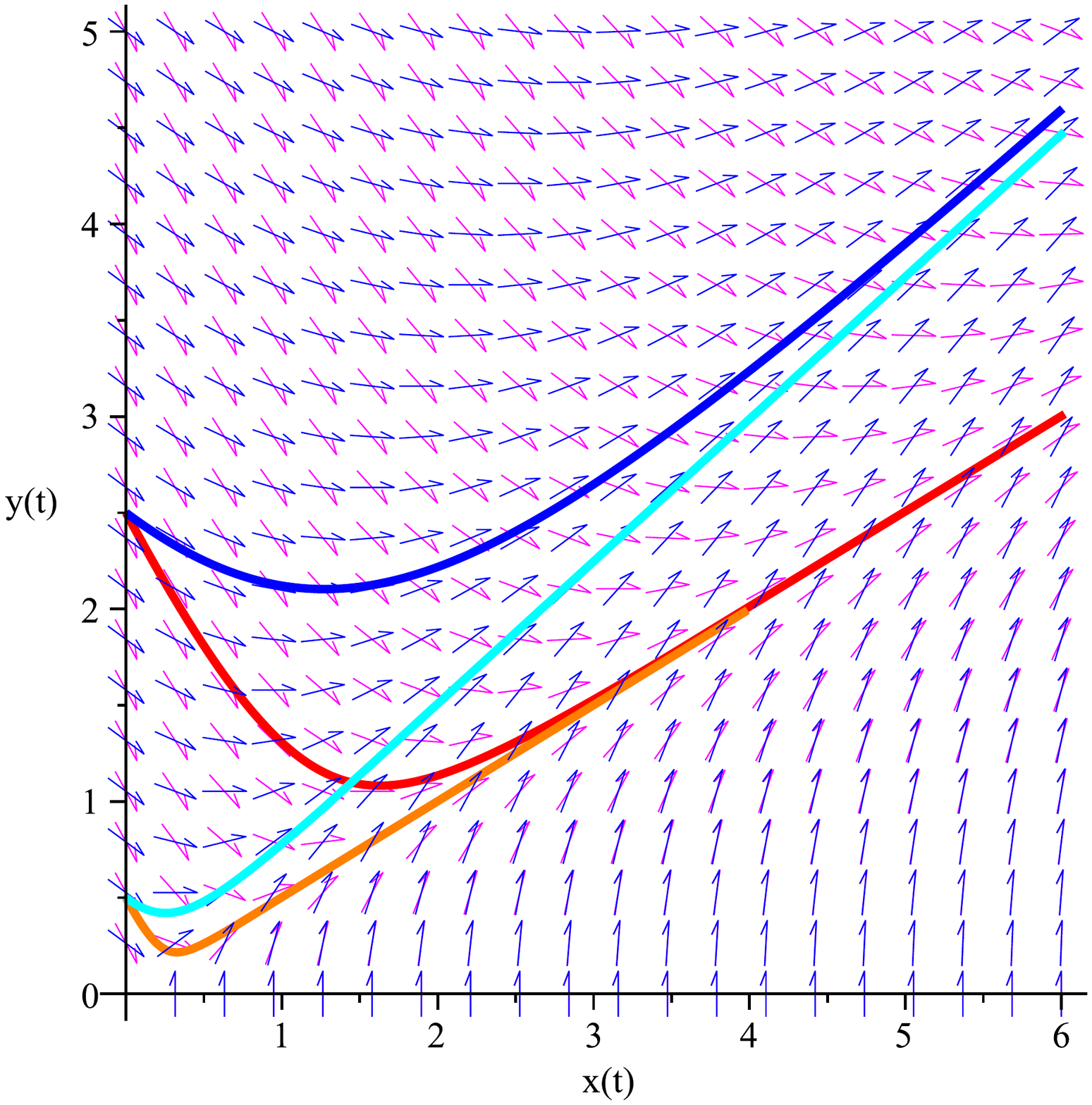}$B.\,\,\,\,\,$
 \caption{(color online) A. The phase portrait of the tachyon   evolving   in  the dS space.
 There are two attractors and two repulsers. B. A comparison of the tachyon phase
 portraits  in the  dS space
 with different  $\Lambda$ (the blue vectors for small $\Lambda$, the  magenta vectors for large $\Lambda$).
 The blue/cyan and red/pink solid lines present the dynamics of the tachyon with the same initial data in the dS space-time with different values of the cosmological
 constant $\Lambda$. The blue/red and cyan/pink lines have the same initial coordinates, but different initial velocities.}
 \label{dS-pp}
 \end{figure}

 In Fig.\ref{dS-pp}.A we present the phase diagram for the tachyon evolution  in the dS space, equation (\ref{EOM-H0}).
We see that there are  two attractors and two repulsers. The phase portraits for dynamics of the tachyon in the dS space with two different
$\Lambda$ are presented in Fig.\ref{dS-pp}.B by blue and magenta vectors (the blue vectors for small $\Lambda$ and  the  magenta vectors for large $\Lambda$).

Equation (\ref{EOM-H0})  has solutions
\be
\label{sol1t}
\phi ( t ) =C_1\,e^{ r_+ t}+C_2\,
e^{ r_- t},
\ee
where
\be
\label{r-pm}
r_{\pm}= -\frac32\,H_{{0}}\pm \frac32\,\sqrt {{H_{{0}}}^{2}+\frac49\,{\mu}^{2}},
\ee
i.e. $r_+>0,\,\,\,\,\,r_-<0$.
We see that for $H_0>0$ the $C_2$-mode decreases faster in comparison with
the $C_1$-mode that increases for large $t$.

For the initial date $\phi(0)=0$ we have
\be
\phi \left( t \right) =C\,e^{-3/2\,H_0 t}\sinh \left(\frac{3H_0}{2}
\,\sqrt {1\,+\frac49\,\frac{\mu^2}{H_0^{2}} }t\right),\,\,\,\,\,\,C_1=C/2
\ee
and
\be
\lim _{t\to \infty}\frac{\dot\phi \left( t \right) }{\phi(t)}=-\frac32\,H_0+\frac{3H_0}{2}
\,\sqrt {1\,+\frac49\,\frac{\mu^2}{H_0^2}}.\ee
This is exactly the same answer if we keep only the first term in (\ref{sol1t}).
One can say that $C_1$-mode is an attractor for all solutions of the
tachyon dS equation
with zero initial coordinates.   $C_1$-mode
is important for large times, but we cannot ignore the $C_2$-mode
for small times. We also see that the $C_1$-mode is a dominating mode for $t>0$
for the case when $H_0=0$.  One can say
that the $C_1$-mode is a viable mode and the $C_2$-mode is
a transient one.

To select the $C_1$-mode one can also assume that one deals with
the solution such as if the tachyon was sitting on the top of the hill at $t=-\infty$.

Let's note that the $C_1$-mode corresponds to the slow-roll trajectory on the interval where we can consider $H$ as an approximate constant. It means that the slow-roll trajectory is a quasi-attractor for the exact trajectories.

\begin{figure}[!h]
 \centering
\includegraphics[width=5cm]{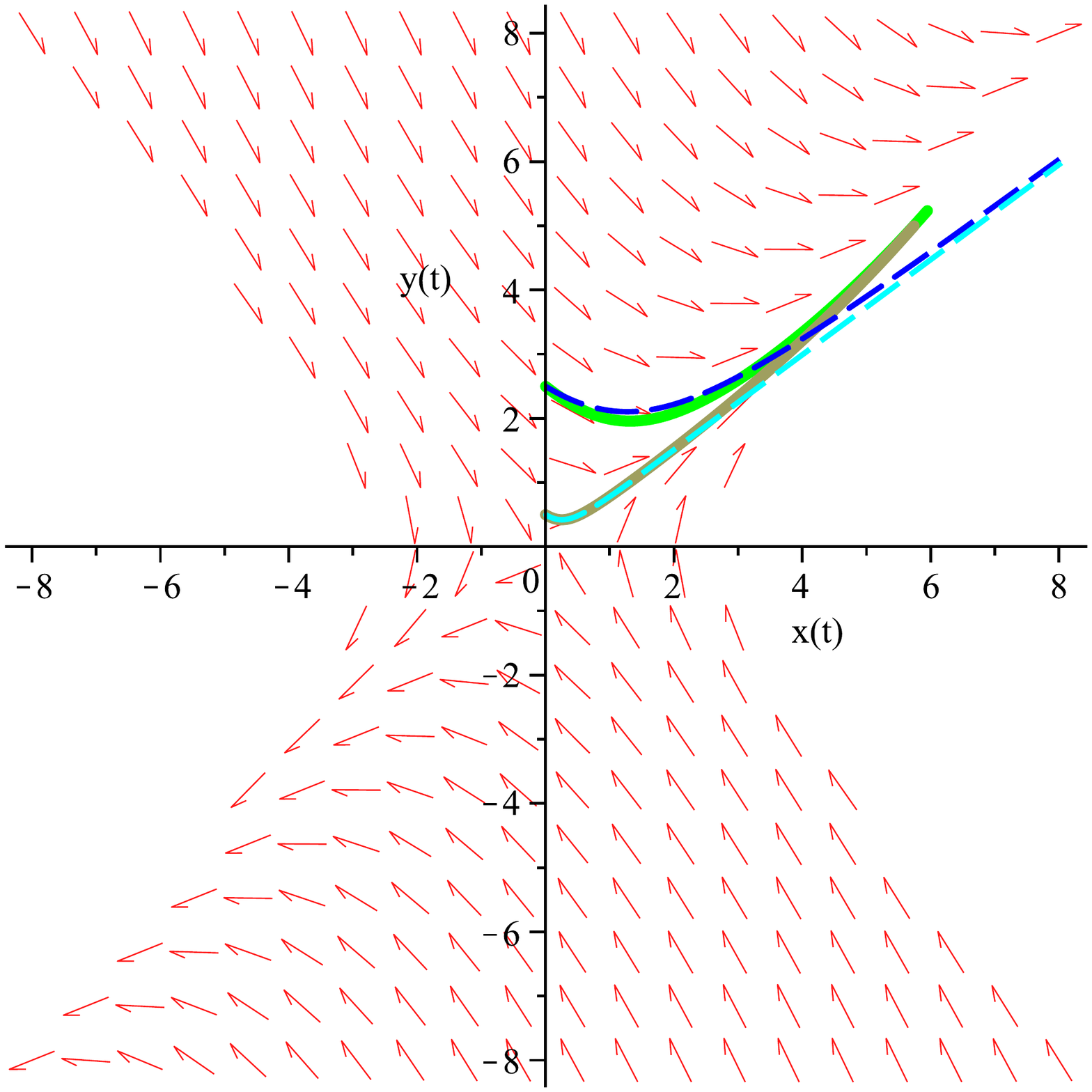}$A.\,\,\,\,$
\includegraphics[width=5cm]{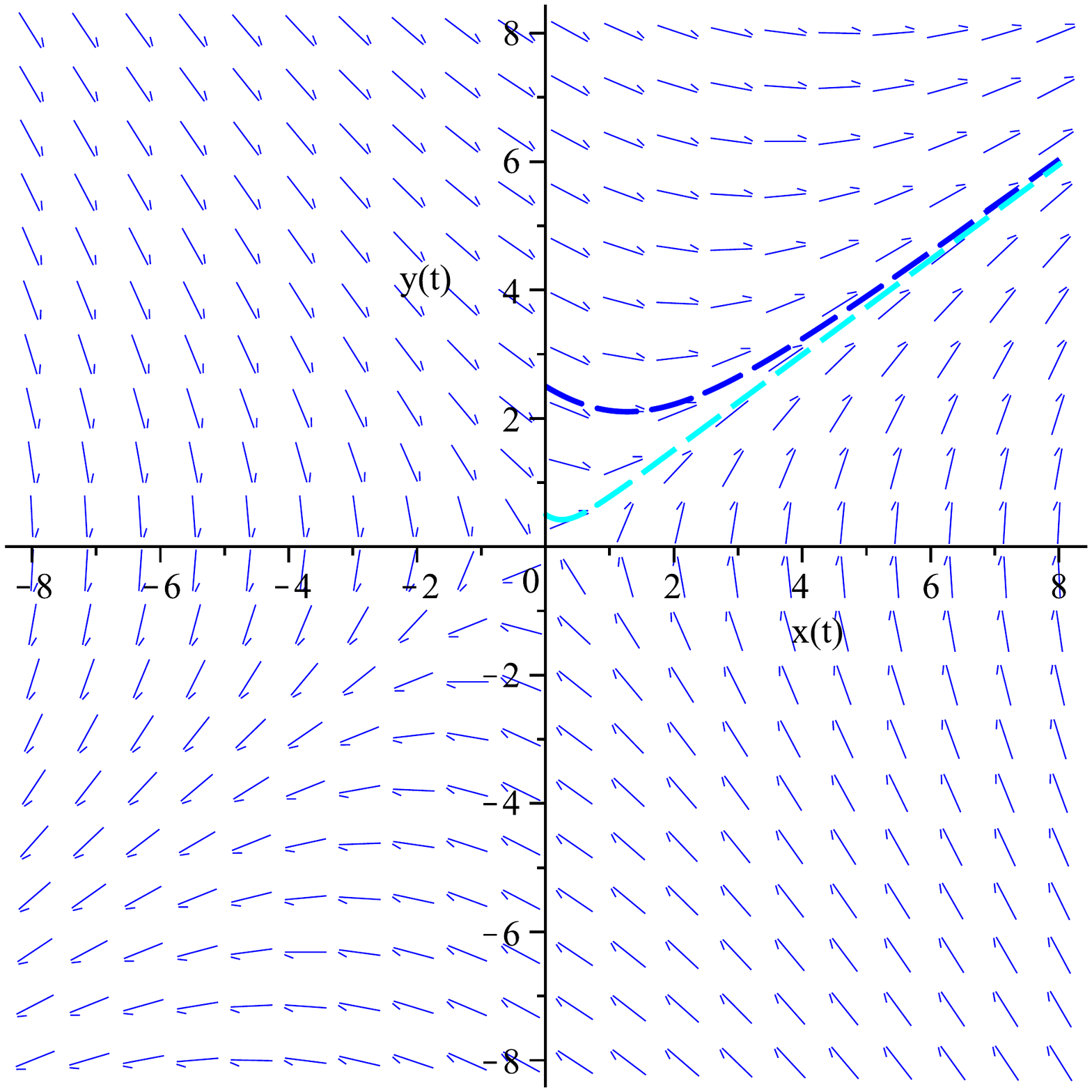}$B.$
\caption{(color online) A.  A comparison of the exact evolution (numerical solution of eq.
 (\ref{EOM-2})) with
an approximate evolution  (\ref{EOM-H0})
 with $H_0$ given by (\ref{H0-Lambda}). The solid green and khaki lines in the left panel represent the exact trajectories
 and the blue and cyan dashed lines  represent the approximate trajectories. Magenta vectors in the left panel represent the phase portrait of the exact equation.  Blue vectors in the right panel represent the phase portrait in the dS space. B. The solution and the phase portraits for the approximate equation.
 }
 \label{dS-FRW}
\end{figure}

To illustrate how  the approximation (\ref{EOM-H0}) works, in Fig.\ref{dS-FRW} we compare  phase portraits  for equations (\ref{EOM-2}) and   (\ref{EOM-H0}).
We see that for small times and  small initial velocities the approximation
 works, but it is failed for large times

To  estimate  a deviation  of $H$  from the initial $H_0$ we calculate
the energy of  solution (\ref{sol1t}) of  the tachyon  equation in the dS space-time.
The energy of this solution  is a sum of the energies of
two modes \cite{AJV}
\be
E=E_{C_1}+E_{C_2},\nonumber\ee
where
\bea
\label{energyC1-t}
E_{C_1}=\frac12\, \dot{\phi}_{C_1}^2-\frac{\mu^2}{2}\phi_{C_1}^2
&=&C_1^2\,\frac94\,H^2_0\left(1
 -\,\sqrt {1+\frac49\,\frac{{\mu}^{2}}{H^2_0}}\right)
\rm e^{ 2r_+ t},\\\label{energyC2}
E_{C_2}=\frac12\, \dot{\phi}_{C_2}^2-\frac{\mu^2}{2}\phi_{C_2}^2&=&
C_2^2\,\frac94\,H^2_0\left(1
 +
\,\sqrt {1+
\frac49\,\frac{{\mu}^{2}}{H_0^2}}\right)
e^{ 2r_- t}.\eea
We see that
$C_1$ gives a negative contribution to the energy and $C_2$-mode
gives a positive one.
Since the $C_1$-mode is more viable mode its negative energy contribution
dominates and  the tachyon loses the initial energy in the FRW space up to zero
value.

%%%%%%%%%%%%%%%%%%%%%%%%%%%
\subsection{Inflation over the hill}
%%%%%%%%%%%%%%%%%%%%%%%%%%%

In this subsection we remind the consideration of a scalar field rolling over the top of a local maximum of a potential following the paper by Tzirakis and Kinney \cite{top-hill}. This evolution needs a special attention as it cannot be considered in framework of the slow-roll approximation. It follows from the fact that in the slow-roll limit
\begin{equation}
\dot\phi \propto V'\left(\phi\right) = 0,\nonumber
\end{equation}
at the maximum of the potential.

The tachyon equation (\ref{EOM-H0}) can be rewritten in terms of the number of
e-foldings N,  $N\equiv \ln a/a_{initial}$ (we omit $a_{initial}$) as
\be
\label{EOM-N}
 \phi^{\prime\prime}+3\phi^{\prime}=\nu^2\phi,\ee
where $\nu^2=\frac{\mu^2}{H_0^2}$ and
$\phi'=\frac{d\phi}{dN}$.
The general solution of (\ref{EOM-N}) is
\be
\label{1}
\phi(N)=\phi_+e^{\nu_+N}+\phi_-e^{\nu_-N},
\ee
with $\nu_{\pm}=-\frac32(1\mp \sqrt{1+\frac49\nu^2})$.  This form of the tachyon evolution is  just the
reparametrization of (\ref{sol1t}).

 If $\nu^2$  is  small, the
constants $\nu_{\pm}$ become in the first approximation
\begin{equation}
\begin{split}
\label{small field approx r+-}
\nu_{+} & = \frac{\nu^2}{3},\\
\nu_{-} & = -3 - \frac{\nu^2}{3}.
\end{split}
\end{equation}

As in the previous consideration the general solution (\ref{1}) can be decomposed into two branches: a  transient branch $\nu_{-}$  which dominates at  early times,
and a ``slow-rolling'' branch $\nu_{+}$ which can  be identified as the
late-time attractor.

If we consider these branches separately we see that the slow-roll parameters are \cite{top-hill}
\begin{equation}
\begin{split}
\label{A SR params}
\varepsilon & = 4 \pi \nu_{\pm}^2\left(\frac{\phi}{m_p}\right)^2,\\
\eta & = -\nu_{\pm}+\varepsilon.
\end{split}
\end{equation}
Near the maximum of the  potential being considered, $\phi \approx 0$ and $\varepsilon \ll
1$, so that
\begin{equation}
\label{A eta 1}
\eta \approx -\nu_{\pm}=constant.
\end{equation}
Even though we assume $\varepsilon$  to be small, it is obvious from eqs.
(\ref{small field  approx r+-}) and  (\ref{A eta 1}) that  the $\nu_{-}$
branch  will never satisfy  the slow -roll conditions,  since $\eta>3$
always.  For the  $\nu_{+}$ branch, the slow-roll  limit is obtained for
$\nu \ll 1$ for which
\begin{equation*}
\label{A eta 2}
|\eta_{SR}| \approx \frac{\nu^2}{3} \ll 1.
\end{equation*}
It was also shown in \cite{top-hill} that $\nu_{-}$
branch corresponds to a field rolling up the potential.

If we consider the whole solution (\ref{1}) we see that in order for the field to be a monotonic function of time, its derivative must not change a sign. It means
\begin{equation}
\label{B dphi dN}
\frac{d\phi}{dN}=\nu_{+}\phi_{+}e^{\nu_{+}N}+\nu_{-}\phi_{-}e^{\nu_{-}N}\neq0,
\end{equation}
we assume that the derivative of the field with  respect to $N$  is a
smooth  function throughout  its evolution.   Since we know the  signs  of the
constants $\nu_{\pm}$,
the above  condition can be achieved  only if one  of the coefficients
$\phi_{\pm}$  is negative. If we  take $\phi_{-}<0$, it will result in a
positive time derivative for the field
\begin{equation*}
\dot \phi=H\frac{d\phi}{dN}>0.
\end{equation*}

If here we consider the field rolling over the hill then at some moment when $N=N_0$ the field must get the value equal to $0$ that corresponds to the maximum of the potential. It means that
\begin{equation*}
\phi_{+}e^{\nu_{+}N_{0}}=-\phi_{-}e^{\nu_{-}N_{0}}.
\end{equation*}
We can rewrite the expression for the general solution as follows
\begin{equation*}
\phi=A\left[e^{\nu_{+}(N-N_{0})}-e^{\nu_{-}(N-N_{0})}\right],
\end{equation*}
where
\begin{equation*}
A=\phi_{+}e^{\nu_{+}N_{0}}=-\phi_{-}e^{\nu_{-}N_{0}}.
\end{equation*}
As we see in the left part of potential the $\nu_{-}$
branch dominates that means that slow-roll approximation cannot be used in this region and after rolling over the top the $\nu_{+}$ branch dominates and slow-roll mode can start if necessary conditions are satisfied.

In the next subsection we consider the next approximation  of the tachyon in the FRW metric.

%%%%%%%%%%%%%%%%%%%%%%%%%%%%%%%%%%%%%
\subsection{Next-to-leading terms in the
approximate dynamics of the tachyon in the FRW metric\label{Next-to-leading}}
%%%%%%%%%%%%%%%%%%%%%%%%%%%%%%%%%%%%%%%

Starting from the dS approximation, $H=H_0$, where $H_0$ is given by
(\ref{H0-Lambda}), and selecting the $C_1$-mode  one can find a
solution of equation (\ref{EOM-2}) performing an analytical
expansion in powers of the $C_1$-mode  \cite{25,BarCline,39},
\bea
\label{s-phi}
  \phi(t) &=& \sum_{n=1}^{\infty} \phi_n e^{n r_+ t}, \\
 \label{s-H}  H(t) &=& H_0 - \sum_{n=1}^{\infty} H_n e^{nr_+t}.
\eea
In the next to the leading approximation
\be
 \label{H-2}
 H(t) = H_0 -  H_2 e^{2r_+t}+...\ee
 with
 \be
\label{H2-C1-mode}H_2=-3\pi G\phi_1^2\,H_0\left(1
 -\,\sqrt {1+\frac49\,\frac{{\mu}^{2}}{H^2_0}}\right).\ee

In this approximation the tachyon equation (\ref{EOM-2})
 has the form
\be
\label{EOM-H2-s}
 \ddot{\phi}+3(H_0-H_2e^{2r_+t})\,\dot{\phi}=\mu^2\phi\ee
 and expanding the solution in the series (\ref{s-phi}) we get
 \be
\label{phi-next-order}
\phi_{pert.sol}(t) \approx  \phi_1 e^{ r_+ t}+\phi_3 e^{ 3r_+ t}+...,\ee
where
\be
\label{phi-3-m}
\phi_3=\frac{3H_2}{8r_++6H_0}\phi_1.
\ee
Note that the denominator in (\ref{phi-3-m}) has no zeros.

Note also that  in the accepted scheme of calculation we get an
one-parametric set of solutions and  we deal only with the more
viable mode.
 We can check this explicitly. To this purpose we note that the general solution to
 (\ref{EOM-H2-s}) has the form
 \be
 \label{maple2-m}
 \phi (t)=\left(c_1{\rm \bf M}(a,a,x)+c_2 {\rm \bf W}(a,a,x)\right)e^{\frac{x}{2}} e^{-2a r_+t},
\ee
where ${\rm \bf M}$ is the Whittaker $M$ function and ${\rm \bf W}$ is the Whittaker $W$
function and
\be
a=\frac {2\,r_++3\,H_0}{4r_+}=
\frac {\sqrt {9\,H_0^2+4\,\mu^2}}{4r_+},\qquad x=\frac {3H_2 e^{2\,r_+t}}{2r_+}.\nonumber
\ee
Using the relation of the Whittaker function with the hypergeometrical function\\
$\Phi(\alpha ,\beta,z)\equiv_1F_1(\alpha ,\beta,z)$
we get
\bea
\phi(t)&=&e^{r_+t}\left(c_1\left(\frac {3H_2 }{2r_+}\right)^{a+\frac12}\Phi\left(\frac12,1+2a,x\right)
+c_2 \frac{\Gamma(-2a)}{\Gamma(\frac12-2a)}\left(\frac {3H_2 }{2r_+}\right)^{a+\frac12}\Phi\left(\frac12,1+2a,x\right)\right.\nonumber\\
&+&\left.c_2
\frac{\Gamma(2a)}{\Gamma(\frac12)}\left(\frac {3H_2 }{2r_+}\right)^{-a+\frac12}e^{-4a\,r_+t}\Phi\left(\frac12-2a,1-2a,x\right)\right)
\eea
and the series expansion
\be
\Phi(\alpha, \gamma,z)=1+\frac{\alpha}{\gamma}\frac{z}{1!}+\frac{\alpha(\alpha+1)}{\gamma(\gamma+1)}\frac{z^2}{2!}+...\nonumber
\ee
we get
\bea
&\,&\phi(t)\approx \left(c_1 +
c_2 \frac{\Gamma(-2a)}{\Gamma(\frac12-2a)}\right)\left(\frac{3H_2}{2r_+}\right)^{a+\frac12}e^{r_+t}
+\left(c_1 +
c_2 \frac{\Gamma(-2a)}{\Gamma(\frac12-2a)}\right)\left(\frac{3H_2}{2r_+}\right)^{a+\frac32}\frac{1}{2(1+2a)}e^{3r_+t}\nonumber\\
&+&c_2 \frac{\Gamma(2a)}{\Gamma(\frac12)}\left(\frac {3H_2 }{2r_+}\right)^{-a+\frac12}e^{(1-4a)\,r_+t}
+c_2 \frac{\Gamma(2a)}{\Gamma(\frac12)}\left(\frac {3H_2 }{2r_+}\right)^{-a+\frac32}\frac{1-4a}{2(1-2a)}e^{(3-4a)\,r_+t}.
\label{approx-maple-m}
\eea
 We see that to select the solution that starts
  from zero at $t=-\infty$ we have to remove the  $c_2$-mode
 \be
 \label{sbc}
\phi_{s.b.c.}(t)=c_1{\rm \bf M}(a,a,x)\approx c_1\left(
\left(\frac{3H_2}{2r_+}\right)^{a+\frac12}e^{r_+t}
+\left(\frac{3H_2}{2r_+}\right)^{a+\frac32}\frac{1}{2(1+2a)}e^{3r_+t}\right),
\ee
here the subscript {\it s.b.c.} means the  special boundary condition that selects the $c_1$-mode. Denoting
\be
\label{phi1}
\phi_1\equiv
c_1\left(\frac{3H_2}{2r_+}\right)^{a+\frac12}\ee
from (\ref{sbc}) we get (\ref{phi-next-order}) where $\phi_3$ is given by (\ref{phi-3-m}).

 Numerical solutions to  equation (\ref{EOM-H2-s}) are presented in
 Fig.\ref{Next-to-leading}.A by dotted red and magenta lines.
 We can also see in
 Fig.\ref{Next-to-leading}.B a comparison of the evolution described by equation
 (\ref{EOM-H2-s}) with the exact evolution (numerical solution to equation (\ref{EOM-2}))
 of the tachyon dynamics in the FRW space (green and khaki solid lines).
 We see that near the top there is no essential difference between the first approximation
 (corresponding solutions are presented by dashed blue and cyan lines) and the next
 approximations as well as a difference between the first approximation and the exact solutions.
 We also see  that for the small initial velocity the next to leading approximation
 (the magenta dotted line) to  the exact
  solution
 (the khaki solid line)  works better in comparison with the same
 approximation (the red  dotted line) to the solution with the large initial velocity (the green solid line) as well as the correction to the first approximation is more
 essential for the case of the small velocity.
 However near the  forbidden region the approximations do not work at all.
 In the next subsection we study the dynamics near the boundary
of the forbidden region in more details.
 \begin{figure}[h!]
\centering
\includegraphics[width=5cm]{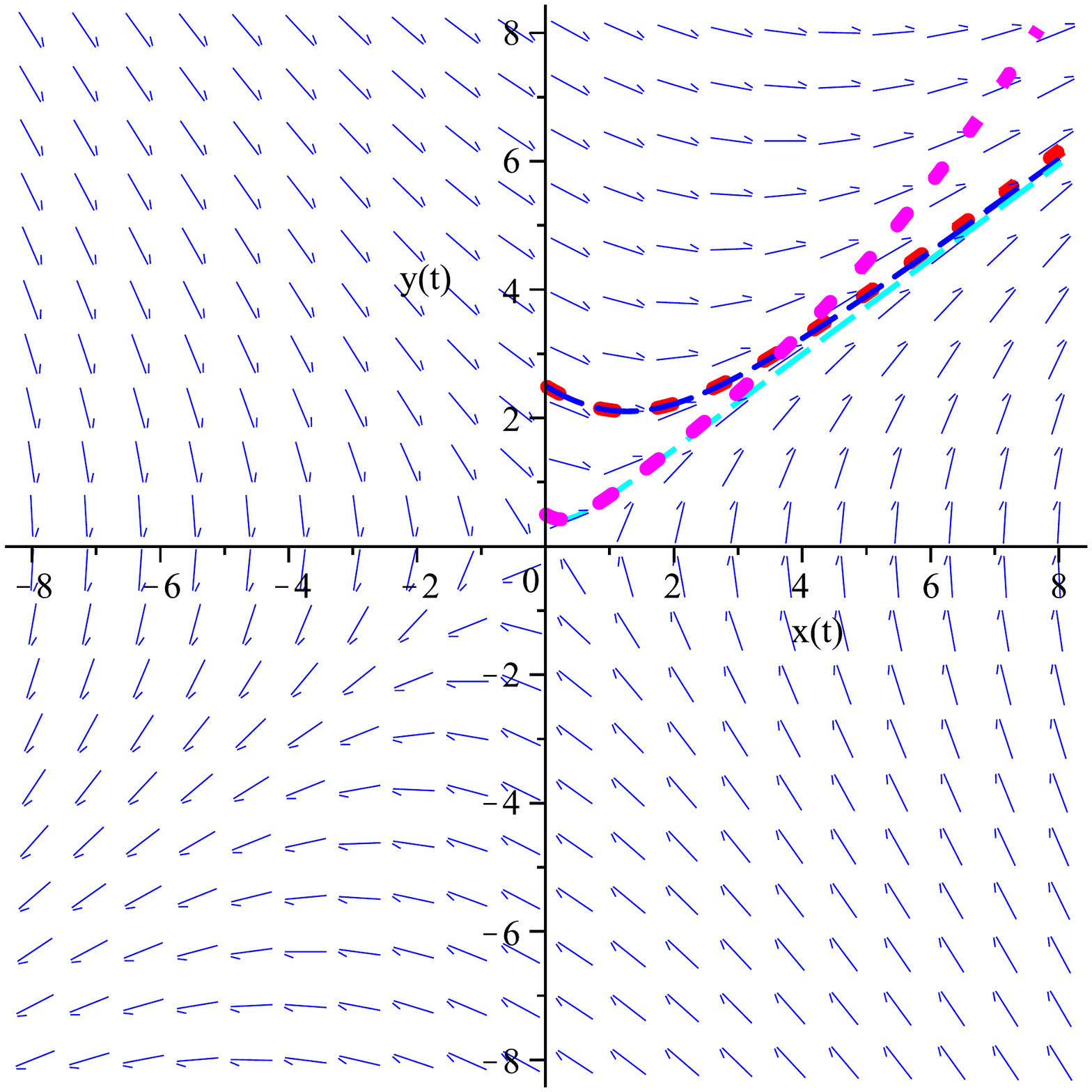}A.
\includegraphics[width=5cm]{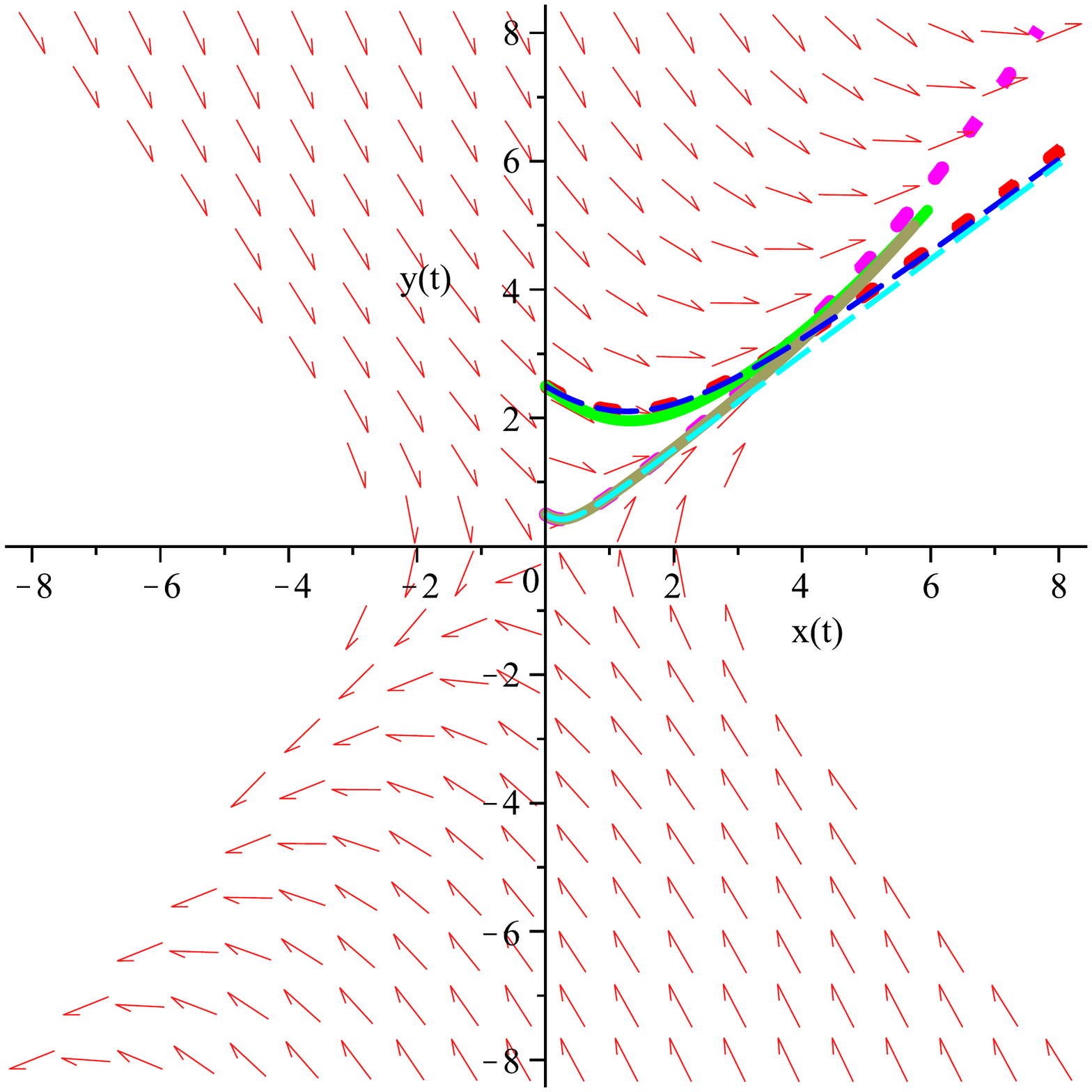}$B.~~~~~$
\caption{(color online) A. First
order approximation (dashed blue and cyan lines) and the third
 order approximation (dotted red and magenta lines) to solutions of (\ref{EOM-2}). The  first order approximation is given by  the dynamics of tachyon  in the dS space  and third
 order approximation is calculated as a solution of  (\ref{EOM-H2-s}) with $H_2$ given by (\ref{H2-C1-mode}). Blue vectors represent the phase portrait of the tachyon in the dS space. B. A comparison with the exact solutions of (\ref{EOM-2}) (khaki and green solid lines). Red vectors represent the phase portrait of the tachyon in the FRW space.}
\label{Next-to-leading}
\end{figure}

%%%%%%%%%%%%%%%%%%%%%%%%%%%%%%%
\subsection{Expansion to contraction \label{2-3-dim}}
\setcounter{equation}{0}
%%%%%%%%%%%%%%%%%%%%%%%%%%%%%%%%%%%%%%%%%%
 \subsubsection{Tachyon dynamics on e-foldings as 3-dim dynamical system}
%%%%%%%%%%%%%%%%%%%%%%%%%%%%%%%%%%%%

 As it has been mentioned before to study the transition from expansion
  to contraction it is convenient  to use the analogue of the coordinates proposed in
  \cite{Liddle:1994dx, Copellan}
\bea
\label{x2th1}x&\equiv& \sqrt{\frac{4\pi G}{3}}\,\,\frac{\dot \phi}{H},\\
\label{y2th1}y&\equiv& \sqrt{\frac{4\pi G}{3}}\,\,\frac{\mu \phi}{H},\\
\label{z2th1}z&\equiv& \frac{\mu}{H},
\eea
and rewrite eq.  (\ref{EOM-2}) as a system of ordinary differential equations on number of
e-foldings $N\equiv \ln a/a_{initial}$ (we omit $a_{initial}$)
\bea
\label{EOM-t-x}
x'&=&3x(y^2-b^2z^2)+zy,\\
\label{EOM-t-y}y'&=&3x^2y+zx,\\
\label{EOM-t-z}z'&=&3x^2z,\eea
here
$
X'=\frac{dX}{dN}$.
Note that due to $\frac{dN}{dt}=\frac{\dot a}{a}=H$ we have the following relation
$
X'=\dot X\,\cdot \frac{1}{H}$.
We can check that
\be
\label{cons-th2}
I(x,y,z)\equiv x^2-y^2+b^2z^2,\ee
where
$
b^2=\Lambda \frac{8\pi G}{3\mu^2}$
is  an integral of motion.
Therefore, if we take the initial conditions $x(0),y(0),z(0)$ such that
\be
x(0)^2-y(0)^2+b^2z(0)^2=1,\nonumber
\ee
this condition is true for ever.
\begin{figure}[h!]
\centering
\includegraphics[width=4.5cm]{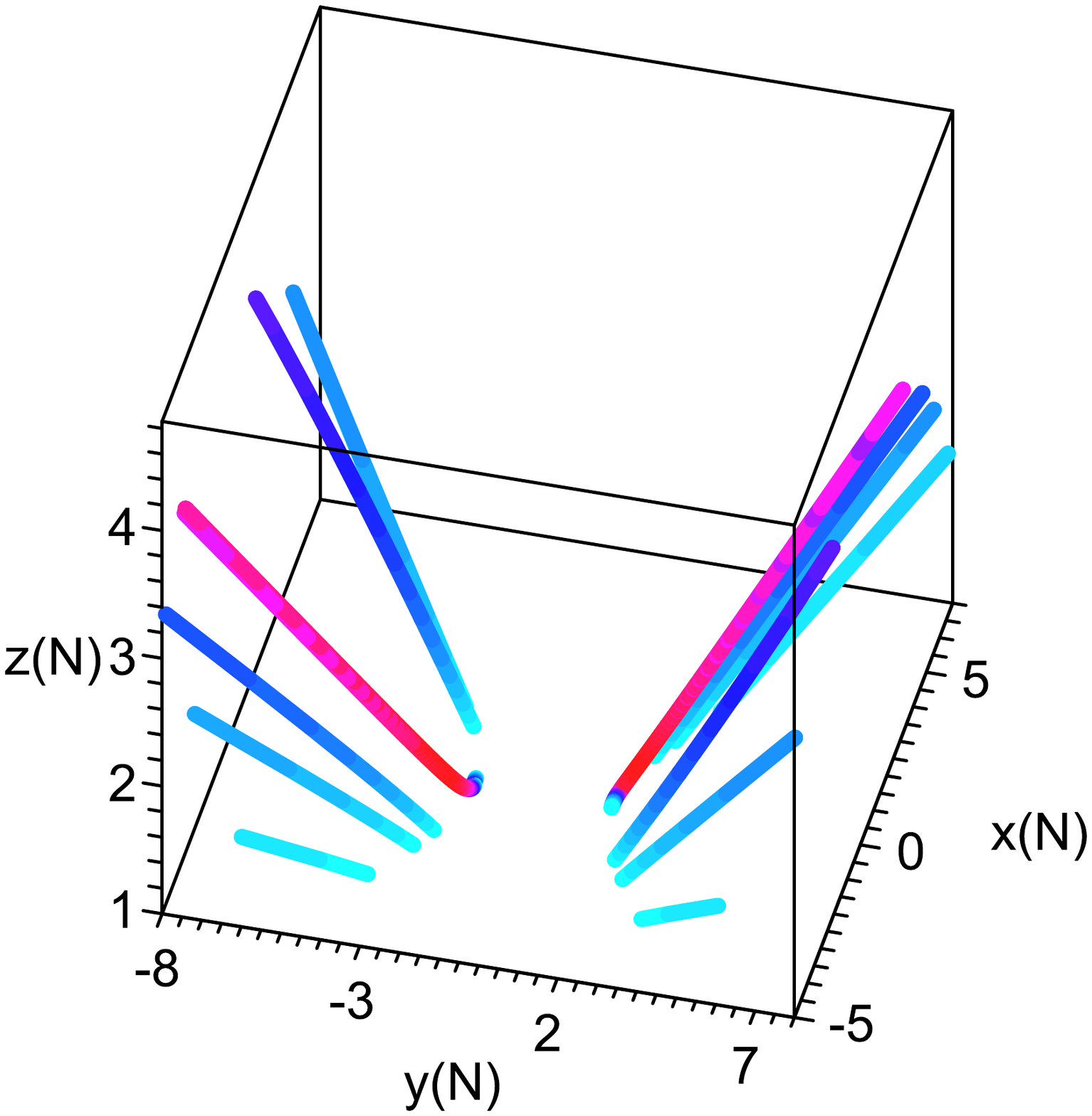}$A.~~~~~$
\includegraphics[width=4cm]{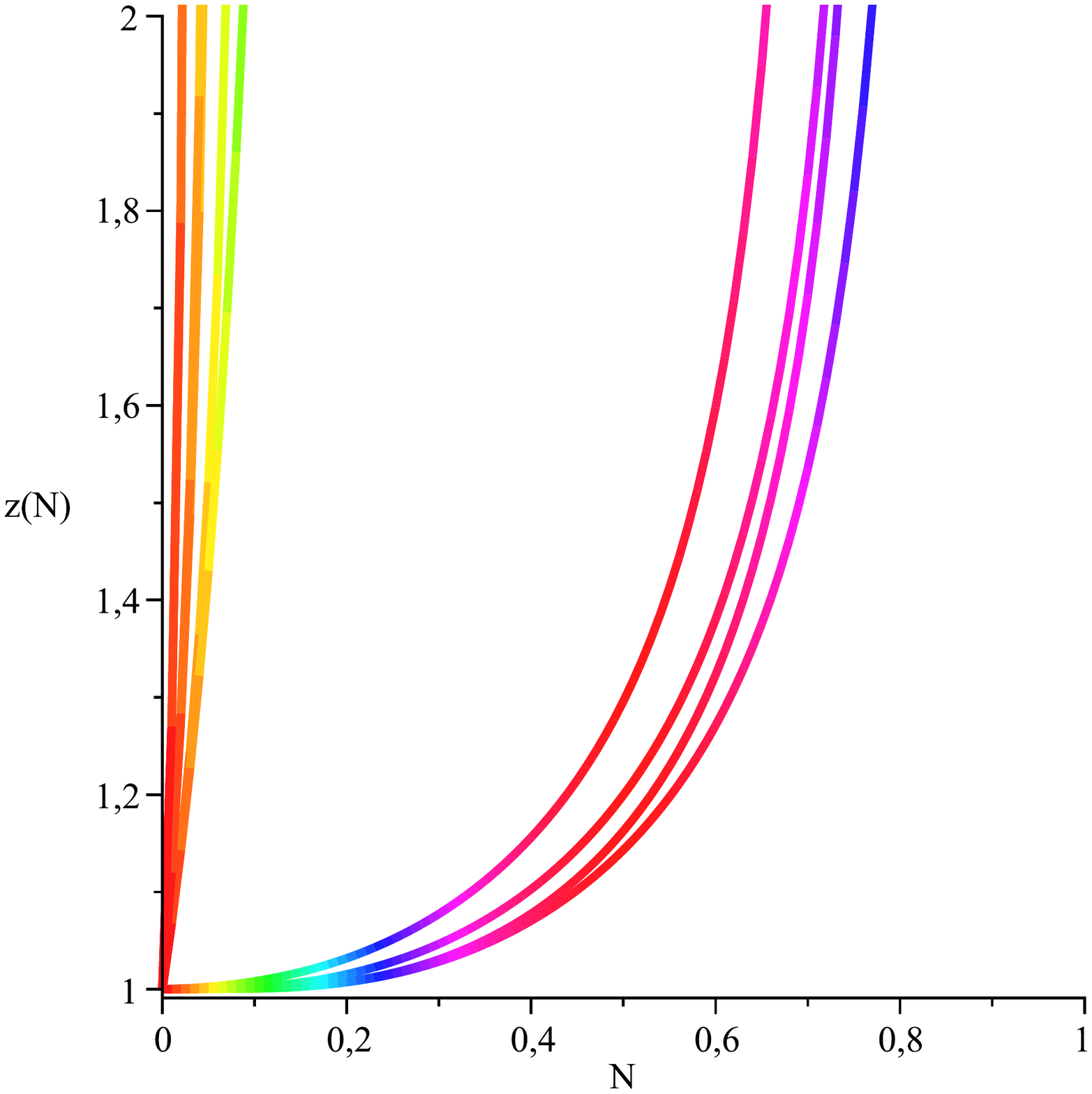}$B.~~~~~$
\includegraphics[width=4cm]{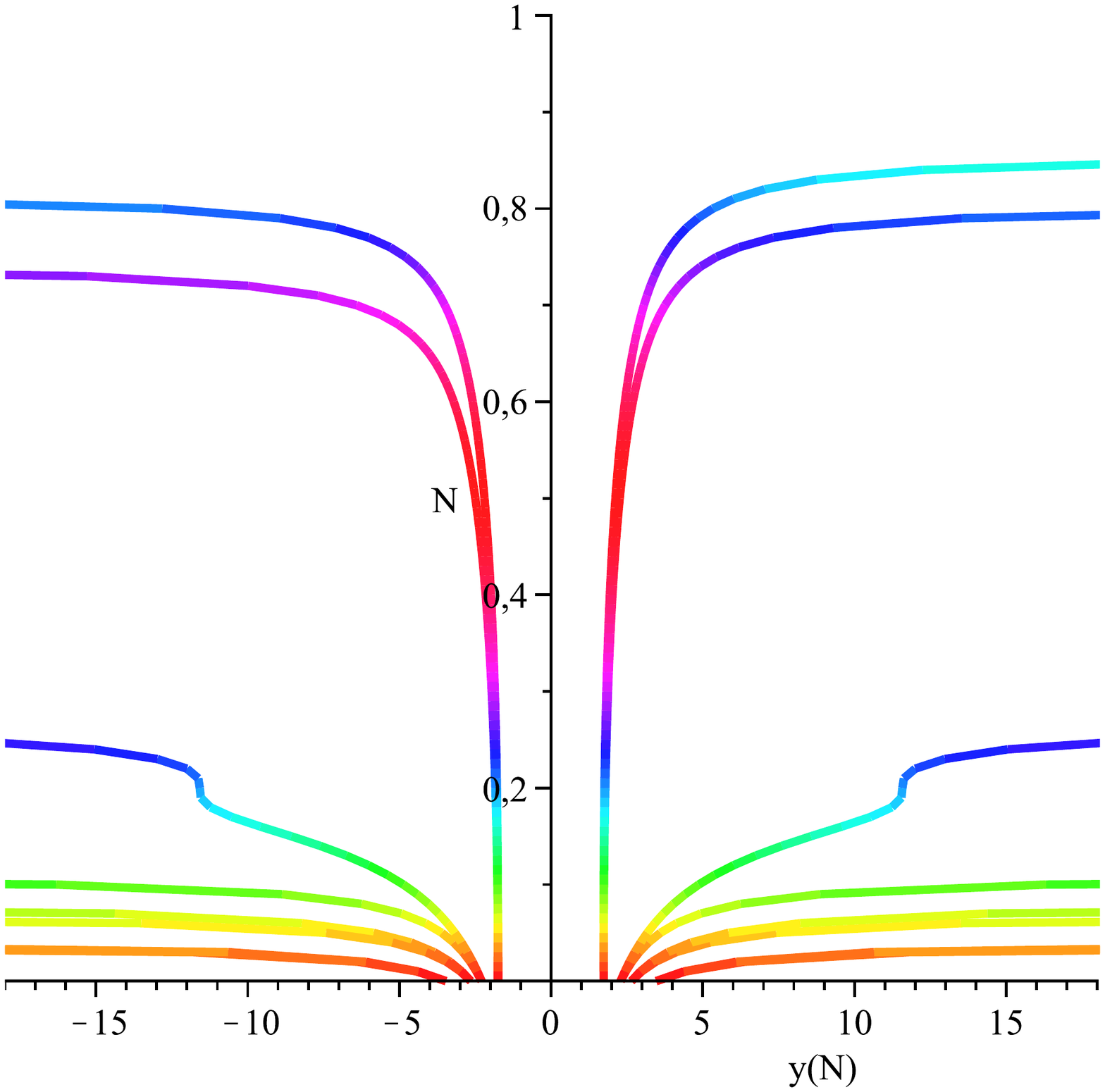}$C.$
\caption{(color online) A. The 3d-phase portrait of system (\ref{EOM-t-x})-(\ref{EOM-t-z}).
 B.  The plot presents dynamics
in the $z-N$ plane.  C. The plot presents dynamics in the $y-N$ plane. }
\label{st}
\end{figure}

We present the 3-dimensional phase portrait of the system of equations (\ref{EOM-t-x})-(\ref{EOM-t-z})
in  Fig.\ref{st}.A.
This is a very remarkable phase portrait. First of all we see the constraint (\ref{cons-th2})
in Fig.\ref{st}.A. We also see, Fig.\ref{st}.C, that for each trajectory there is a fixed number of maximal e-foldings, $N_{max}$, that can make the tachyon with given initial data.
Each trajectory reaches t his number when $z\to \infty$.

%%%%%%%%%%%%%%%%%%%%%%%%%%%%%%%%%
\subsubsection{Tachyon dynamics on e-foldings as 2-dim dynamical system}
%%%%%%%%%%%%%%%%%%%%%%%%%%%%%%%%%

It is instructive to solve firstly the constraint (\ref{cons-th2}) and then to study the dynamics of the reduced 2-dimensional system. Let us introduce the following parametrization
\bea
\label{x-cosh-sin-m}x&=&\sin \psi\cosh \vartheta, \\
\label{x-sinh-sin-m}y&=&\sinh \vartheta, \\
\label{z-cos-m}z&=&\frac1{b}\cos \psi\cosh \vartheta, \eea
where
\bea
-\pi<&\psi&<\pi, \nonumber\\
-\infty<&\vartheta&<\infty.\nonumber
\eea
Let us note that expansion solutions, $H>0$, correspond to $-\pi/2<\psi<\pi/2$.
The contracting solution corresponds to  $-\pi<\psi<-\pi/2$ and $\pi/2<\psi<\pi$.
Relations between $\psi$ and $\vartheta$ and initial $\phi$, $\dot \phi$ and $H$
are given by the following formulae
\bea
\label{phi-dot-trig1}\phi^\prime&=& \frac{\sqrt{2}}{l}\sin \psi\cosh \vartheta,\\
\label{phi-dot-trig2}\dot \phi&=&\sqrt{2\Lambda }\,\tan \psi,\\
\label{y2th-trig1}\phi&=&
=\frac{\sqrt{2}\,b}{l}\,\frac{\tanh \vartheta}{\cos \psi },\\
\label{z2th-trig2}H&=& \frac{\mu b}{\cos \psi\cosh \vartheta},
\eea
here
$
l\equiv\frac{b\sqrt{\mu }}{\Lambda}$.

Equations (\ref{EOM-t-x}) -- (\ref{EOM-t-z}) in these new variables have the form
\bea
\label{EOM-vartheta}\vartheta'&=&\frac32\sin^2 \psi\sinh 2\vartheta+\frac1{2b}
\sin 2\psi\cosh \vartheta,\\
\label{EOM-psi-1}\psi'&=&\frac1{b}\cos^2 \psi\sinh \vartheta
-\frac3{2}\sin 2\psi.\eea

%%%%%%%%%%%%%%%%%%%%%%%%%%%%%%%%%%%%
\subsubsection{The phase portraits of the system of equations (\ref{EOM-vartheta}) and (\ref{EOM-psi-1})}
%%%%%%%%%%%%%%%%%%%%%%%%%%%%%%%%%%%%

\begin{figure}[h!]
\centering
\includegraphics[width=5cm]{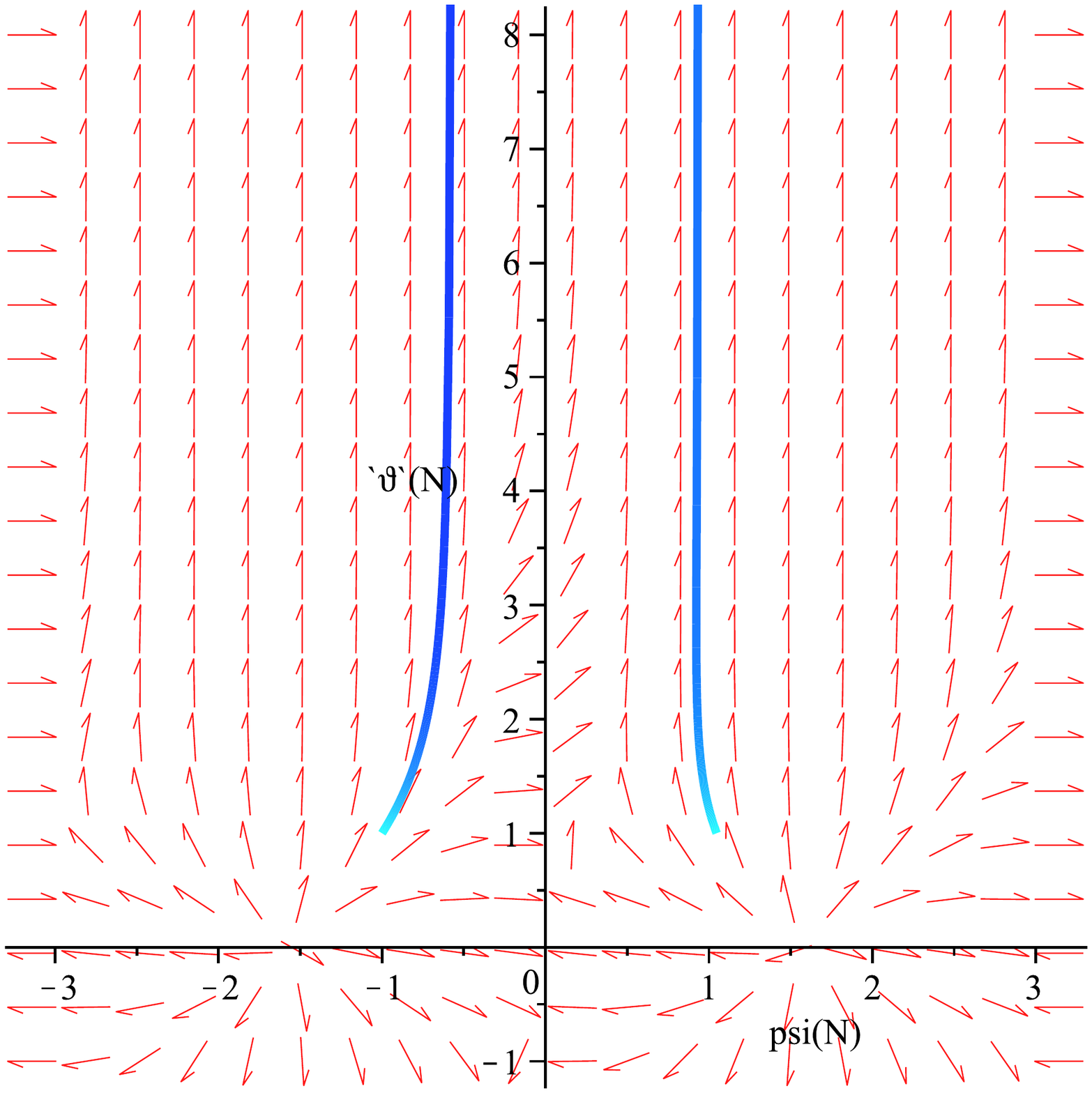}$A.\,\,\,\,\,\,\,\,\,\,\,$
\includegraphics[width=4cm]{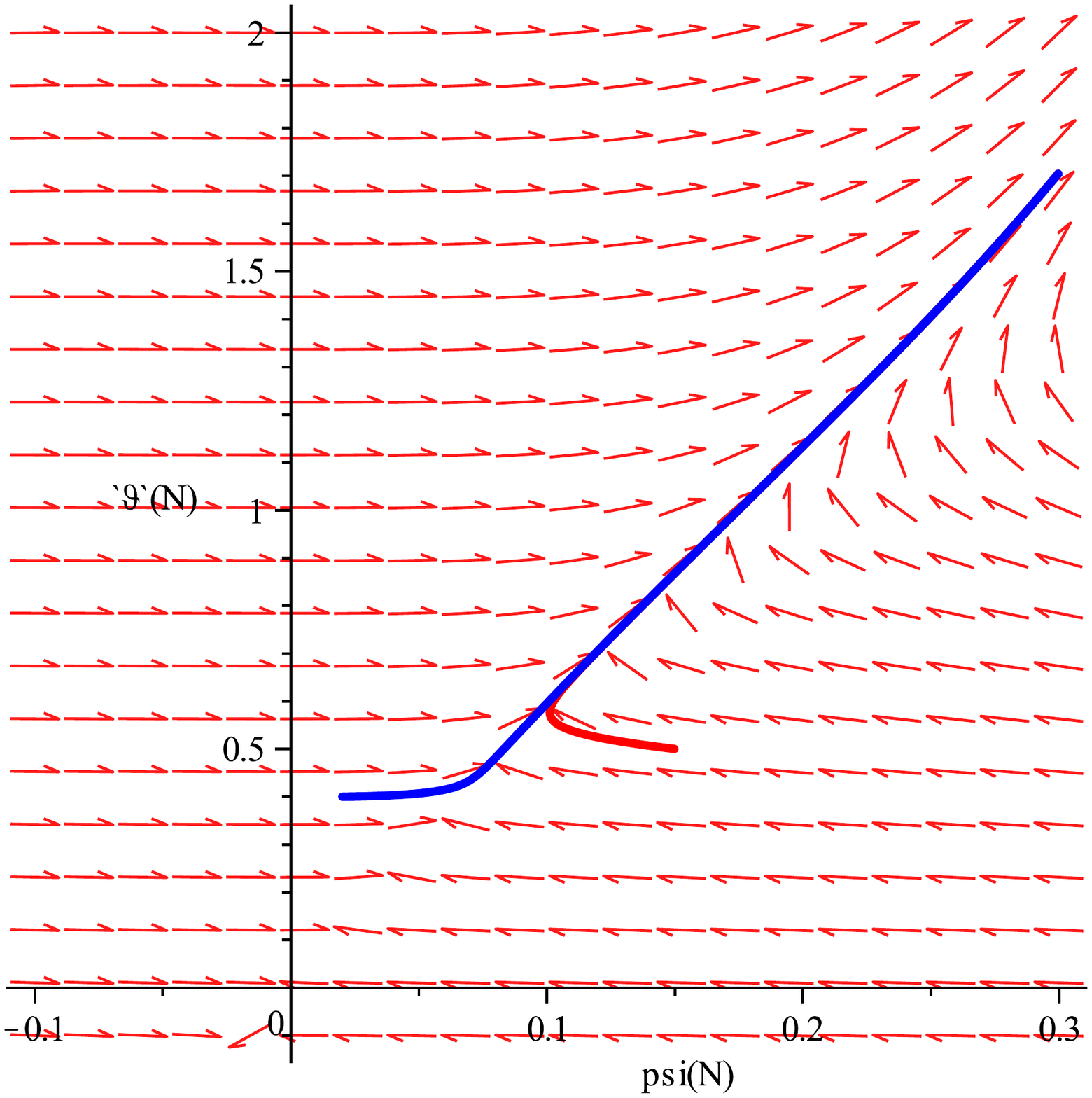}$B.\,\,\,\,\,\,\,\,\,\,\,$
\includegraphics[width=4cm]{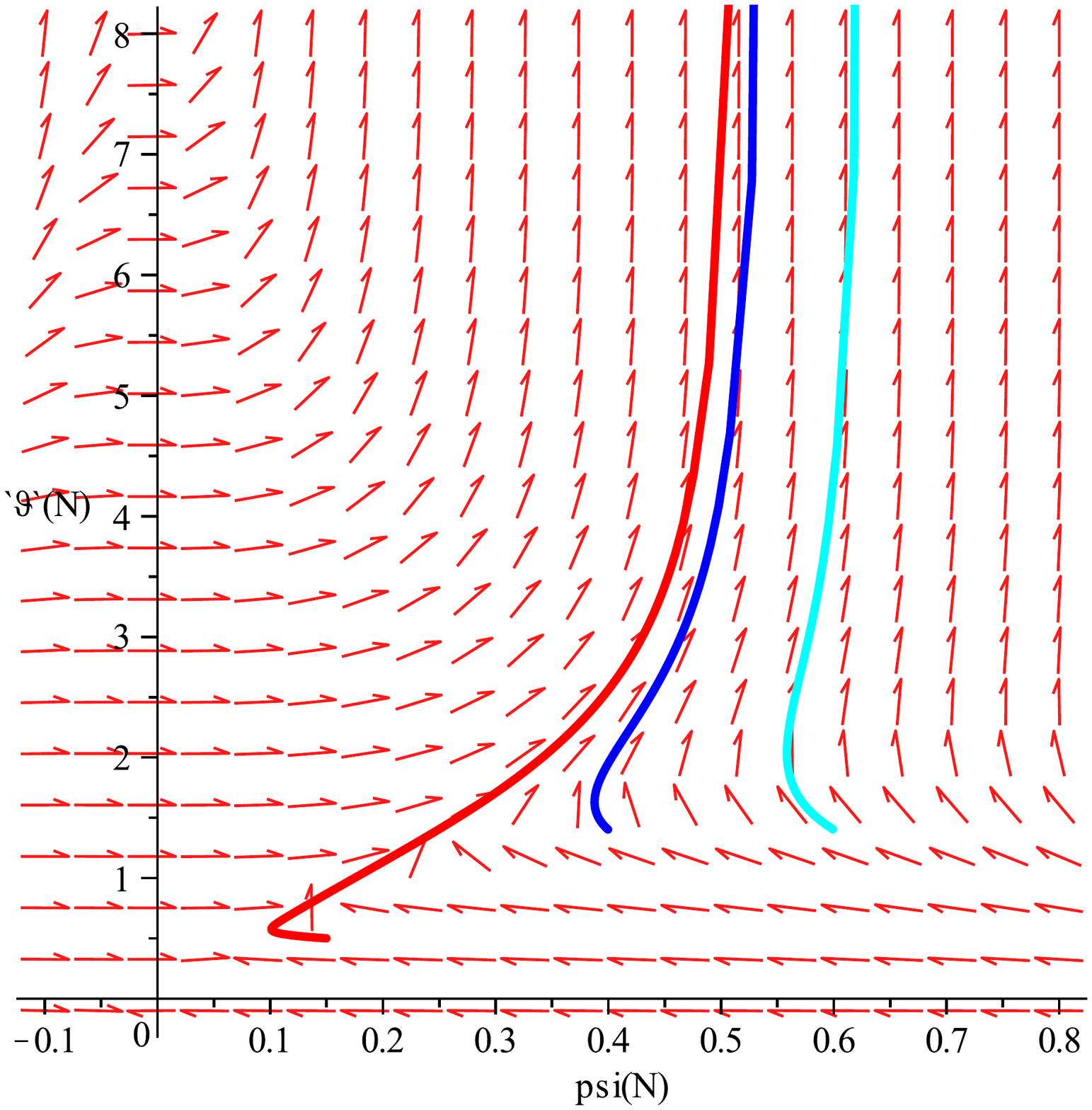}C.
\caption{ A. (color online) The  phase portrait of the system of
equations (\ref{EOM-vartheta}) and (\ref{EOM-psi-1}) for
$-\pi<\psi<\pi$ and $-2<\vartheta<2$. B. The  phase portrait of the
same system for small and positive $\psi$ and small $\vartheta $. We
see an attractor and trajectories go to the attractor from
 both sides of the part of the blue line. C.  The  phase portrait of system (\ref{EOM-vartheta})-(\ref{EOM-psi-1}) for $0<\psi<\pi/2$ and  large $\vartheta$. We see  that starting at  enough large values of $\vartheta $  the value of $\psi$
 is almost constant along the trajectories.
   }
\label{sinh-cos-new-det}
\end{figure}

\begin{figure}[h!]
\centering
\includegraphics[width=5cm]{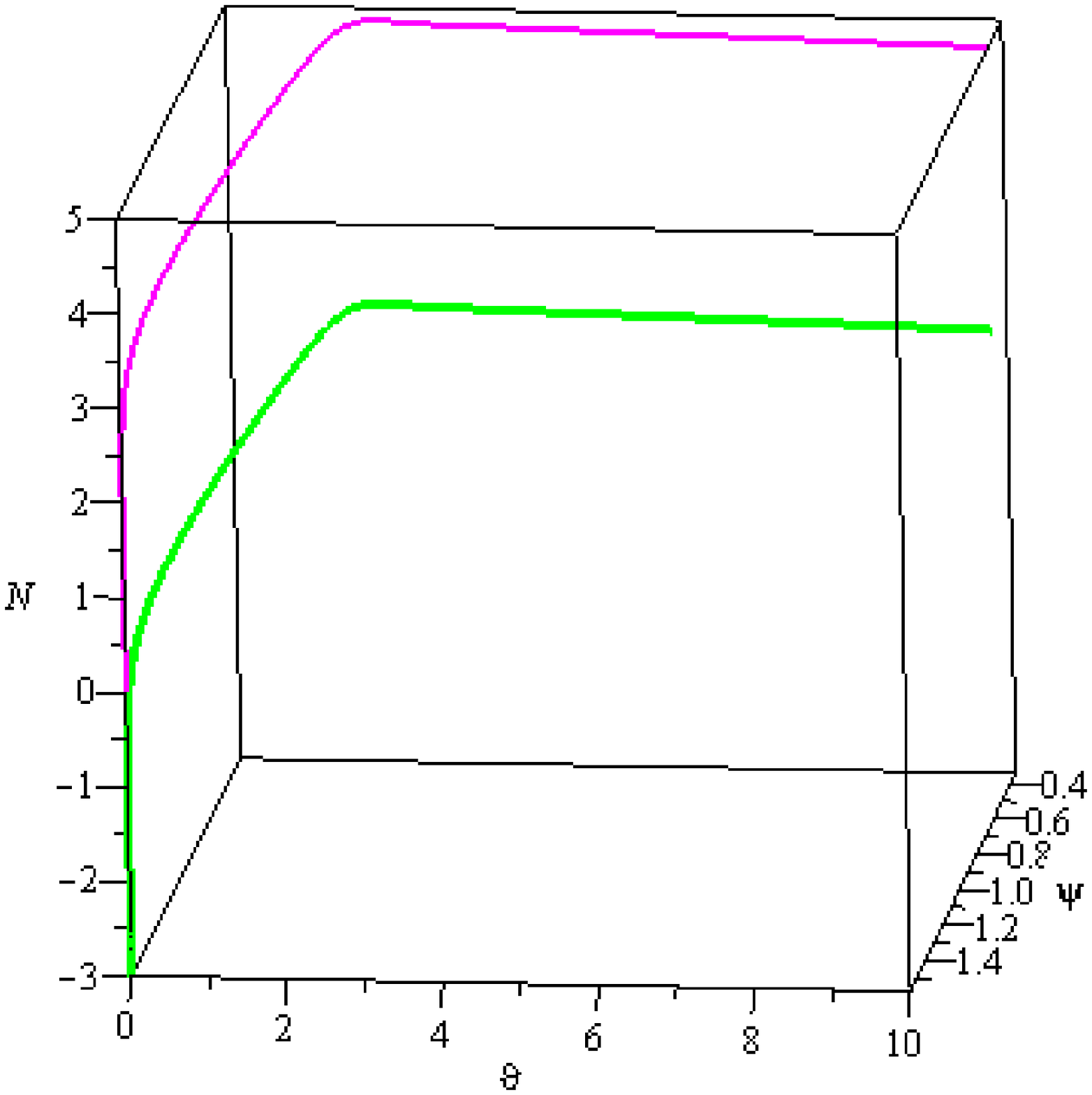}$\,\,\,$
\caption{ (color online) A solution to
(\ref{EOM-vartheta}) and (\ref{EOM-psi-1}). }
\label{psi-vartheta-Nn}
\end{figure}

The phase portrait of the system  of eqs. (\ref{EOM-vartheta}) and (\ref{EOM-psi-1}) is presented in Fig.\ref{sinh-cos-new-det}. The 3d plot of $\psi$ and $\vartheta$ as functions of $N$
 found by numerical solution of the system of eqs.  (\ref{EOM-vartheta}) and (\ref{EOM-psi-1})
 is presented in Fig.\ref{psi-vartheta-Nn}.

In Fig.\ref{sinh-cos-new-det}.A and \ref{sinh-cos-new-det}.C we
see that  $\psi$ is almost constant for large $ \vartheta$, i.e.
for $ \vartheta (N)>\vartheta_0=
\vartheta(N_0)$ we have $\psi(N)\approx \psi(N_0)$.

%%%%%%%%%%%%%%%%%%%%%%%%%%%%%%%%%%%%%%%
\subsubsection{Critical points}
%%%%%%%%%%%%%%%%%%%%%%%%%%%%%%%%%%%%%%

We can see in the phase portrait Fig.\ref{sinh-cos-new-det} that there are the critical points. They solve the equations
\bea
\label{fist-eq}
\frac32\sin^2 \psi_0\sinh 2\vartheta_0+\frac1{2b}
\sin 2\psi_0\cosh \vartheta_0&=&0,\\
\label{second-eq}
\frac1{b}\cos^2 \psi_0\sinh \vartheta_0
-\frac3{2}\sin 2\psi_0&=&0.\eea

There are only the following real solutions to the system of equations (\ref{fist-eq})
and (\ref{second-eq})
\bea\label{psi0-theta0}
\psi_0 &=& \pm \frac{\pi}{2},\,\,\,\,\, \vartheta_0= 0;\\
\label{psi0-theta0}\psi_0 &=& 0,\,\,\,\,\,\,\,\,\,\, \vartheta_0= 0;\\
\label{psi0-theta0-0-pi}\psi_0 &=& \pi,\,\,\,\,\, \vartheta_0= 0.
\eea

%%%%%%%%%%%%%%%%%%%%%%%%%%%%%%%%%%%%%%%%%%%%%
\subsubsection{Near critical points}
%%%%%%%%%%%%%%%%%%%%%%%%%%%%%%%%%%%%%%%%%%%%

Let us examine the system near critical points $\psi_0=\pm\frac\pi2$, $\vartheta_0=0$. Let us take, for example, $\psi\approx\frac\pi2 -\Delta\psi$, $\vartheta\approx\Delta\vartheta$ and linearize (\ref{EOM-vartheta}) and (\ref{EOM-psi-1}). We
get \bea
\label{Delta-vartheta}\Delta\vartheta'&=&3\Delta\vartheta+\frac1{b}\Delta\psi,\\
\label{Delta-psi}\Delta\psi'&=&3\Delta\psi.
\eea
Solutions to (\ref{Delta-vartheta}) and  (\ref{Delta-psi}) are given by
\bea
\label{sol-delta1}\Delta\vartheta&=&\frac12(C_2N+2C_1b)e^{3N},\\
\label{sol-delta2}\Delta\psi&=&C_2\frac{b}{2}e^{3N}.
\eea
Let us write $\phi,\dot\phi$ and $H$ near this point.
Taking into account  (\ref{phi-dot-trig1})-(\ref{z2th-trig2}) we get
\bea
\label{phi-dot-trig}\phi^\prime&\approx&
\frac{\sqrt{2}}{l}+{\cal O}(\Delta\psi^2),\\
\label{phi-dot-trig}\dot \phi&\approx&\sqrt{2\Lambda }\left(\frac{2}{C_2be^{3N}}-\frac{b}{6}C_2e^{3N}\right),
\\
\label{y2th-trig}\phi&\approx&
\frac{\sqrt{2}\,b}{l}\left(\frac12(C_2N+2C_1b)e^{3N}\right)
\left(\frac2{C_2be^{3N}}+\frac{b}{12}C_2e^{3N}\right),\\
\label{z2th-trig}H&\approx&
b\mu\,
\left(\frac2{C_2be^{3N}}+\frac{b}{12}C_2e^{3N}\right)\left(1-\frac12\left(\frac12(C_2N+2C_1b)e^{3N}\right)^2\right).
\eea

%%%%%%%%%%%%%%%%%%%%%%%%%%%%%%%%%%%%%
\subsubsection{Solution for the large $\vartheta$ limit}
%%%%%%%%%%%%%%%%%%%%%%%%%%%%%%%%%%%%%%

\begin{figure}[h!]
\centering
   \includegraphics[width=6cm]{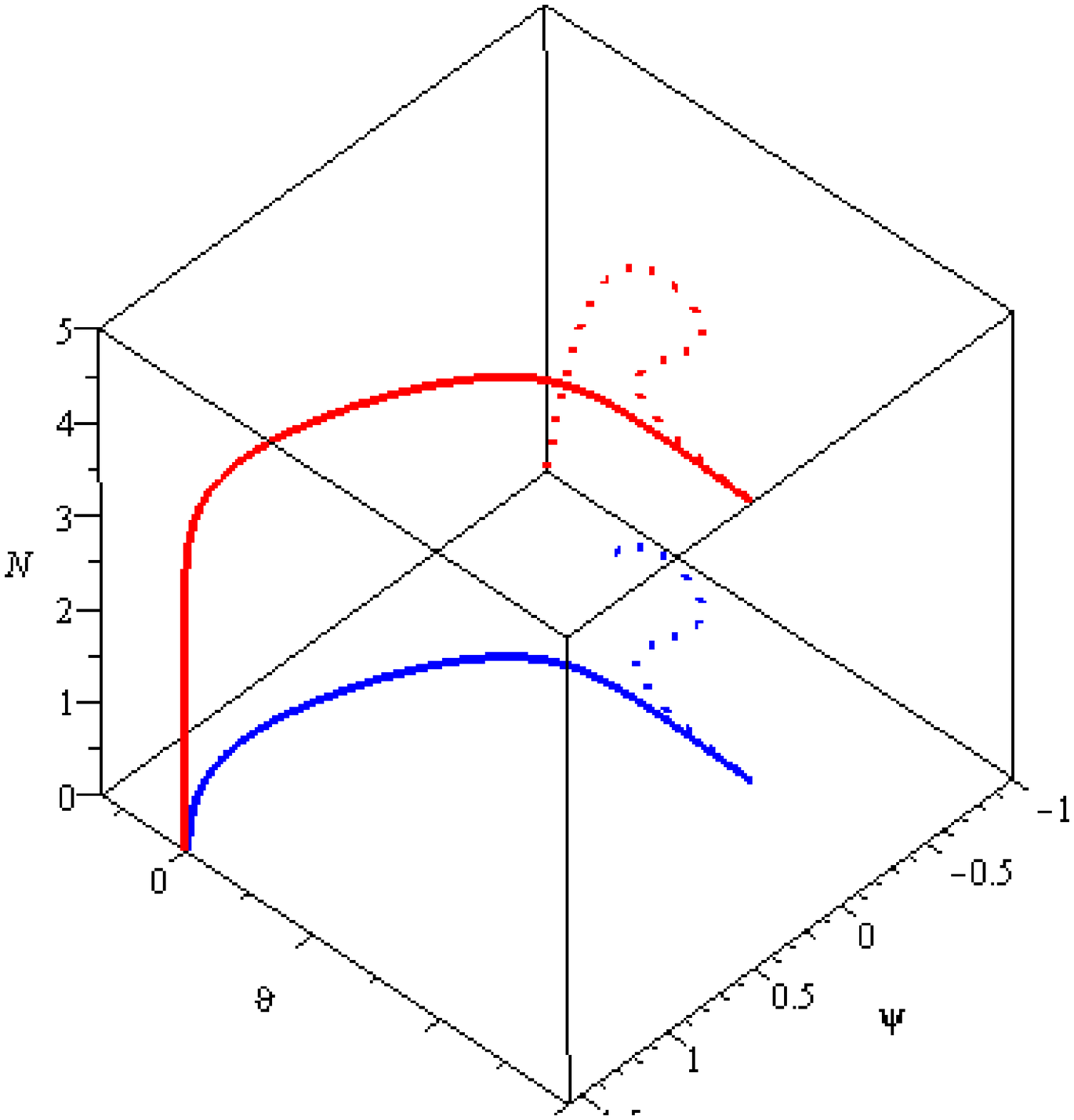}
     \caption{ (color online) The comparison
     of the exact solution  to (\ref{EOM-vartheta}),  (\ref{EOM-psi-1}) (think lines) with the approximate solution (thick dotted lines).}
\label{large-com}
\end{figure}

Let us keep in (\ref{EOM-vartheta}) and  (\ref{EOM-psi-1}) only the terms
that dominate in  the large
$\vartheta$ region. We have
\bea
\label{EOM1l}\vartheta'&=&\frac34\sin^2 \psi \,e^{
2\vartheta},\\
\label{EOM2l}\psi'&=&\frac1{2b}\cos^2 \psi e^{\vartheta}.
\eea
In Fig.\ref{large-com} we can compare the phase portraits for the system of equations (\ref{EOM-vartheta})
and (\ref{EOM-psi-1})
with that of equations (\ref{EOM1l}) and (\ref{EOM2l}).
From these plots we see that for the large $\vartheta$ limit  the approximation
(\ref{EOM1l}), (\ref{EOM2l}) works rather well and for small it does not works at all.

From (\ref{EOM1l}) and (\ref{EOM2l}) we have
\be
\frac{d\vartheta}{e^{\vartheta}}=\frac{3b}{2}\frac{\sin^2 \psi}{\cos^2 \psi}\,d\psi.\nonumber
\ee
That gives
\be
\label{var-psi-m}
e^{-\vartheta}=\frac{3b}{2}( \psi-\tan\psi)+C,\ee
where
\be
C\equiv\frac{3b}{2}(\tan \psi_0-\psi_0)+e^{-\vartheta_0}.\nonumber\ee

Let us now substitute (\ref{var-psi-m}) into (\ref{EOM2l}). We get
\be
\label{psi-n}\psi'=\frac1{2b} \,\frac{\cos^2 \psi}{\frac{3b}{2}( \psi-\tan\psi)+C}.\ee
We see that due to possible zeros of the denominator in the RHS of (\ref{psi-n}) singular points
can appear. Since  $\psi-\tan \psi$ is negative, a positive $C$  leads to a singularity.

\begin{figure}[h!]
\centering
  \includegraphics[width=8cm]{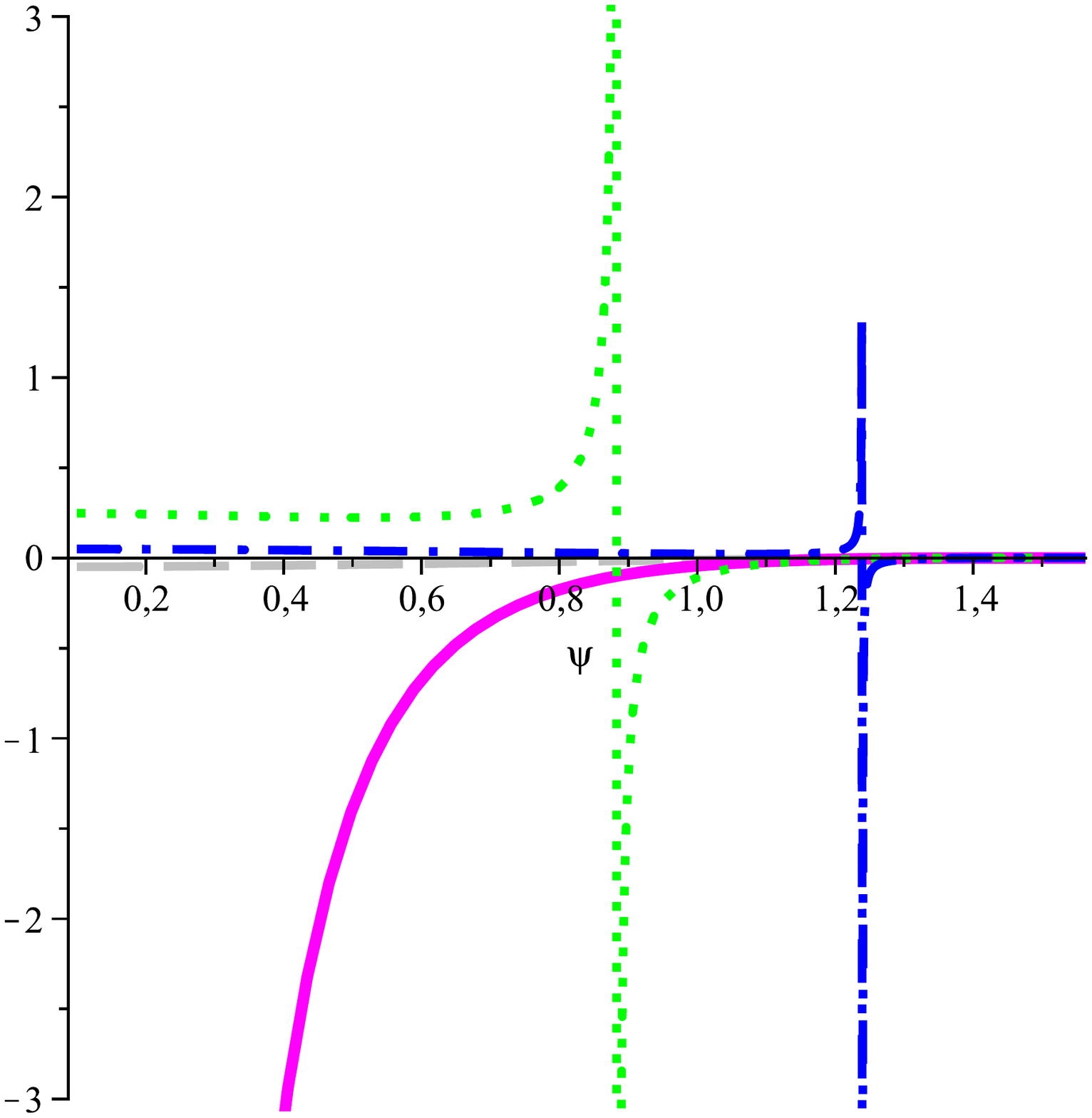}
\caption{(color on line) 2d plot of  the RHS of (\ref{psi-n}) for
  $b=2$ and  $0<\psi<\pi/2$ and different values of $C$:  C=1 (green pointed line), C=-5
 (grey dashed line), C=5 (blue dashed-pointed line) and C=0 (magenta solid line).}
\label{derivatives-m}
\end{figure}

For us it is important that $N$ increases during the dynamics (this corresponds to expansion), meanwhile $\psi$
can increase or decrease, this does not matter. We can say,
that if the RHS of (\ref{psi-n}) is positive, then $\psi$ increases  with increasing $N$,
or if the RHS of (\ref{psi-n}) is negative, then $\psi$ decreases  with increasing $N$.
In the case of a singularity, before the singularity $\psi$ increases,
and after $\psi$ decreases, see Fig.\ref{derivatives-m}.

 We can solve  (\ref{psi-n}) explicitly to get
  an implicit form of a dependence  of $\psi$ on $N$
\be
\label{Upsilon-N}
\Upsilon(\psi,C,b)-\Upsilon(\psi_0,C,b)=N-N_0,\ee
where
\be
\Upsilon(\psi,C,b)\equiv
3\,{b}^{2}\psi\,\tan \psi +3\,{b}^{2}\ln  \left( \cos
 \psi  \right) -{\frac {3{b}^{2}}{ 2\left( \cos \psi  \right) ^{2}}}+2Cb\tan \psi.\nonumber\ee
\begin{figure}[h!]
\centering
$\,\,\,\,\,\,\,\,$
  \includegraphics[width=4cm]{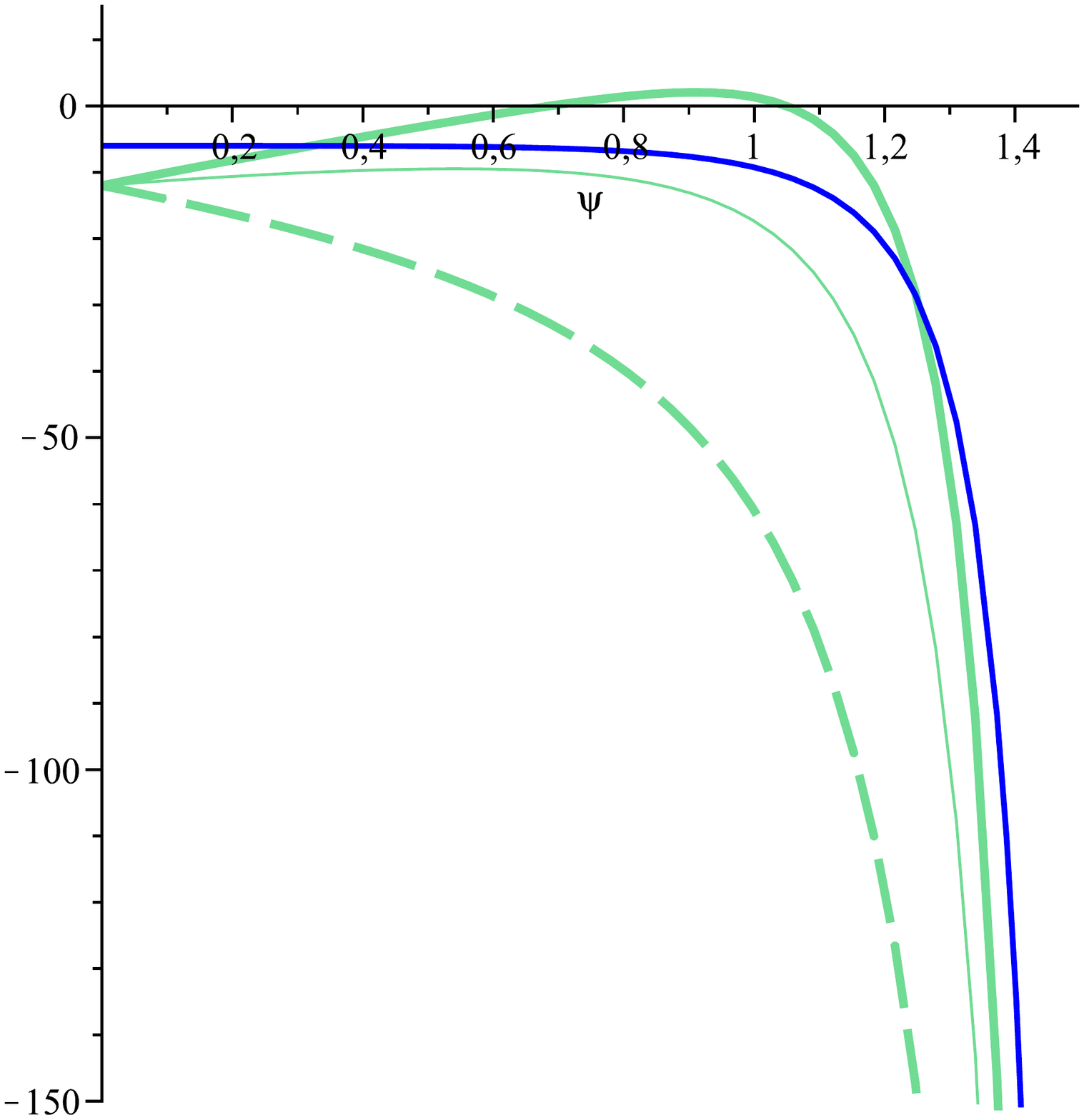}$A.\,\,\,\,$
  \includegraphics[width=4cm]{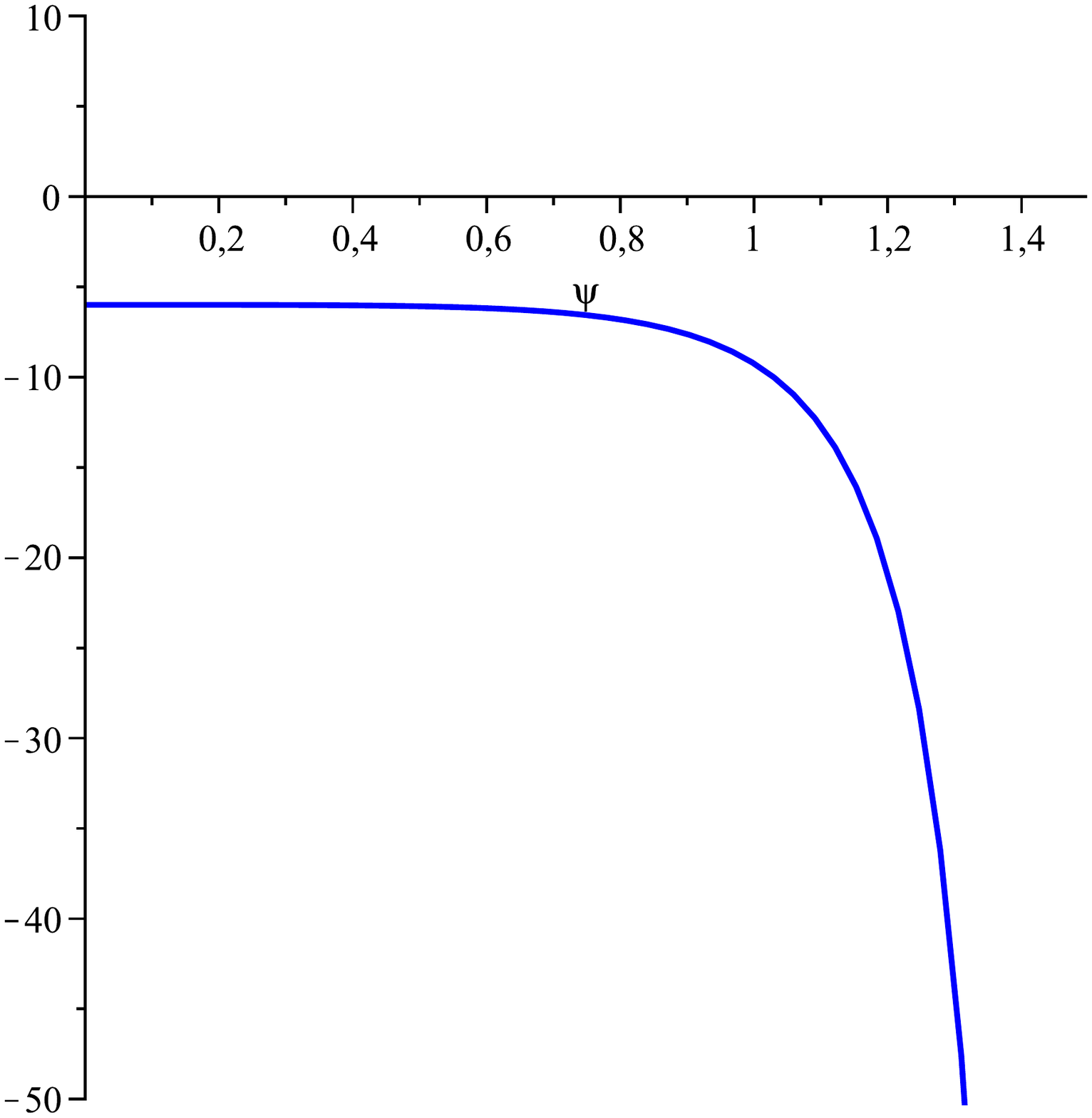}$B.$
              \caption{ (color online)
 A. 2d plot of  $\Upsilon(\psi,C,b)$ for different values of $C$ and
$b=2$ (the solid aquamarine thick line corresponds to $C=5$,
the solid think line corresponds to $C=2$,  the dashed thick line corresponds to $C=-5$,
the solid
blue line corresponds to $C=0$).
B. 2d plot of  $\Upsilon(\psi,C,b)$ for  $C=0$ and
$b=2$.}
\label{psi-nd-mm}
\end{figure}

From Fig.\ref{psi-nd-mm} we see that for positive values of $C$ there are maximal values of $\Upsilon(\psi,C,b)$. These maxima are reached at some point $0<\psi _{max}<\pi/2$,
that satisfies the relation
\be
\label{tan-psi}
3b( \psi_{max}-\tan\psi_{max})+2C=0\,\,\,\,\,\Rightarrow\,\,\,\,\,
\tan\psi_{max}-\psi_{max}=\frac{2C}{3b}.\ee
If $\psi_{max}$ is close to zero, we can expand the LHS of the second formula in (\ref{tan-psi}) in the Taylor expansion and get
$\psi_{max}\approx(\frac{2C}{b})^{1/3}$. This obviously works for $\frac{2C}{b}<1.$

For negative values of $C$  the maxima are at $\psi _{max}=0$ and these solutions
correspond to  decreasing $N$ for increasing $\psi$.

Let us now examine what this approximation gives for the variables $\phi,\dot\phi$.
In the accepted approximation (\ref{phi-dot-trig1})-(\ref{z2th-trig2}) gives
\bea
\label{phi-dot-trig}\phi^\prime&=&  \approx\frac{\sqrt{2}}{2l}\sin \psi e^\vartheta,\\
\label{phidot-trig}
\dot\phi&=&\sqrt{2\Lambda }\,\tan \psi,\\
\label{phi-trig}\phi&\approx&
\frac{\sqrt{2\Lambda }}{\mu}\,\frac{1}{\cos \psi },\\
\label{H-trig}H&\approx&
\frac{2l\sqrt{\Lambda }}{\cos \psi }e^{-\vartheta}.\eea

Note that  instead of the Friedmann equation we  have
\bea
\label{H-2m}
\frac{8\pi G}{3}\frac{\mu^2}{2}\phi^2-
\frac{8\pi G}{3} \,\frac12\, \dot{\phi}^2\approx\frac{8\pi G}{3}\Lambda.
\eea

From formulae (\ref{phidot-trig}), (\ref{phi-trig})  we see that
increasing $\psi$ within $ (0<\psi<\pi/2)$ corresponds  to increasing $\dot\phi$
and $\phi$. From
 the phase portrait Fig.\ref{pp-tach-2} we see that for increasing time keeping on the
part of the solution corresponding to $H>0$ (but very small) we increase $\dot\phi$ and
$\phi$. From this consideration we conclude that to know the behavior near
the ``separatrix'' with positive $H$ we can take the regime near $\psi_{max}$ from the left.
We can also say that the field and its time derivative reach the values
\bea
\label{phi-boundary}
\phi_{boundary}&=&\frac{\sqrt{2\Lambda }}{\mu}\,\frac{1}{\cos \psi_{max} },\\
\label{phi-dot-boundary}\dot\phi_{boundary}&=&\sqrt{2\Lambda }\,\tan \psi_{max},\eea
(that belongs to the boundary of the forbidden region) and then dynamics becomes contracting.

So, we have the following picture:
We start from some $N_0$ and some $\psi_0$.
Suppose that we take a suitable $\vartheta_0$ (let us remind that we assume that
$\vartheta_0$ is large enough) to guarantee that
   \begin{itemize}
   \item  $C>0$. Since  we study dynamics with $H>0$, i.e. the number of foldings should
    increase,  $N_0<N_{max}(C)$. In this case we have increasing  $N$ during the time evolution
    till $N$ reaches $N_{max}(C)$. $N_{max}$ can be found from (\ref{Upsilon-N})
    \be
\label{Upsilon-N-max}
N_{max}=\Upsilon(\psi_{max},C,b)-\Upsilon(\psi_0,C,b)+N_0.\ee
\item   $C=0$. In this case the number of foldings decreases with increasing $\psi$, and
as we have seen before, increasing $\psi$ corresponds to increasing $\phi$.
    \item   $C<0$, here we have also a decreasing number of e-foldings.
     \end{itemize}

We can estimate the dynamics near $\psi=\psi_{max}$.
For this purpose let us expand $\Upsilon$ near $\psi=\psi_{max}$. We have
  \bea
  \label{Upsilon-appr}
  \Upsilon(\psi,C,b)&=&\Upsilon(\psi_{max},C,b)\\
 &+&\frac{3b}{2}\,\frac {\tan \psi_{max}}{ \cos ^2\psi_{max} }
   \left( 2\,b\psi_{max}  +\frac{4}{3}\,C  -
3\,b\tan \psi_{max}  \right)
(\psi-\psi_{max})^2+{\cal O}((\psi-\psi_{max})^3) \nonumber
\eea
 From the condition that we are near the maximum and $0<\psi_{max}<\pi/2$ it follows that
 \be
2\,b\psi_{max}  +\frac43\,C  -
3\,b\tan \psi_{max}=-b\tan \psi_{max}<0\nonumber\ee
Now eq. (\ref{Upsilon-N}) takes the form
\be
\label{Upsilon-N-mmm}
\upsilon
(\psi-\psi_{max})^2\approx-\Upsilon(\psi_0,C,b)+\Upsilon(\psi_{max},C,b)- N+N_0,\ee
where
\be
\upsilon\equiv-\frac12\,\frac {\tan \psi_{max}}{ \cos ^2\psi_{max} }
   \left( 6\,{b}^{2}\psi_{max}  +4\,b C  -
9\,{b}^{2}\tan \psi_{max}  \right)>0\nonumber\ee
and we get
\be
\label{psi-N-explis}
\psi=\psi_{max}-\frac{1}{\upsilon^{1/2}}\sqrt{\Upsilon(\psi_{max},C,b)-
\Upsilon(\psi_0,C,b)- N+N_0}.\ee
As $\psi_{max}$ and $C$ are functions of the initial data denoting
\be
{\cal N}_0\equiv\Upsilon(\psi_{max},C,b)-
\Upsilon(\psi_0,C,b)+N_0,\nonumber\ee
we can rewrite (\ref{psi-N-explis})
as
\be
\label{psi-N-cal}
\psi(N)=\psi_{max}-\frac{1}{\upsilon^{1/2}}\sqrt{{\cal N}_0- N}.\ee

Let's now find the dependence on the time.
Due to (\ref{H-trig}) and  (\ref{var-psi-m}) we have
\bea
\frac{dt}{dN}\approx
\frac{1}{2l\sqrt{\Lambda }}\frac{\cos \psi }{\frac{3b}{2}( \psi-\tan\psi)+C}.\nonumber
\eea
Substituting here the explicit dependence of $\psi$ on $N$, eq. (\ref{psi-N-cal}),
we get
\bea
t-t_0\approx
\int _{N_0}^N\frac{dN}{2l\sqrt{\Lambda }}\frac{\cos \psi(N) }{\frac{3b}{2}( \psi(N)-
\tan\psi(N))+C},\nonumber
\eea
where
$\psi(N)$ is given by (\ref{psi-N-cal}). Making the change of variables
\be
u\equiv\frac{1}{\upsilon^{1/2}}\sqrt{{\cal N}_0- N},\nonumber\ee
we get
\begin{equation*}
 t-t_0\approx
-\int _{\frac{\sqrt{{\cal N}_0- N_0}}{\upsilon^{1/2}}}^{\frac{\sqrt{{\cal N}_0- N}}{\upsilon^{1/2}}}\frac{uvdu}{l\sqrt{\Lambda }}\,\frac{\cos (\psi_{max}-u) }{\frac{3b}{2}( \psi_{max}-u-
\tan(\psi_{max}-u))+C}.
\end{equation*}
Expanding the integrand near $\psi_{max}$ up to $u$ we get
\begin{equation}
\label{t-N-approx}
\begin{split}
t-t_{0}&\approx\frac{1}{l \sqrt{\Lambda} \sqrt{\frac{3}{2}}\tan\psi_{max}}(\sqrt{{\cal N}_0-N_0}-\sqrt{{\cal N}_0-N})\\
&+\frac{1}{l \sqrt{\Lambda} }\frac{(\sin{\psi_{max}}\cos{\psi_{max}}-1)}{3b\sin^3{\psi_{max}}}\cos^2\psi_{max}(N-N_0).
\end{split}
\end{equation}
Let's keep only the leading term (linear in $u$).
Then equation (\ref{t-N-approx}) can be rewritten as
\be
\label{t-N-approx-m}
t\approx t_0-\nu \sqrt{ {\cal N}_0-N},\,\,\,\,{\cal N}_0-N\approx\frac{(t-t_0)^2}{\nu^2},
\ee
where $\nu$ is the function of the initial data and has the form
\bea
\nu=\frac{1}{l \sqrt{\Lambda} \sqrt{\frac{3}{2}}\tan\psi_{max}}.\nonumber
\eea
Substituting this expression into the asymptotic for $\phi$ near $\phi_{boundary}$ given by
(\ref{phi-boundary})
we get
\be
\label{phi-t-approx}
\phi(t)\approx\frac{\sqrt{2\Lambda }}{\mu}\,\frac{1}{\cos (\psi_{max}-\frac{ t_0- t}{\upsilon^{1/2}\nu})}.
\ee

%%%%%%%%%%%%%%%%%%%%%%%%%%%%
\section{Stretched Higgs in FRW}
\setcounter{equation}{0}
%%%%%%%%%%%%%%%%%%%%%%%%%%%%%%%%%%%%%%%%%%%
\subsection{Stretched coupling constant}
%%%%%%%%%%%%%%%%%%%%%%%%%%%%%%%%%%%%%%%%%

In this section we consider the case of non-positively defined Higgs potential. As it was mentioned in Introduction due to non-locality of the fundamental theory the effective stretch of a coupling constant occurs. We will write the equation of evolution in the following form
\bea
\label{EOM-SH}
 &\,&e^{\alpha^2_{eff}}\left(\ddot{\phi}+3H\,\dot{\phi}-\mu^2\phi\right)
 =-\epsilon\phi^3,\\
 H^2&=&\frac{8\pi G}{3}\left(\frac{e^{\alpha^2_{eff}}}{2}
 \,(\, \dot{\phi}^2-\mu^2\phi^2)+\frac{1}{4}\epsilon\phi^4+\Lambda\right).
 \eea
Therefore, we have the ordinary theory with redefined constants
\bea
\label{phi-eff}
&\,&\ddot{\phi}+3H\,\dot{\phi}-\mu^2\phi =- \epsilon _{eff}\phi^3,\\
\label{H-eff}
H^2&=&\frac{8\pi G_{eff}}{3}\left(\frac12
\, \dot{\phi}^2-\frac12\,\mu^2\phi^2+\frac{\epsilon _{eff}}{4}\,\phi^4+
 \Lambda_{eff}\right),
\eea
where
\bea
G_{eff}&=&G\,e^{\alpha^2_{eff}},\\
\epsilon _{eff}&=&\epsilon e^{-\alpha^2_{eff}},\\
\Lambda_{eff}&=&\Lambda e^{-\alpha^2_{eff}}.
\eea
We have also
\be
\label{H-dot-eff}
\dot H=-4\pi G_{eff}\dot \phi ^2.\ee
%%%%%%%%%%%%%%%%%%%%%%%%%%%%%%%%%%%%%%%%%%%%%%%%%%%%%%%%%%%%
\subsection{Phase portraits for different $\Lambda_{eff}$}
%%%%%%%%%%%%%%%%%%%%%%%%%%%%%%%%%%%%%%%%%%%%%%%%%%%%%%%%5

In this subsection we consider the phase portraits for the Higgs field for different values of the total cosmological constant. The Friedmann equation has the form (here and below we write $G$, $\epsilon$ instead $G_{eff}$ and $\epsilon_{eff}$)
 \bea
\label{H-eff}
H^2&=&\frac{8\pi G}{3}\left(\frac12
\, \dot{\phi}^2+\frac{\epsilon}{4}\,(\phi^2-a^2)^2+
 \Lambda\right),
\eea
where
\bea
\Lambda=\Lambda_{eff}-\frac{\epsilon a^4}{4}.\nonumber\eea
 We will consider the cases $\Lambda>0$, $\Lambda=0$ and $\Lambda<0$. As we just mentioned the unusual case $\Lambda<0$ appears due to the effect of the stretch of coupling constant in nonlocal theory.

%%%%%%%%%%%%%%%%%%%%%%%%%%%%%%%%%%%%%%%%%%%%%%%%%%%%%%%%%%%
\subsubsection{The case of Higgs potential without extra cosmological constant, $\Lambda_{eff}=\frac{\epsilon a^4}{4}$}
%%%%%%%%%%%%%%%%%%%%%%%%%%%%%%%%%%%%%%%%%%%%%%%%%%%%%%%%%%%

\begin{figure}[!h]
 \centering
\includegraphics[width=3.5cm]{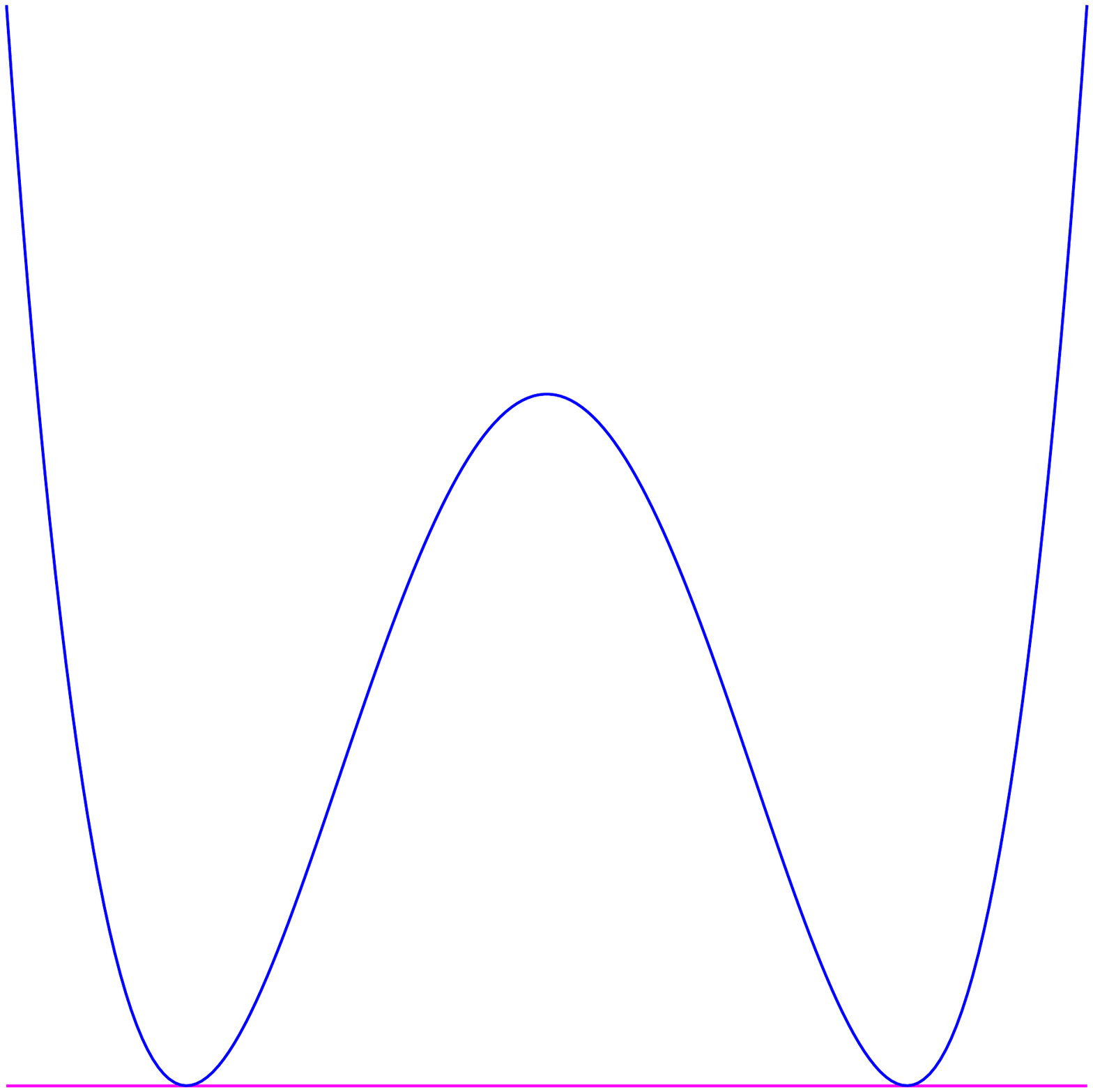}$A.$
\includegraphics[width=3.5cm]{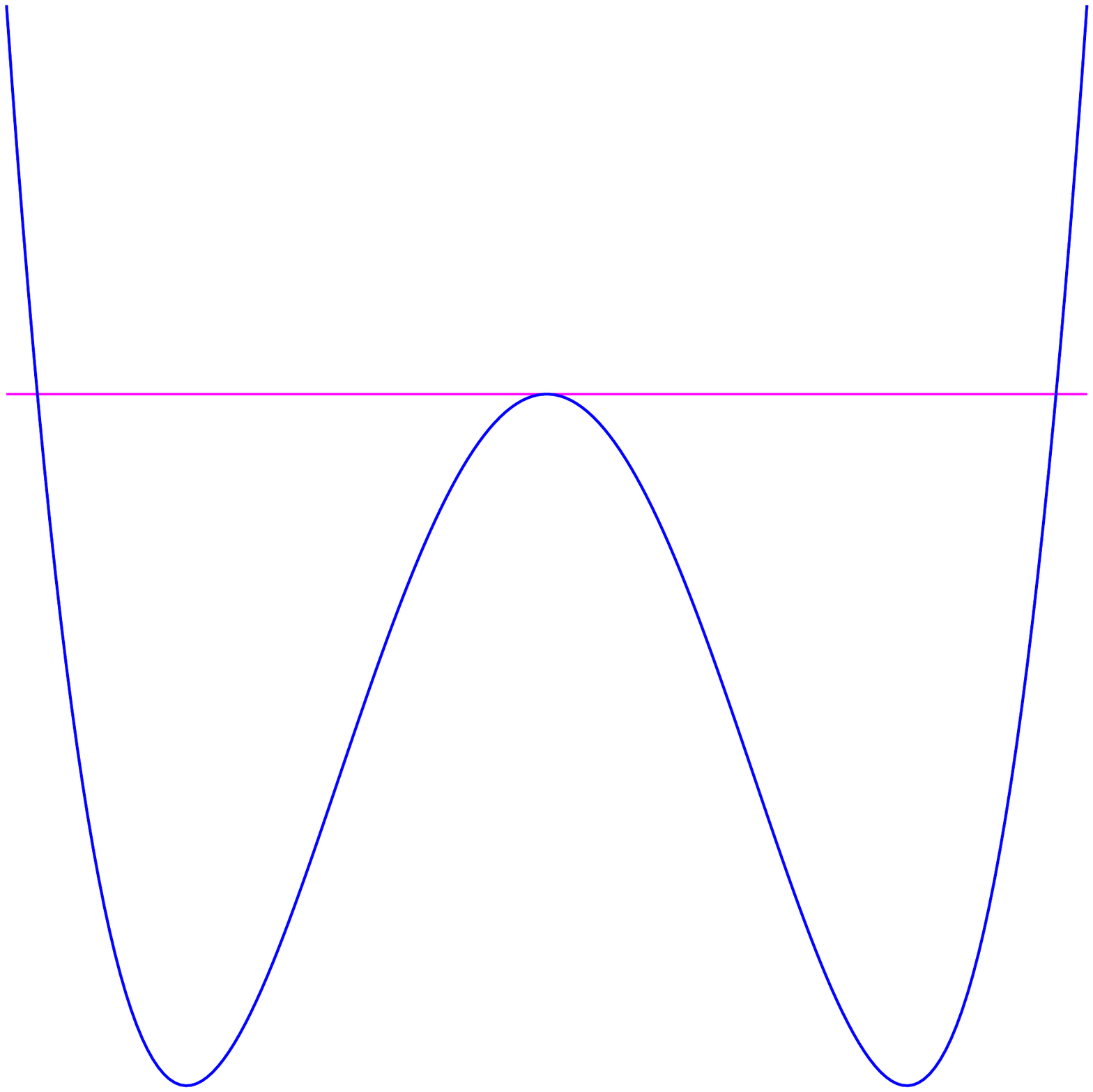}$B.$
\includegraphics[width=3.5cm]{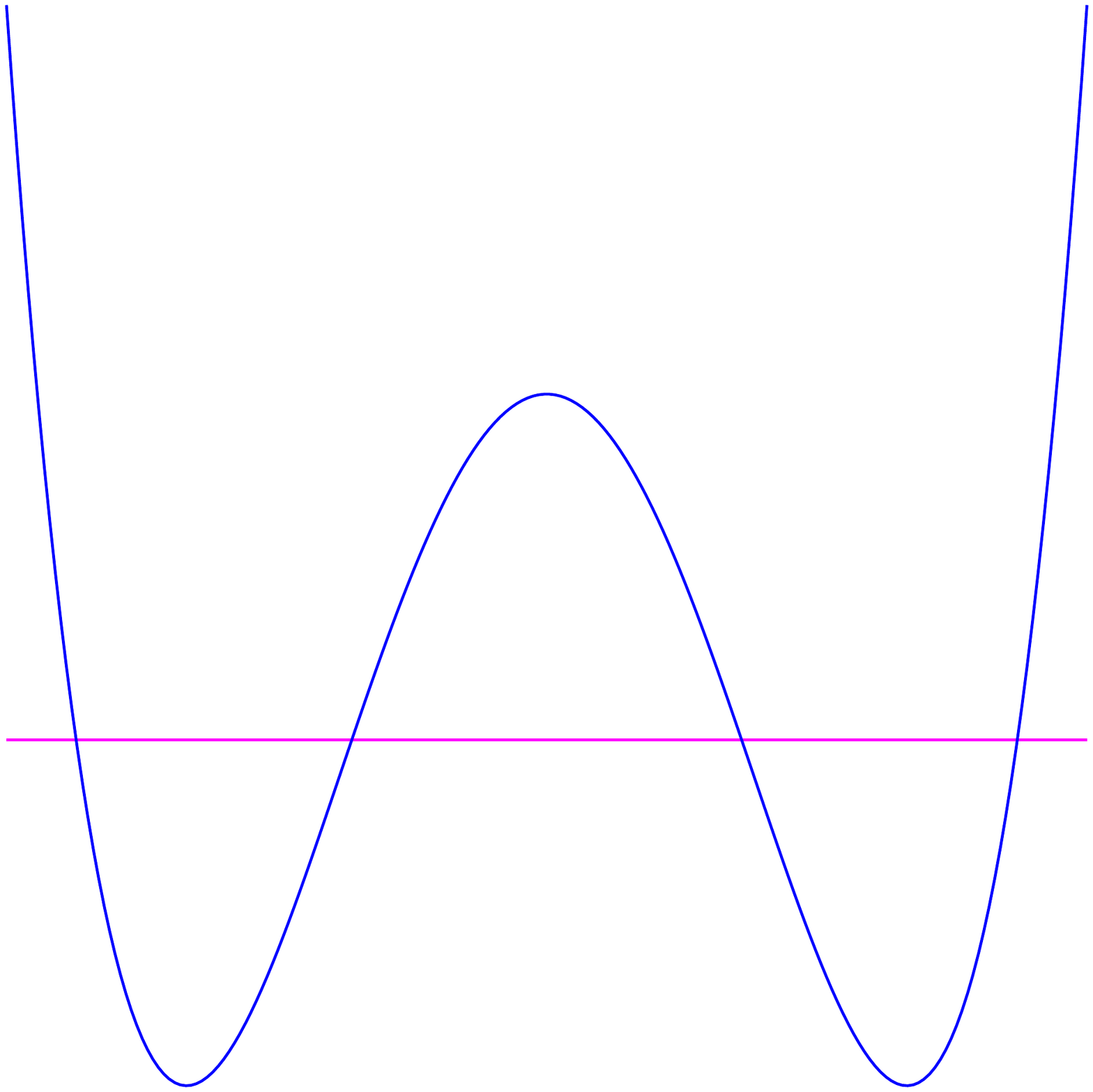}$C.$
\caption{ A. The potential in the case
 $\Lambda_{eff}=\frac{\epsilon a^4}{4}$.
 B. The potential in the case $\Lambda_{eff}=0$.
 C. The potential in the case $0<\Lambda_{eff}<\frac{\epsilon a^4}{4}.$
}
\label{potential-1}
 \end{figure}

In this case we deal with the potential presented in Fig.\ref{potential-1}.A.  and the phase portrait doesn't have a forbidden domain. This case has been considered in many works and we discuss it in  Sect.2.

%%%%%%%%%%%%%%%%%%%%%%%%%%%%%%%%%%%
\subsubsection{$\Lambda_{eff}=0$}
%%%%%%%%%%%%%%%%%%%%%%%%%%%%%%%%%%%

In the case $\Lambda_{eff}=0$, see Fig.\ref{potential-1}.B., the phase portrait does have the forbidden region and the reheating domain is totally or partially removed (dependently on the trajectory). The characteristic feature of this case is an appearance of the contraction regime.   To see this let's consider this case in more details.

\begin{figure}[!h]
 \centering
 $\,\,\,\,\,\,\,\,\,\,\,\,\,$
 $\,\,\,\,\,\,\,\,\,\,\,\,\,$
 \setlength{\unitlength}{1mm}
\begin{picture}(50,70)
\put(-25,-2){\includegraphics[width=7cm]{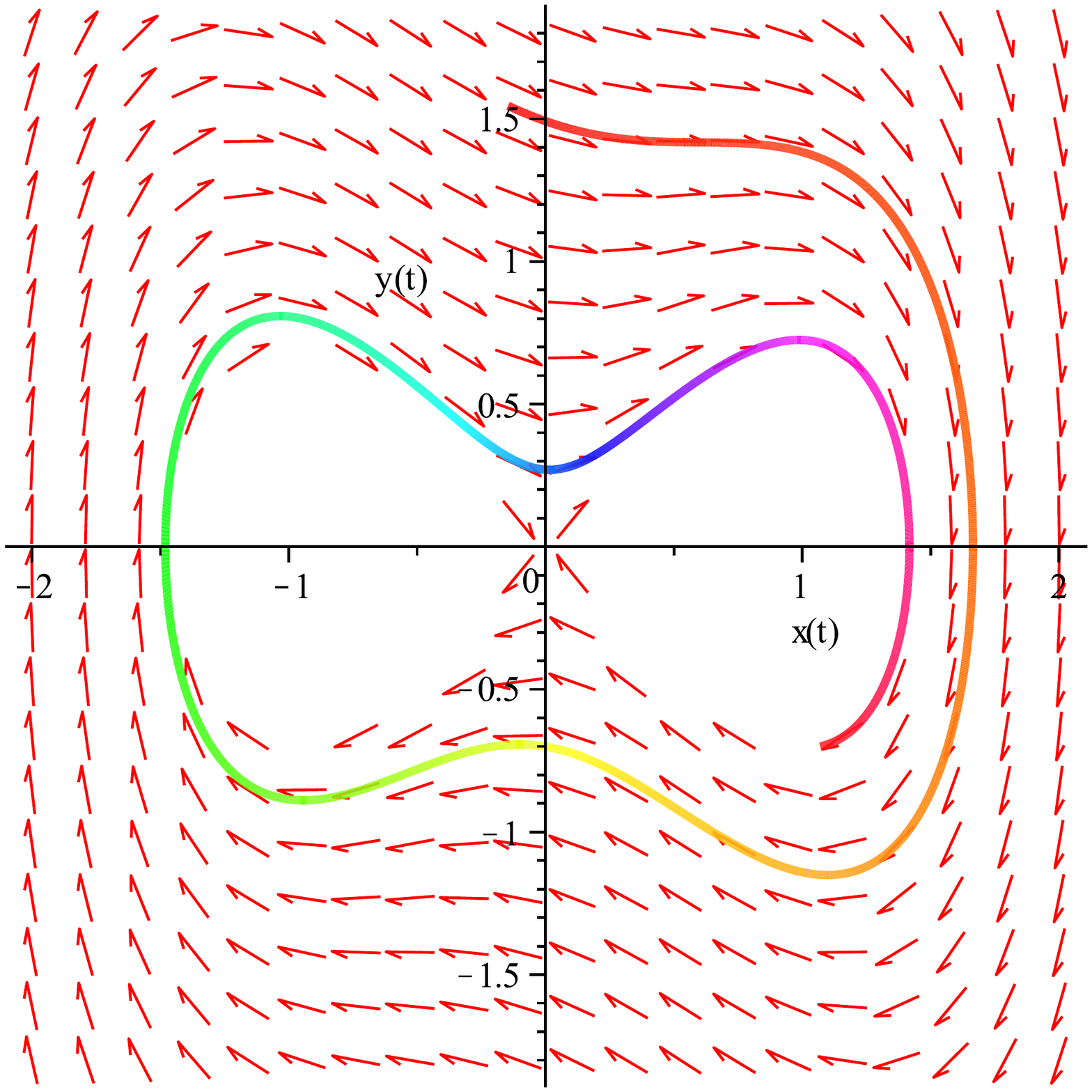} }
%\put(0,0.5){\line(1,0){5}} \put(0.5,1){\circle*{0.2}}
%\put(25,61){${\bf S_1}$}\put(26,65){\circle*{1}} \put(25,22){${\bf
%S_2}$}\put(27,20){\circle*{1}} \put(-6,22){${\bf
%S_3}$}\put(-3.5,20.5){\circle*{1}} \put(-3,64){{\bf 1}}
%\put(25,67){{\bf 2}} \put(35,67){{\bf 3}}
\put(40,35){$\phi$}
\put(5,68){$\dot\phi$}
\end{picture}
 $A.\,\,\,\,\,\,\,\,\,\,\,\,$
\begin{picture}(50,70)
\put(-10,2){\includegraphics[width=7cm]{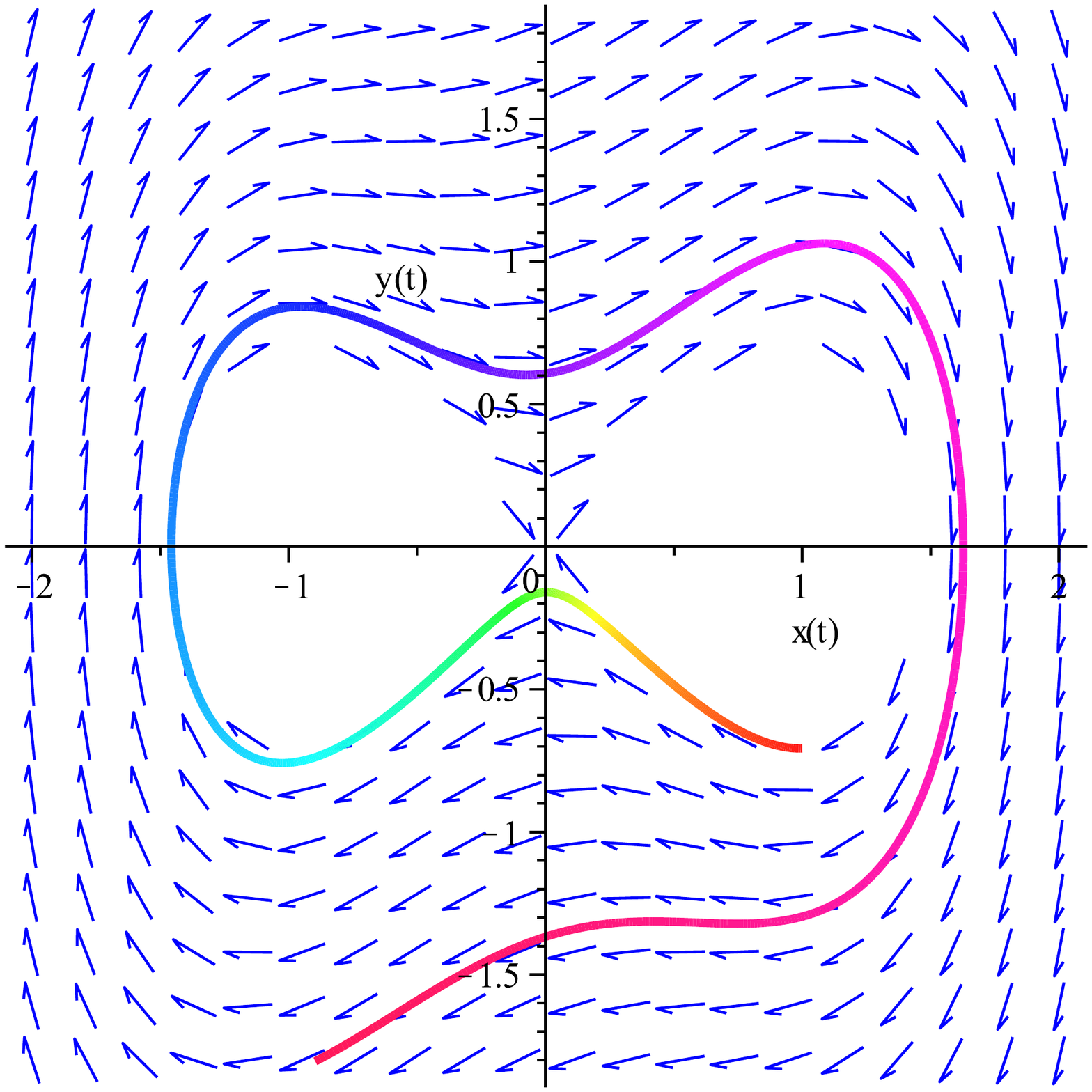}}
 %\put(42,40){$1^\prime$} \put(33,29){$2^\prime$}
%\put(3,31){$3^\prime$} \put(38,46){${\bf
%S_1}$}\put(43,49){\circle*{1}} \put(38,25){${\bf
%S_2}$}\put(40.5,24){\circle*{1}} \put(7,26){ ${\bf
%S_3}$}\put(10,24){\circle*{1}}
\put(55,35){$\phi$}
\put(25,71){$\dot\phi$}
\end{picture}$B.$
\caption{ (color online)
 A. The phase portraits of  equation (\ref{EOM-T4-zero-Lambda}).
 The forbidden region  consists of two merged   holes.
Colored line presents the trajectory starting from some point of the phase diagram.
The trajectory reaches the border of the forbidden area.
B. The phase portrait of equation (\ref{EOM-T4-zero-contr}).
Colored line presents trajectory that is  continuation of the  trajectory
presented in the left panel.
 It  is a solution to the Friedman
equation with negative values of $H$.  }
\label{pp-tach-4-0}
 \end{figure}

Our goal is to see the future and the past singularities of the Higgs scalar field in the FRW metric.
 EOM describing inflation is
\be
\label{EOM-T4-zero-Lambda}
 \ddot{\phi}+3\sqrt{\frac{8\pi G}{3} \left(\,\frac12\, \dot{\phi}^2-\frac{\mu^2}{2}\phi^2+\frac{\epsilon}{4}\phi^4\right)}\,\dot{\phi}
 =+\mu^2\phi-\epsilon\phi^3.\ee
 The phase portrait of this system is presented in Fig.\ref{pp-tach-4-0}.A.
The contraction phase satisfies equation
\be
\label{EOM-T4-zero-contr}
 \ddot{\phi}-3\sqrt{\frac{8\pi G}{3} \left(\,\frac12\, \dot{\phi}^2-\frac{\mu^2}{2}\phi^2+\frac{\epsilon}{4}\phi^4\right)}\,\dot{\phi}
 =+\mu^2\phi-\epsilon\phi^3\ee and its phase portrait is presented in Fig.\ref{pp-tach-4-0}.B.
To study the infinite phase space following \cite{BGXZ} let us
use dimensionless variables ($X, Y, Z$) (\ref{X})-(\ref{Z}).
In these variables the dynamical system has the form
\bea
\label{EOM-X-mu-g}X_\tau&=&Y,\\
\label{EOM-Y-mu-g}Y_\tau&=&X-3ZY-gX^3,\\
\label{EOM-Z-mu-g}Z_\tau&=&-X^2-2Y^2-Z^2+\frac{g}{2}X^4,
\eea
where $g=\epsilon\frac34\frac{m_p^2}{\pi\mu^2}$.

The integral of motion is
\be
-X^2+Y^2-Z^2+\frac{g}{2}X^4.\nonumber\ee
To describe $\Lambda_{eff}=0$ we take
\be
\label{IM-g}-X^2+Y^2-Z^2+\frac{g}{2}X^4=0.\ee
The Friedmann equation keeps the dynamics of the system on the surface (\ref{IM-g}). This
surface  has two parts: one corresponds to expansion ($H>0$)
 and the second one corresponds to contraction.

The phase portrait of  the dynamical system (\ref{EOM-X-mu-g})-(\ref{EOM-Z-mu-g}) is presented in Fig.\ref{pp-tach-4-0}.

To study the infinite phase space  we use the
spherical  coordinates $(\rho, \theta, \psi)$ (\ref{X-t})-(\ref{Z-t}), see Section 3.
In these coordinates EOM have the form
\bea
\label{EOM-rho}
\rho_{\sigma} &=&\rho(1-\rho)^2\left[\sin2\psi\sin^2\theta-\frac{\rho}{1-\rho}\cos\theta(1+4\sin^2\psi\sin^2\theta)\right.\nonumber\\
&&-\left.g\frac{\rho^2}{(1-\rho)^2}\cos^3\psi\sin^3\theta\left(\sin\psi\sin\theta-\frac12\sin2\theta\cos\psi\right)\right] \\
\label{EOM-psi2}
\psi_{\sigma}&=&-\frac32\rho\sin2\psi\cos\theta+(1-\rho)\cos2\psi-g\frac{\rho^2}{1-\rho}\sin^2\theta\cos^4\psi\\
\label{EOM-theta}
\theta_\sigma &=&\frac12(1-\rho)\sin2\psi\sin2\theta+\rho\sin\theta(1+\sin^2\psi-4\sin^2\psi\cos^2\theta)\nonumber\\
&&-\frac{g}{2}\frac{\rho^5}{(1-\rho)^4}\sin^5\theta\cos^4\psi
\eea

\begin{figure}[!h]
 \centering
 $\,\,\,\,\,\,\,\,\,\,\,\,\,$
 $\,\,\,\,\,\,\,\,\,\,\,\,\,$
 \setlength{\unitlength}{1mm}
\begin{picture}(25,70)
\put(-25,-2){
 \includegraphics[width=4cm]{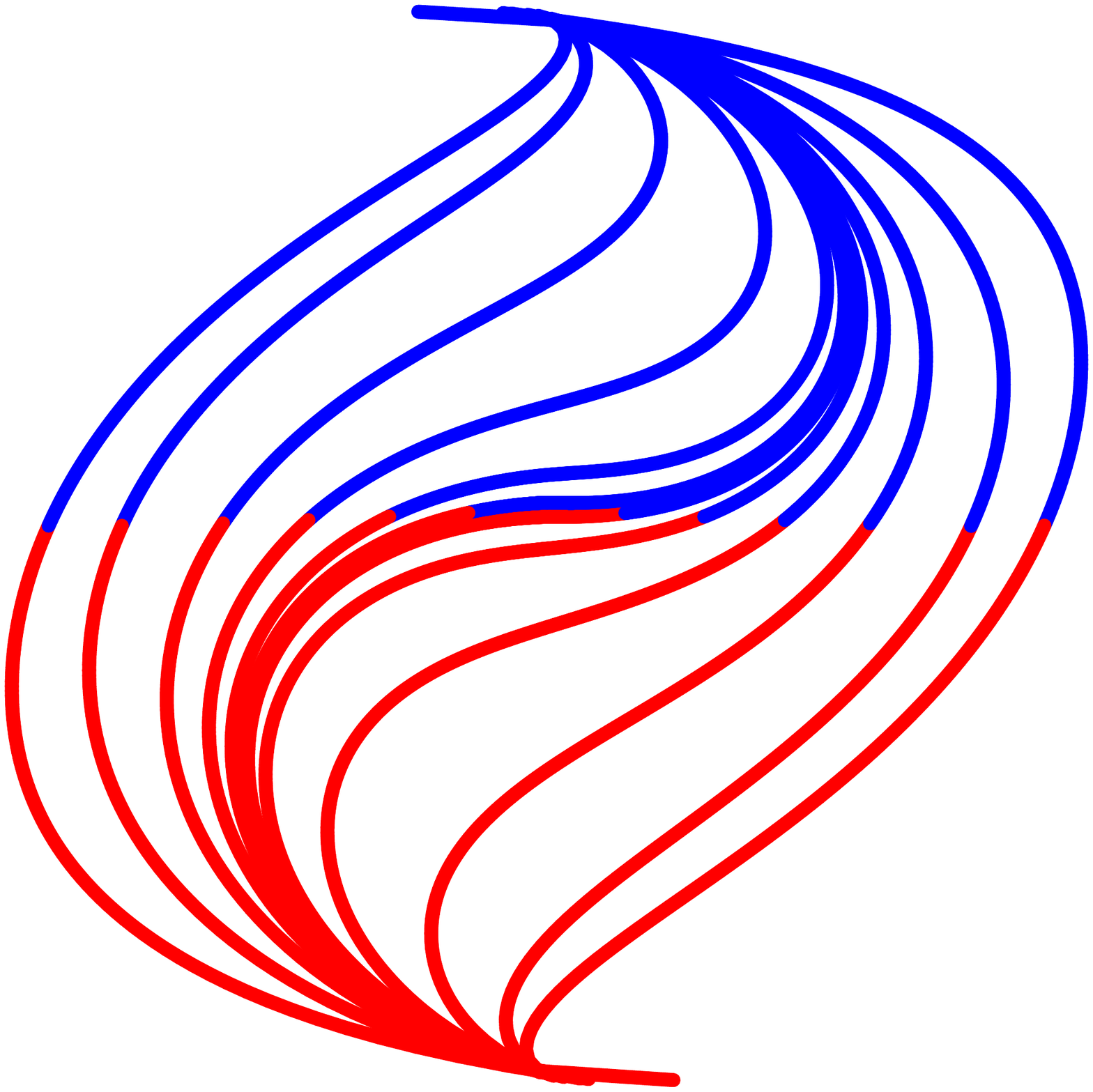}
 }
\put(0.5,6){\circle*{1.5}} \put(-28,25.5){\line(1,0){45}}
\put(-8,47){{\bf FS}} \put(0,8){{\bf PS}}
\put(-9.5,43.5){\circle*{1.5}}
\end{picture}$A.\,\,\,\,\,\,\,\,\,\,\,\,\,\,\,\,\,\,$
\begin{picture}(20,70)
\put(0,-2){
\includegraphics[width=4cm]{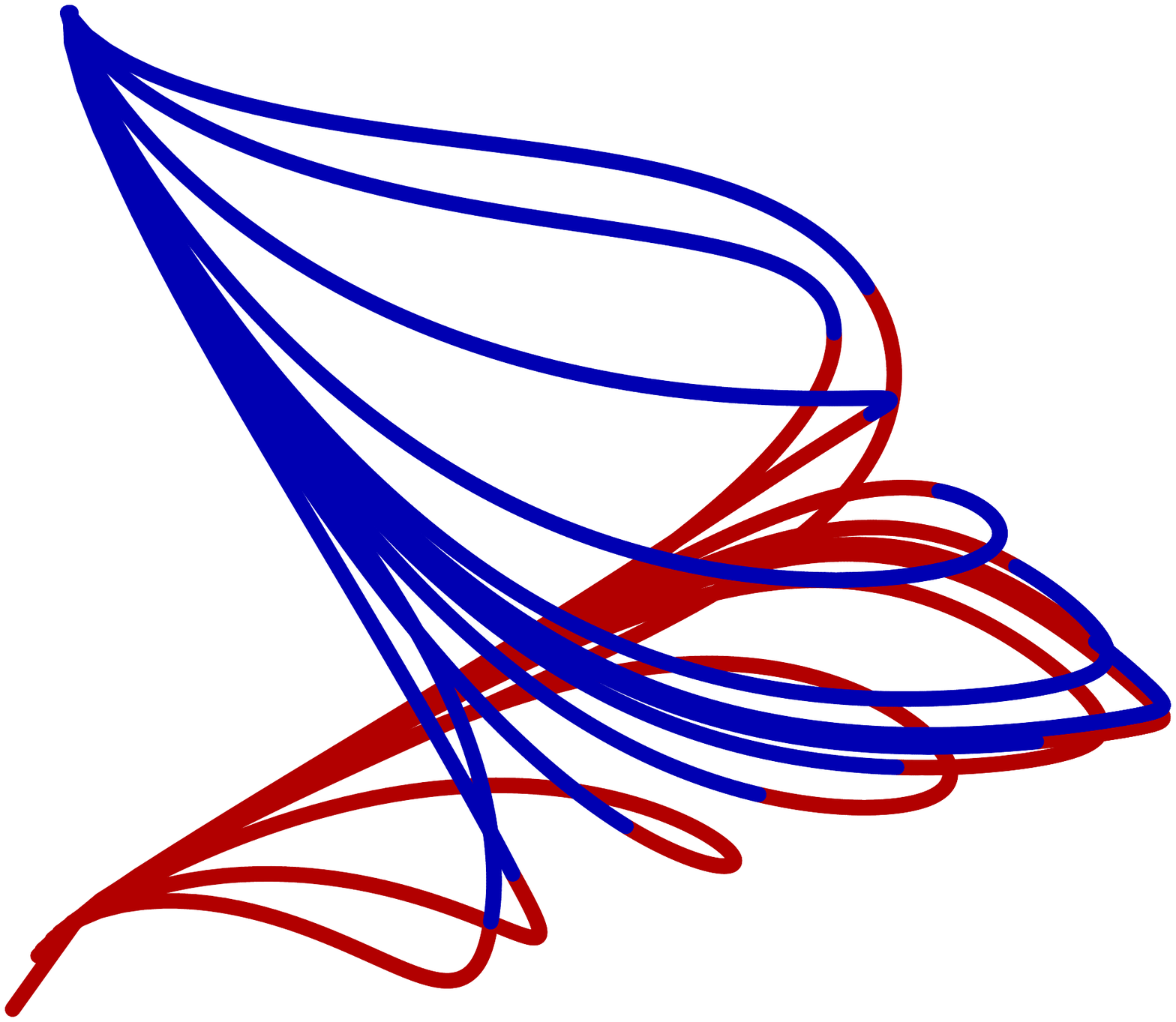}
}
\put(-8,48){{\bf FS}} \put(-3,7){{\bf PS}}
\end{picture}$B.$
\caption{ (color online)  A. The phase portrait of equations (\ref{EOM-rho}) -- (\ref{EOM-theta})
 in the $(\rho,\theta)$-plane. The black thin  line corresponds to
$\theta=\pm \pi/2$.
 Blue parts of lines represent contraction and red
parts represent expansion. We see the symmetry with respect to replacing the red parts by the blue parts
of lines. {\bf FS} and {\bf PS} are future and past cosmological singularities. B. The 3-dimensional plot in $(\rho,\theta, \psi)$ coordinates
of the same phase portrait.}
\label{pp-4-sphere}
\end{figure}

The phase portrait of this system of equations is presented in Fig.\ref{pp-4-sphere}.
We see that trajectories are symmetric under the reflection with respect to the boundary of the forbidden
area. We also see the two singular points -- at the past (cosmological singularity)
and the same in the future.

%%%%%%%%%%%%%%%%%%%%%%%%%%%%%%%%%%%%%%%%%
\subsubsection{$0<\Lambda_{eff}<\frac{\epsilon a^4}{4}$}
%%%%%%%%%%%%%%%%%%%%%%%%%%%%%%%%%%%%%%%%%

In this case the potential is presented in Fig.\ref{potential-1}.C., the forbidden domain contains two non-connected parts.

Near the top of the potential the Higgs potential can be approximated with the tachyon potential
\bea
\label{EOM-SH}
 V(\phi)\approx-\frac{\epsilon a^2}{2} \phi^2 +\Lambda_{eff}.
 \eea

\begin{figure}[!h]
 \centering
\includegraphics[width=5cm]{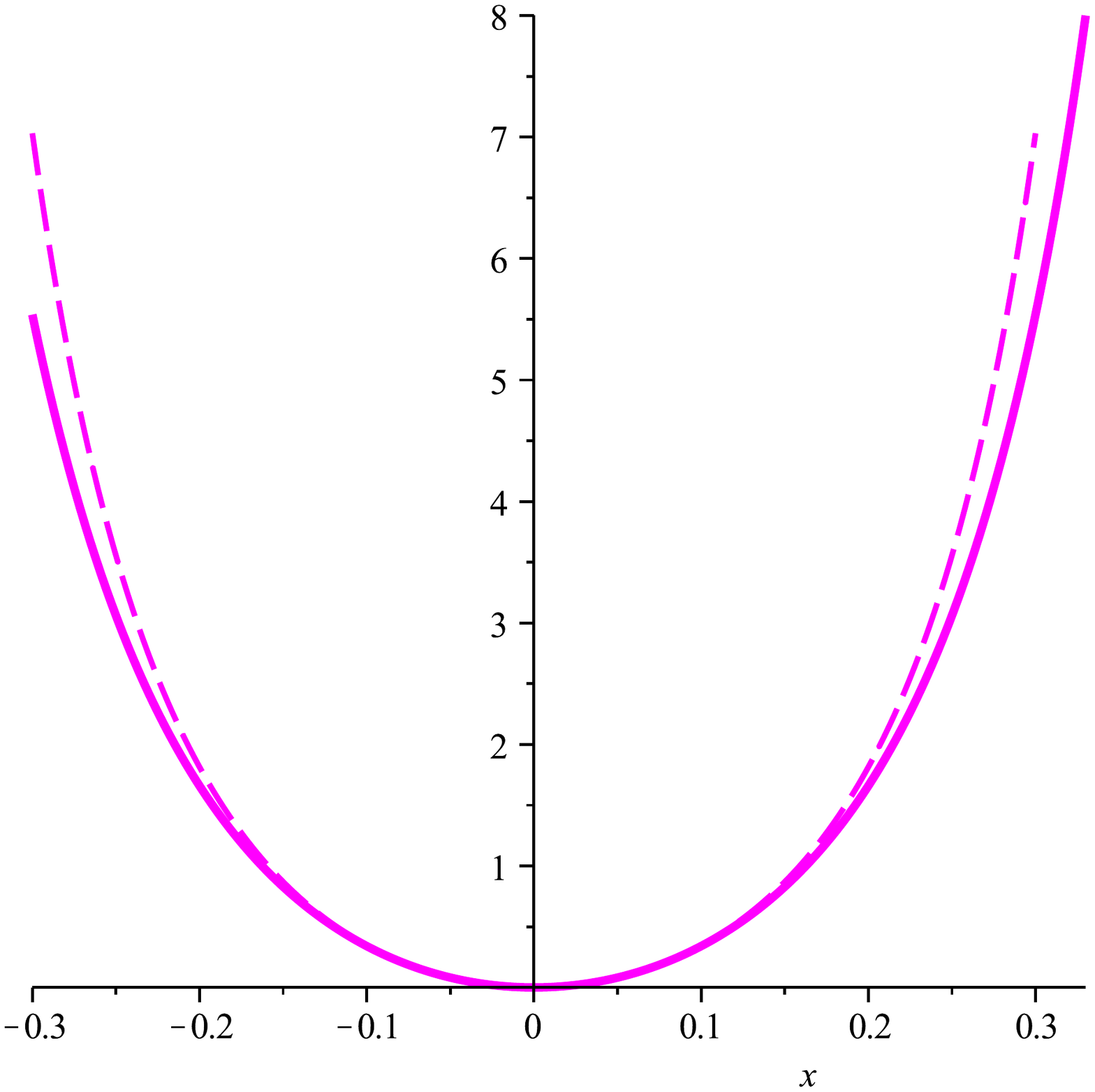}$A.$
\includegraphics[width=5cm]{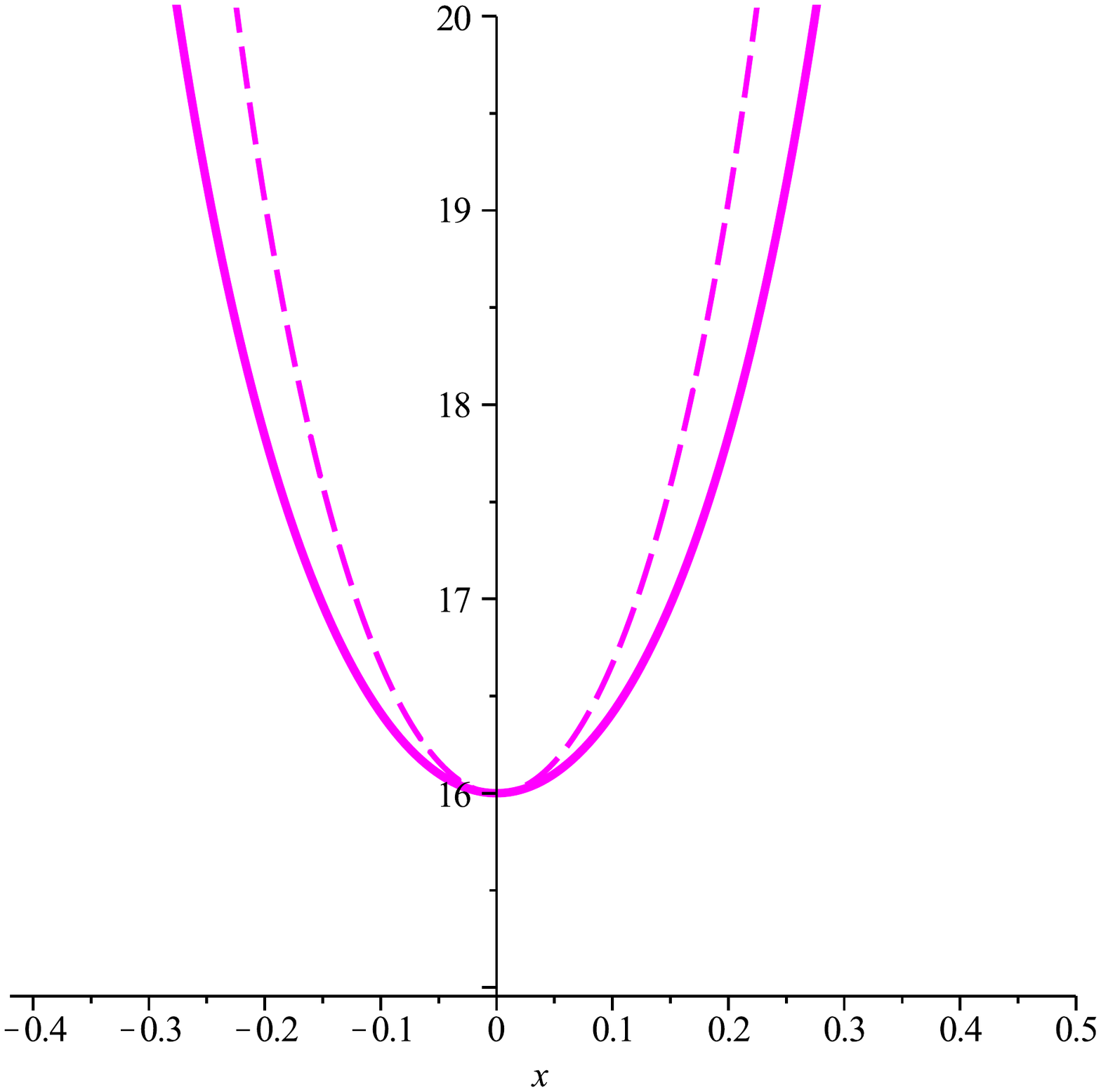}$B.$
\caption{ (color online) A. The slow-roll parameter $\varepsilon$ depending on the value of $\phi$. The solid line represents the plot in the case of the exact Higgs potential, the dashed line represents the plot in the case of the tachyon approximation. B. The slow-roll parameter $|\eta|$ depending on the value of $\phi$. The solid line represents the plot in the case of the exact Higgs potential, the dashed line represents the plot in the case of the tachyon approximation.
}
\label{epsilon and eta}
 \end{figure}

 We can note that near the top of the hill the slow-roll conditions can be satisfied and the scenario of ``new inflation'' is possible. To illustrate this we present the plot for slow-roll parameters $\varepsilon$ and $|\eta|$ depending on the value of $\phi$ in Fig.\ref{epsilon and eta} A and B. As we see near the top the tachyon approximation can represent the exact model.

To get some estimation in the region near
the top one can use the same technique as in the Section 3.

%%%%%%%%%%%%%%%%%%%%%%%%%%%%%%%%%%%%%%%%
\section{Conclusion}
%%%%%%%%%%%%%%%%%%%%%%%%%%%%%%%%%%%%%

 To conclude, let us remind that our logic was the following.
  We started from    nonlocality.  Using the ``stretched" arguments presented by \cite{Lidsey,BarCline} we have got non-positive defined potential. There exist a region where we can use the tachyon approximation.
  For free tachyon there are  4 eras of evolution: i)
 from the past cosmological singularity to reaching the hill,
  ii) rolling from the hill,
  iii) transition from expansion to contraction,
  iv) contraction to the  future cosmological singularity.

  For the Higgs model with an extra negative cosmological constant and small coupling constant
  produces 5 eras of evolution:
  i)
    the ultra-hard regime starting from the past cosmological
    singularity,
  ii) reaching the hill  (here we can use the free tachyon
  approximation),\\
  iii) rolling from the hill (here we can use the free tachyon
  approximation),
  iv) the transition from expansion to contraction
  (here we can use the free tachyon approximation),\\
v) contraction to the  future cosmological singularity.

Let us note that  there are the following differences between dynamics for
 positively defined  and non-positively defined potentials in the FRW cosmology.
 For non-positively defined potential  there is a stage of transition from expansion to
 contraction
and there is no the oscillation regime ending with the zero field configuration.
     To obtain the era of reheating for the Higgs potential with an extra negative cosmological constant one can
     introduce an extra matter interacting with the Higgs field.

%%%%%%%%%%%%%%%%%%%%%%%%%%%%%%%
\section*{Acknowledgements}
%%%%%%%%%%%%%%%%%%%%%%%%%%%%%%

We would  like to thank I.V. Volovich and S.Yu. Vernov for the
helpful discussions. The work is partially supported by grants RFFI
11-01-00894-a and NS -- 4612.2012.1

%%%%%%%%%%%%%%%%%%%%%%%%%%%%%%%%%%

\setcounter{equation}{0}
\end{document}